\documentclass[twocolumn,twocolappendix]{aastex63}

\usepackage{multirow}
\usepackage{lineno}
\usepackage{graphicx}  % needed for figures
\usepackage{dcolumn}   % needed for some tables
\usepackage{bm}        % for math
\usepackage{amssymb}   % for math
\usepackage{amsmath}
\usepackage{longtable}
\usepackage{lipsum}
\usepackage[bottom]{footmisc}
\shorttitle{MCMC lanthanide properties and $r$-process dynamics}
\shortauthors{Vassh et al.}
\begin{document}
%\linenumbers

% Use the \preprint command to place your local institutional report
% number in the upper righthand corner of the title page in preprint mode.
% Multiple \preprint commands are allowed.
% Use the 'preprintnumbers' class option to override journal defaults
% to display numbers if necessary
%\preprint{}

%Title of paper
\title{Markov Chain Monte Carlo Predictions of Neutron-rich Lanthanide Properties \\ as a Probe of $r$-process Dynamics}

\correspondingauthor{Nicole Vassh}
\email{nvassh@nd.edu; nvassh@gmail.com}

\author{Nicole Vassh}
\affiliation{Department of Physics, University of Notre Dame, Notre Dame, IN 46556, USA}

\author{Gail C. McLaughlin}
\affiliation{Department of Physics, North Carolina State University, Raleigh, NC 27695, USA}

\author{Matthew R. Mumpower}
\affiliation{Theoretical Division, Los Alamos National Laboratory, Los Alamos, NM 87545, USA}

\author{Rebecca Surman}
\affiliation{Department of Physics, University of Notre Dame, Notre Dame, IN 46556, USA}

%\date{\today}

\begin{abstract}
Lanthanide element signatures are key to understanding many astrophysical observables, from merger kilonova light curves to stellar and solar abundances. To learn about the lanthanide element synthesis that enriched our solar system, we apply the statistical method of Markov Chain Monte Carlo to examine the nuclear masses capable of forming the $r$-process rare-earth abundance peak. We describe the physical constraints we implement with this statistical approach and demonstrate the use of the parallel chains method to explore the multidimensional parameter space. We apply our procedure to three moderately neutron-rich astrophysical outflows with distinct types of $r$-process dynamics. We show that the mass solutions found are dependent on outflow conditions and are related to the $r$-process path. We describe in detail  the mechanism behind peak formation in each case. We then compare our mass predictions for neutron-rich neodymium and samarium isotopes to the latest experimental data from the CPT at CARIBU. We find our mass predictions given outflows that undergo an extended (n,$\gamma$)$\rightleftarrows$($\gamma$,n) equilibrium to be those most compatible with both observational solar abundances and neutron-rich mass measurements.\vspace{1.2cm}
\end{abstract}

% insert suggested PACS numbers in braces on next line
%\pacs{}
% insert suggested keywords - APS authors don't need to do this
%\keywords{}

%\maketitle

\section{Introduction}\label{sec:intro}

Understanding the nucleosynthesis of lanthanides, at $57\leq Z \leq 71$, is important in order to interpret many astrophysical observables such as the abundances of metal-poor stars \citep{Sneden,Frebel} and merger kilonova light curves \citep{Kasen,Barnes+16}. In order to use lanthanide signatures to probe the origins of heavy element production in our solar system, it is crucial to consider abundances derived from nucleosynthesis calculations. Such calculations connect to production sites by considering the possible outflow conditions present in an astrophysical environment. These efforts are challenged by the uncertain properties of the neutron-rich nuclei synthesized during the rapid neutron capture process ($r$ process). Such nuclear physics uncertainties generate large ranges in the $r$-process nucleosynthetic yields of key lanthanide elements such as Eu ($Z=63$) \citep{CoteGW170817}. Therefore gathering further observational information is alone insufficient to develop a comprehensive picture of lanthanide production since nuclear physics advancements are also needed.

In this work we approach the uncertainties surrounding the synthesis of lanthanide elements in astrophysical environments by taking advantage of statistical methods that consider both nuclear and observational data. Methods to gather data that will illuminate previously unprobed physics are making notable advancements. Nuclear physics serves as a testament to such advancement with facilities such as CARIBU at Argonne National Laboratory, RIKEN, and the upcoming Facility for Rare Isotope Beams (FRIB) expanding the precision and reach of studied properties. 

To take full advantage of current and anticipated data, Bayesian and Monte Carlo statistical methods are expanding their influence in nuclear physics theory. Using these statistical methods to treat complex processes has gained traction for a wide variety of applications \citep{MCMCbookCRobert,MCMCbookBerg,MCMCHandbook}. In nuclear physics theory, these methods have been applied to the determination of properties of superhadronic matter from heavy-ion collisions \citep{SangalinePratt}, to nuclear emission spectra \citep{GulamRazul+2003, Barat+2007},
and to the estimation of thermonuclear reaction rates \citep{Iliadis+2016} and reaction rate uncertainties \citep{deBoer+2014}. Quantum Monte Carlo methods can be coupled with effective field theory to examine a variety of nuclear structure and interaction properties of light nuclei \citep{Carlson+2015} such as explorations of the interactions at play during weak transitions \citep{Pastore+2018} as well as energy levels and level ordering \citep{Piarulli+2018}. Bayesian techniques have also been applied to use mass data as well as mass model predictions to extrapolate properties of exotic nuclei such as the masses of neutron-rich species far from stability \citep{UtamaPiekarewiczMass, Neufcourt+extrap}, even at the neutron dripline \citep{Neufcourt+drip}. Approaches such as these can provide valuable insights to astrophysics since they can inform nucleosynthesis calculations.

In addition to the wealth of current and advancing experimental data, there exist opportunities to take advantage of observational data. For instance, telescope data have been coupled with Markov Chain Monte Carlo (MCMC) methods to infer gas circulation processes in dwarf galaxies from stellar abundances \citep{BenoitMCMC}, to derive cosmological parameters from anisotropies of the cosmic microwave background \citep{PlanckMCMC}, and to infer the probability that Type Ia supernovae will occur in a population of stars \citep{DTDSn1MCMC}. Bayesian methods have also been applied in examining  the crustal composition of neutron stars \citep{UtamaPiekarewiczCrust} as well as the high-density equation of state applicable to neutron stars \citep{Drischler+2020}.

In the era of multi-messenger astrophysics there exist some striking testaments to how observational data can be coupled with statistical methods. Statistical approaches can been used alongside LIGO/VIRGO gravitational wave data to learn about the tidal deformability and mass-radius relation of neutron stars \citep{AbbottPRL,Capano+2020,MillerLamb+2020}, which are connected to the equation of state of dense matter. Bayesian and MCMC methods have also been coupled with X-ray observations \citep{SteinerLattimer+2010,NattilaSteiner+2016,SteinerGlobClus,Goodwin+2019,RaaijmakersNICER,MillerLambNICER,RileyNICER} to learn about the mass, radius, thermal emission, and other properties of neutron stars. A global analysis that takes into account both gravitational wave and X-ray data can also be performed \citep{Zimmerman2020}. Further opportunities for the applications of statistical methods presented by multi-messenger science are exemplified by studies of the GW170817 gravitational wave event coupled with its electromagnetic counterpart \citep{AbbottApJL,AbbottGW170817,Cowperthwaite2017,Villar}. For instance, MCMC methods have been applied to model the ejection dynamics of this event, such as the jet structure \citep{WuMacFadyen}. MCMC methods were also used in some cases to model the kilonova light curve \citep{Cowperthwaite2017} which indicated the synthesis of $r$-process elements through the signatures of high-opacity lanthanides.   

Since interpretations of $r$-process observables are affected by the unknown properties of neutron-rich nuclei, here we apply statistical methods to invert the problem by using solar data to find the nuclear masses that are able to reproduce lanthanide abundances. We focus on the $A\sim164$ enhancement of the solar $r$-process residuals referred to as the rare-earth abundance peak. Note that although the rare-earth elements include all lanthanides as well as scandium ($Z=21$) and yttrium ($Z=39$) here we only include the lanthanides in our analysis. Specifically, we employ the statistical techniques of the Metropolis-Hastings algorithm and MCMC to connect nuclear properties and rare-earth abundances, as first introduced in \cite{REMM1,REMM2}. Here we present how we have evolved this ``reverse engineering" approach and describe the procedure we find to be most suited to our particular MCMC problem. Since the mass predictions given by this method are tied to the astrophysical environment considered, we obtain results given several distinct outflow conditions. By considering distinct outflows, and cross-checking against recent Penning trap mass measurement data for neutron-rich nuclei found by the Canadian Penning Trap (CPT) at CARIBU \citep{jonprc,jonprl,OrfordVassh2018}, this method has the potential to implicate the dominant outflow conditions responsible for the production of rare-earth nuclei observed in our solar system. This work therefore approaches the problem of the uncertain outflow conditions within astrophysical environments from a nuclear physics perspective by examining which outflow dynamics are both consistent with the latest experimental measurements and able to produce not only some lanthanide elements, but the proper elemental and isotopic ratios we observe in our Sun and many other stars.

This paper is organized as follows: in Section~\ref{sec:uncertpeak} we describe why rare-earth abundances are of particular interest, their connection to nuclear physics properties, and the current state of rare-earth element abundance predictions. In Section~\ref{sec:meth}, we describe the MCMC approach implemented to explore neutron-rich masses and show the diagnostic criterion used to gain confidence in our results. Section~\ref{sec:explaincond} describes the three astrophysical outflow conditions considered here and their distinct dynamics. In Section~\ref{sec:results} we present our MCMC results given these three distinct outflows and describe in detail the mechanisms at play in finalizing the abundances in each case. In Section~\ref{sec:compareresults} we compare MCMC predictions for the three cases to the latest mass measurements. We conclude in Section~\ref{sec:conc}.

\section{The uncertain origin of the $r$-process rare-earth abundance peak}\label{sec:uncertpeak}

There exists a well-established connection between the second and third peaks of the solar $r$-process abundance residuals inferred from observational data at $A\sim130$ and $A\sim195$ and the magic numbers at $N=82$ and $N=126$. Since nuclear states at magic numbers have an enhanced stability, during nucleosynthesis $r$-process production sees a `pile-up' of such nuclei, i.e. a tendency of nuclei to capture into such states but then wait for longer timescales to $\beta$-decay or capture out of these states. A question then naturally arises as to whether other features in $r$-process abundances can be linked to pile-up. For instance, the origin of the rare-earth peak, the subtle enhancement in lanthanide abundances between $150\lesssim A \lesssim 180$ with its peak at A$\sim164$, is currently uncertain. In this case, rather than relying on a shell closure, pile-up from an enhanced stability of nuclei could occur due to the presence of a sub-shell closure or nuclear deformation of lanthanide species. In fact, some nuclear mass models predict deformation to be prevalent in many neutron-rich lanthanides; however, the deformation strength, as well as which nuclei are most affected, varies (e.g. \cite{DFTSkM,DFTSLy4,DFTUNEDF0}).

A possible link between the rare-earth peak and the local nuclear structure in neutron-rich rare-earth elements has been the subject of many investigations \citep{Reb97,Matt12}. Since the exact pile-up mechanism capable of dynamically forming the peak during the $r$ process depends on the astrophysical outflow conditions \citep{Matt12,REMM1,REMM2}, the rare-earth peak has the diagnostic power to shed light on the nature of outflows from $r$-process sites. For this dynamical mechanism to operate it is only required that the synthesis produces a main $r$ process, reaching the third peak and not necessarily beyond. Should the astrophysical outflows be either of high enough entropy or sufficiently neutron-rich for synthesis to proceed past the third $r$-process peak and into the actinide region, fission deposition could significantly influence rare-earth abundances. However, since local nuclear features could also affect how fission daughter products settle into their final abundances, neutron-rich rare-earth properties are of relevance in a wide range of outflow conditions. Thus to focus on the influence of solely local nuclear properties in the formation of the rare-earth peak, in this work we consider astrophysical outflows that produce a main $r$ process but do not significantly populate fissioning nuclei. The investigation of outflow conditions that see abundances impacted by fission daughter products will be considered in future work. 

\begin{figure}%
    \centering
     \includegraphics[width=8.4cm]{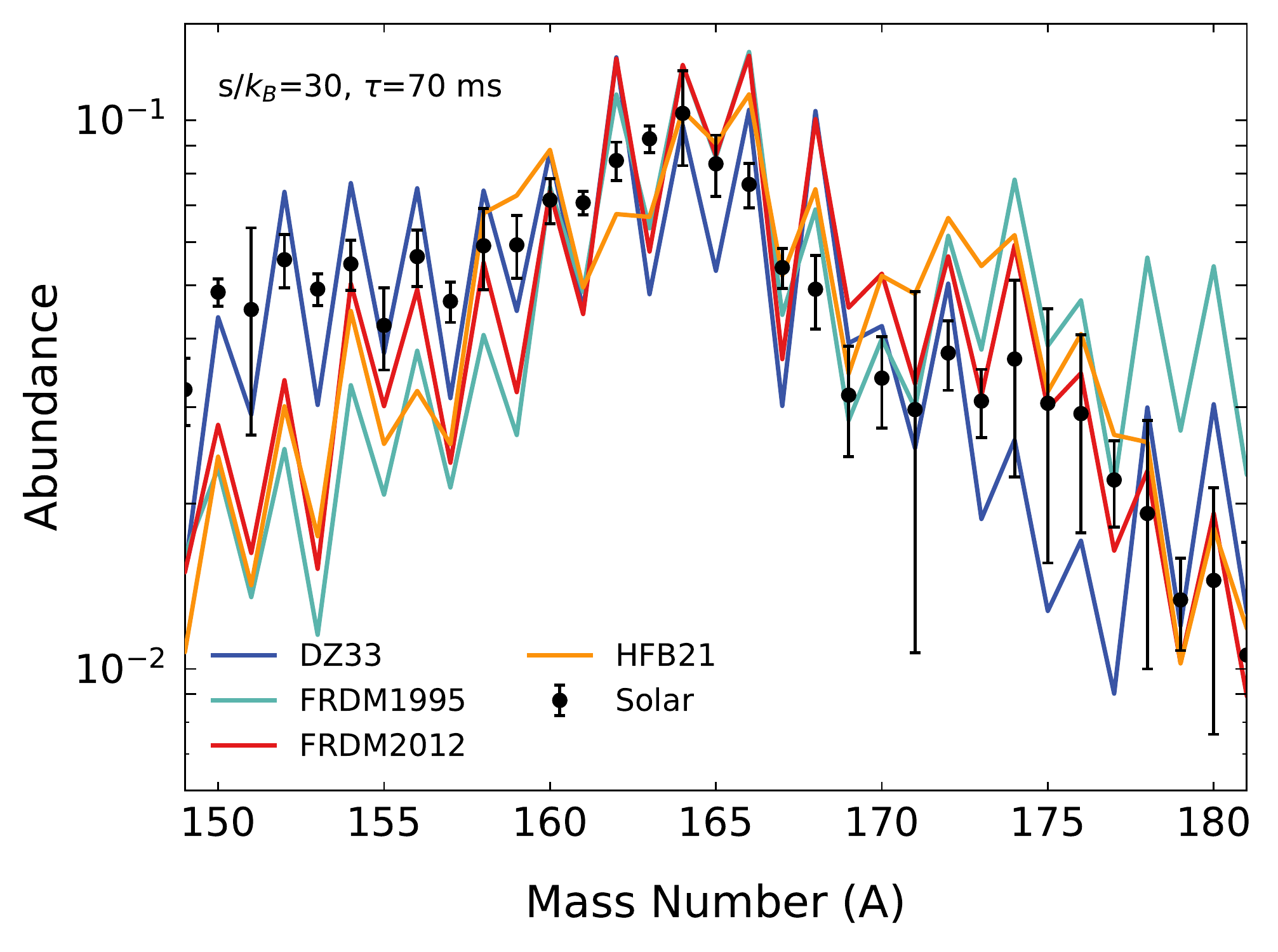}%
     \hspace{0.5cm}
     \includegraphics[width=8.4cm]{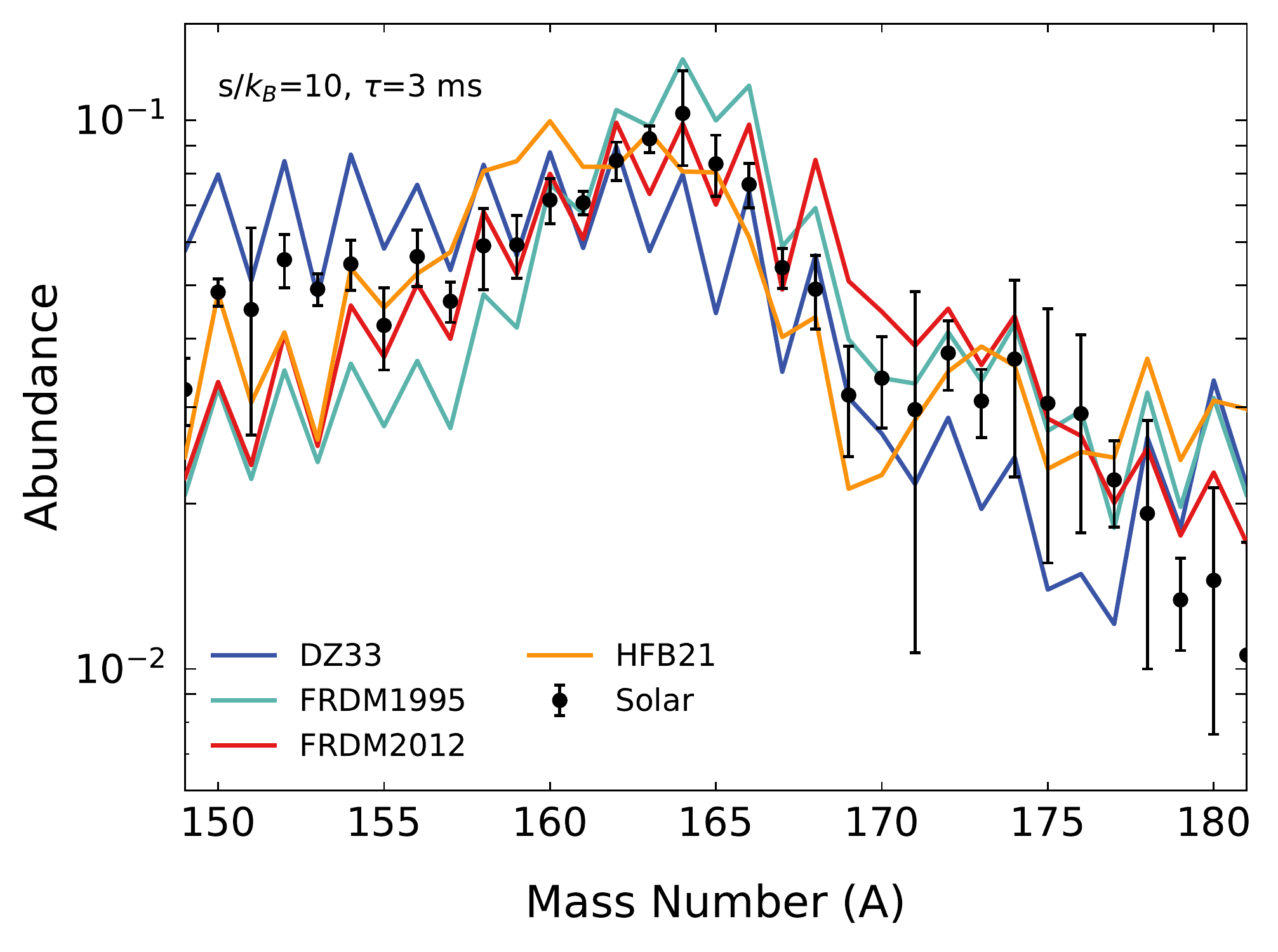}%
   \caption{The predicted $r$-process, rare-earth peak abundances given two distinct astrophysical outflows ($s/k_B=30$, $\tau=70$ ms, and $Y_e=0.2$ (top) and $s/k_B=10$, $\tau=3$ ms, and $Y_e=0.2$ (bottom)) using several mass models commonly applied in $r$-process calculations. The solar data and uncertainties considered in this work are also shown for comparison (see Appendix A).}
\label{fig:massmodelsab}%
\end{figure}

The local nuclear features that can dynamically form the peak could be reflected in the nuclear masses, which ultimately determine the nucleosynthetic outcome via their influence on reaction and decay rates. We therefore focus on nuclear masses in our aim to understand the properties of rare-earth nuclei needed to accommodate rare-earth peak abundances. In Figure~\ref{fig:massmodelsab} we show the calculated abundances of the rare-earth peak using four mass models commonly considered in $r$-process calculations: Duflo-Zuker (DZ, \citealt{DufloZuker}), two versions of the Finite Range Droplet Model (FRDM1995, \citealt{MollerSd0}; FRDM2012, \citealt{FRDM2012}), and Hartree-Fock-Bogoliubov (HFB-21, \citealt{HFB21}). The distinct dynamics of the two types of outflow conditions considered in Figure~\ref{fig:massmodelsab}, which are characterized in terms of their entropy ($s$), expansion timescale ($\tau$), and electron fraction ($Y_e$), are described in detail in Section~\ref{sec:explaincond}.

Although the application of some models can produce a rough rare-earth peak, the models often do not produce a strong enough enhancement, leave the rare-earth peak in the wrong position, and/or miss the more subtle smooth behavior between the abundances of neighboring isotopes. We note that the masses predicted by FRDM1995 are most successful in producing a peak in roughly the correct location for both types of outflows due to a kink in one-neutron separation energies at $N=104$ which creates pile-up \citep{Reb97}. FRDM2012 mass predictions see a significant reduction in this feature predicted by FRDM1995, and therefore a less robust of a peak is produced with the updated model. The DZ model shows a case where predicted deformation in the lanthanides is entirely absent, as evidenced by rare-earth abundances that, on average, are flat. Fortunately, precision nuclear physics measurements are pushing into the lanthanide region relevant for dynamical peak formation \citep{jonprc,jonprl,OrfordVassh2018,JYFLTRAP} with mass data now known for neodymium up to neutron number $N=100$ and for samarium up to $N=102$. Therefore efforts such as those presented in this work aiming to understand astrophysical outflows via the link between mass data and rare-earth abundances are timely.

\section{Method}\label{sec:meth}

Here we outline our method to explore masses capable of rare-earth peak formation and demonstrate the diagnostic metrics used by showing results obtained with the astrophysical outflow condition considered in \cite{OrfordVassh2018}. Our algorithm employs MCMC to perform mass corrections to the DZ mass model. We choose this model as our baseline because of its lack of predicted nuclear deformation in the lanthanide region, which results in calculated rare-earth abundances that, on average, are flat, 
as described in \cite{REMM1,REMM2} as well as the previous section. We make predictions for mass corrections to the DZ mass model with the following mass parameterization \citep{REMM1,REMM2}:
\begin{equation}
M(Z,N) = M_{DZ}(Z,N) + a_N e^{-(Z-C)^2/2f}
\label{eq:param}
\end{equation}

\noindent applied exclusively to the nuclear masses in the rare-earth region. This choice was motivated by past work that considered various nuclear physics inputs and found that it was local features at isotonic values that were most successful at trapping material to form the rare-earth peak \citep{Matt12}. As nuclear theory has progressed, additional work has lent support to the prospect of local features in the lanthanides that manifest in neighboring isotones in a similar way to a mid-shell closure or deformation maximum. For instance, the quadrupole deformation predicted in the lanthanides by some density functional theory models (e.g. Figure~6 of \cite{HorowitzRIB2018}) shows this qualitative behavior where deformation is maximum at a particular $Z$ and falls off at lower and higher proton numbers. Additionally, as can be seen in Figure~\ref{fig:getShifty} in Appendix E, the available experimental mass data for the isotopic chains in the lanthanide region support the trend assumed by Eqn.~\ref{eq:param}. With this parameterization, we randomly vary 28 $a_N$ parameters for neutron numbers $N=93-120$ by generating Gaussian-distributed random numbers with a relative scaling $\sigma\sim0.0095$ MeV. We find that this step size yields an acceptance rate of $\sim20-50\%$ which is ideal for exploring our large, multidimensional parameter space \citep{MCMCbookCRobert}. We fix $C$ based on preliminary runs in which we float this quantity (in this work we use either $C=60$ or $C=58$ depending on the outflow conditions) and set $f=10$ to ensure only local features in the mass surface are produced in order to avoid modifying mass trends near stability (such an $f$ value also works best to reproduce current data trends (see Appendix E)).

After the adjustments to nuclear masses, we then calculate the astrophysical rates for the nuclei near the rare-earth region with $45 \leq Z \leq 69$ at $93 \leq N \leq 120$. Updates to separation energies, $Q$-values, $\beta$-decay rates, and neutron capture rates for the roughly 300 nuclei in this region are performed in a self-consistent manner as in \cite{Mumpower+15, REMM1,REMM2}. We first calculate $\beta$-decay $Q$-values and $\beta$-delayed neutron emission probabilities, $P_n$, using the code BeoH (version 3.3.3) \citep{Mumpower+16}. The majority of the nuclear data calculations are spent updating $P_n$ values. We note that we use experimental data for decays from NUBASE2016 \citep{NUBASE2016} where available in place of the $\beta$-decay predictions derived from our MonteCarlo (MC) masses, but for separation energies we use only the values from our MC procedure in our nucleosynthesis calculations.

Following the updates to separation energies and $\beta$-decay rates, we update photodissociation and neutron capture rates at each time step before calculating the corresponding abundance prediction. We utilize the neutron capture rates predicted by the CoH code \citep{Kawano2016} and perform the fits introduced in \cite{REMM1,REMM2}, which apply the functional form
\begin{equation}
\lambda_{n,\gamma} (Z,N) = \exp[a(N,T)+b(N,T)S_n +c(N,T)S_n^2]
\end{equation}

\noindent where $a$, $b$ and $c$ parameters are dependent on temperature, $T$, and neutron number, and $S_n$ is the one-neutron separation energy, $S_n (Z,N)= M(Z,N-1)-M(Z,N)+M_n$, with $M_n$ being the mass of the neutron. This gives $\lambda_{n,\gamma}$ in units of cm$^3$mol$^{-1}$s$^{-1}$. Photodissociation rates are calculated from neutron capture rates by detailed balance:
\begin{equation}
\lambda_{\gamma,n} (Z,N) \propto T^{3/2} \exp\left[-\frac{S_n(Z,N)}{k_B T}\right] \lambda_{n,\gamma} (Z,N-1)
\end{equation}

\noindent where $T$ is the temperature and $k_B$ is Boltzmann's constant. After the relevant rates have been updated, we write input files for our nucleosynthesis code (PRISM) (as used in \citealt{BDFrp,VasshGEF2019,Cfpaper,TrevorDFT}). Updating nuclear rates to reflect the MC mass values is critical to ensure our predicted abundance pattern reflects as realistic of a set of nuclear data as is possible. 

We then evaluate how well the corresponding abundance pattern fits observed solar data. Before we can compare our calculated abundances to the solar values, an overall scaling must be performed to either the solar data or the predicted abundances, as is standard practice.
Such a scaling is needed since many factors such as site mass ejection lie between solar data and nucleosynthetic predictions. Since the values reported for the solar data are themselves relative numbers often scaled according to the observed amount of silicon, it is the relative abundances that are meaningful when comparing to nucleosynthesis calculations. Thus here we scale calculated abundances by determining the ratio of summed solar abundances in the range $A=150-180$ and summed calculated abundances for the same mass number range. We make use of solar abundances, $Y_{\odot}(A)$, and uncertainties, $\Delta Y_{\odot}(A)$, derived from those given in \cite{goriely99,Arnould07} (see Appendix A). 

To consider the fit to the observational abundance data, we use the Metropolis-Hastings algorithm where the agreement between calculated abundances, $Y(A)$, and solar data, evaluated as $\chi^2 = \sum_{A=150}^{180}\left(Y(A)-Y_{\odot}(A)\right)^2/\Delta Y^2_{\odot}(A)$, guides the evolution of the Markov chain. We note that the number of degrees of freedom used to determine the $\chi^2$ normalization depends on the number of correlations introduced by the parameterization \citep{MCMCbookBerg}, which depends on how these parameters propagate to the final guiding data, the abundances. In our MCMC application masses are propagated to rates and decays, which will affect abundances in a nonlinear, correlated fashion. Therefore since the number of correlations introduced by our parameterization cannot be determined in a physically meaningful way, we use an unnormalized $\chi^2$ when evaluating the likelihood function, $\mathcal{L}\sim e^{-\chi^2/2}$. Since it is the likelihood ratio, $R = \frac{\mathcal{L}_j}{\mathcal{L}_i}$, that determines the acceptance or rejection of a new step, $j$, relative to the previous step, $i$, it is important to recognize that the common factor of the $\chi^2$ normalization will not affect the MCMC evolution. 

\begin{figure}[h!]%
    \centering
     \includegraphics[width=8.75cm]{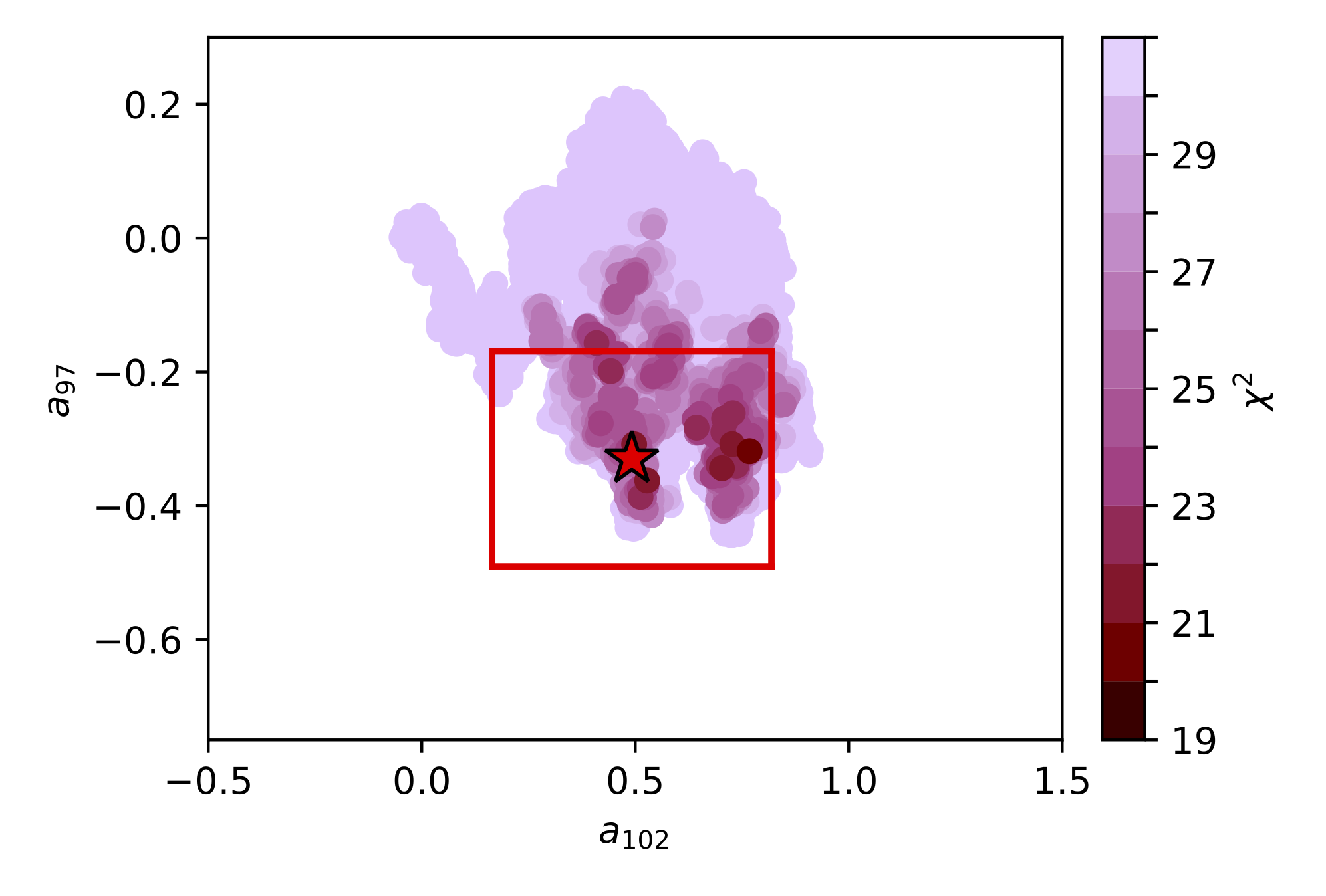} %
     \hspace{0.5cm}
    \includegraphics[width=8.75cm]{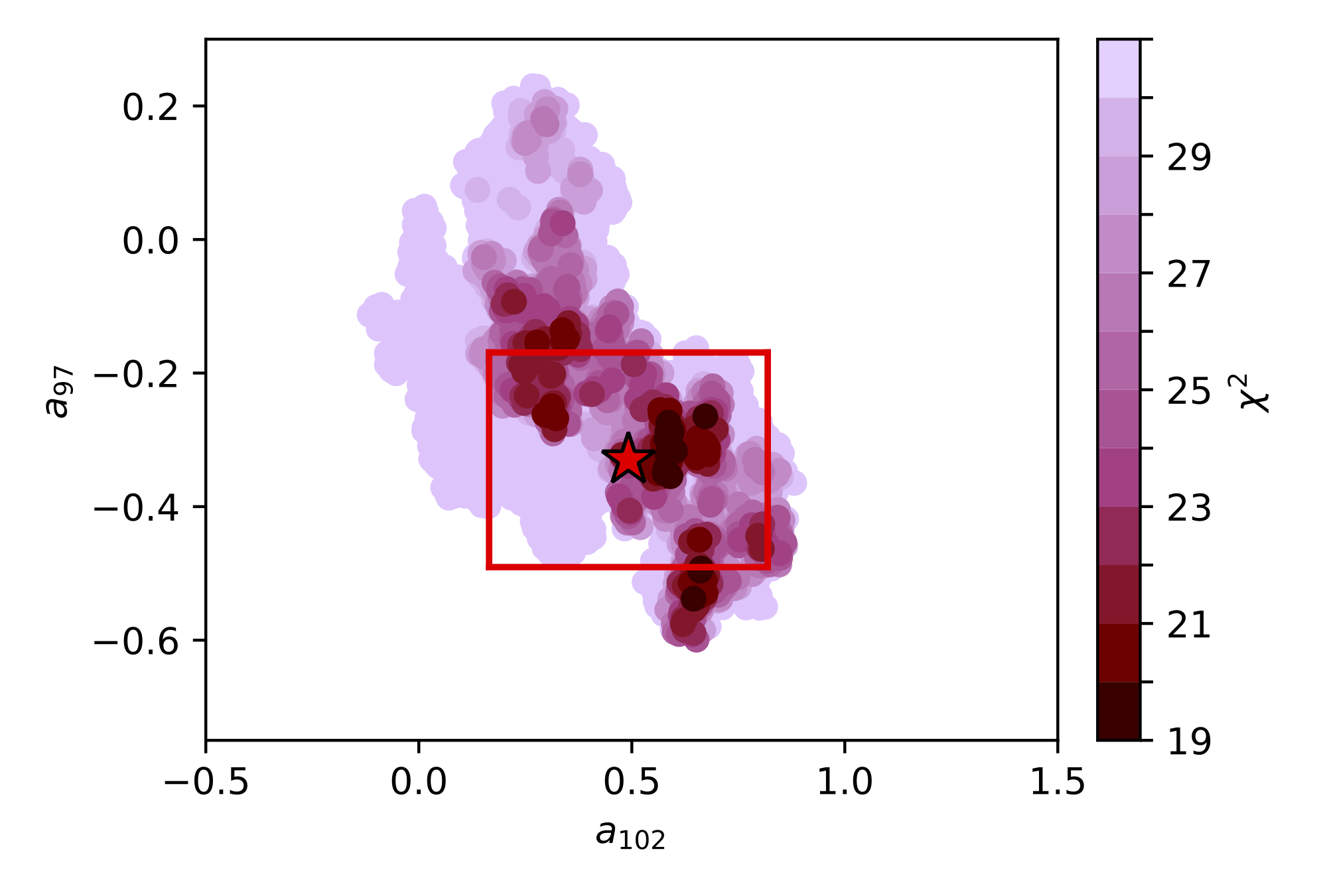} %
    \hspace{0.5cm}
    \includegraphics[width=8.5cm]{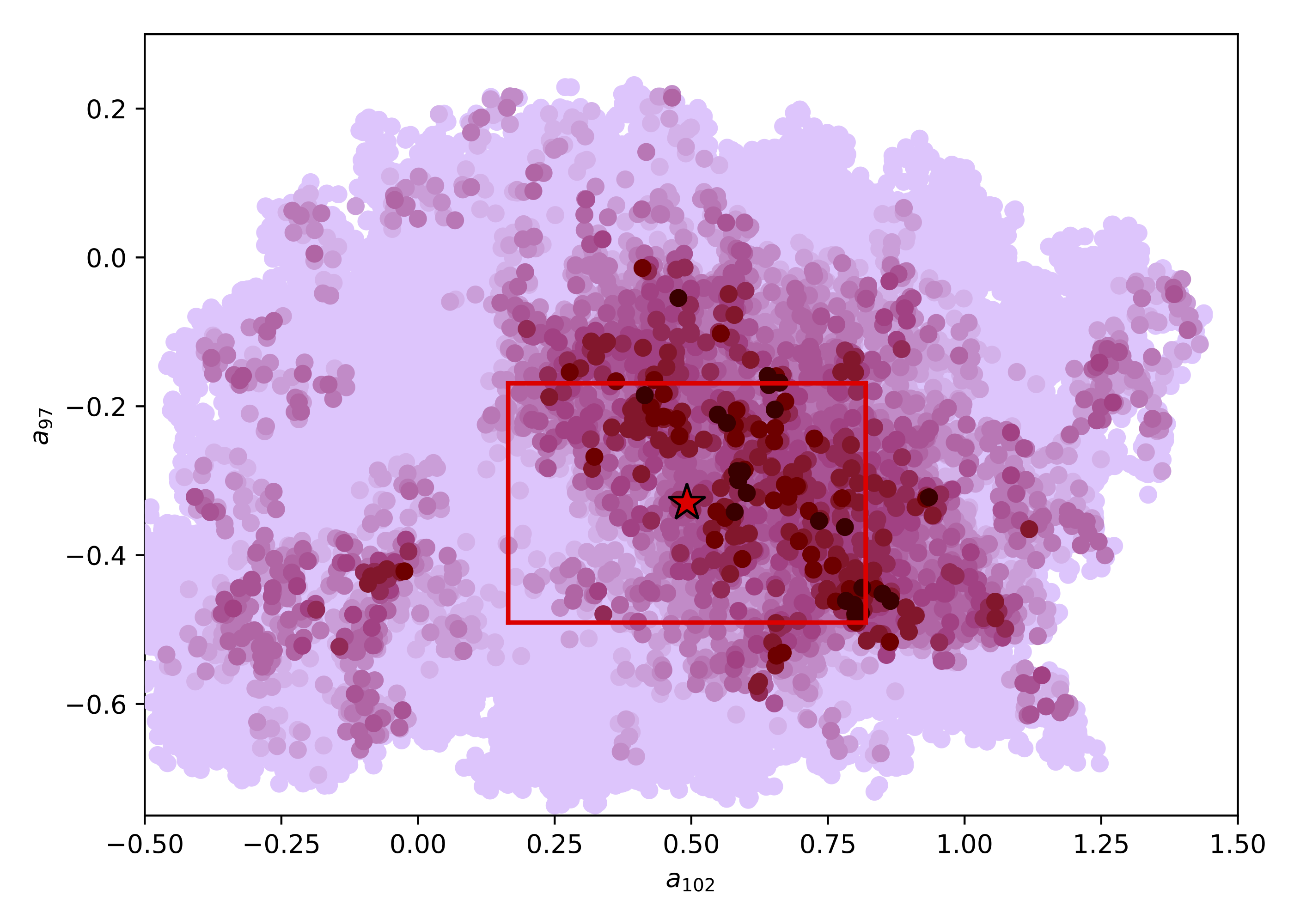}%
   \caption{Every step taken by two representative MCMC runs in the $a_N$ parameter space for $N=102$ and $N=97$ (top and middle panels) colored by the $\chi^2$ of a given step. The complete space searched is demonstrated in the bottom panel, which shows every fourth step for all of the 50 parallel, independent runs used to find the average (red star) and standard deviation (red outlined box) that define the final solution.}
\label{fig:paramspace}%
\end{figure}

The calculation begins from DZ masses such that our MC parameters evolve away from zero to then explore the parameter space for tens of thousands of steps. We then take the solution with the lowest $\chi^2$, i.e. best step, found as the solution from a single MCMC run. Although many solutions with a $\chi^2$ which is significantly lower than that of the initial DZ prediction are readily found, steps with a $\chi^2$ similar to the best step are more unique. This can be seen in the first two panels of Figure~\ref{fig:paramspace} which show all steps taken during the MCMC evolution colored by their $\chi^2$ for two independent MCMC runs. Since this case starts with the DZ masses giving a $\chi^2>200$, we see that many steps find solutions with a $\chi^2$ significantly lower than this, but steps that have a $\chi^2$ comparable to the best step are found in a similar region of parameter space.

\begin{figure}[!h]
\begin{center}
\includegraphics[scale=0.48]{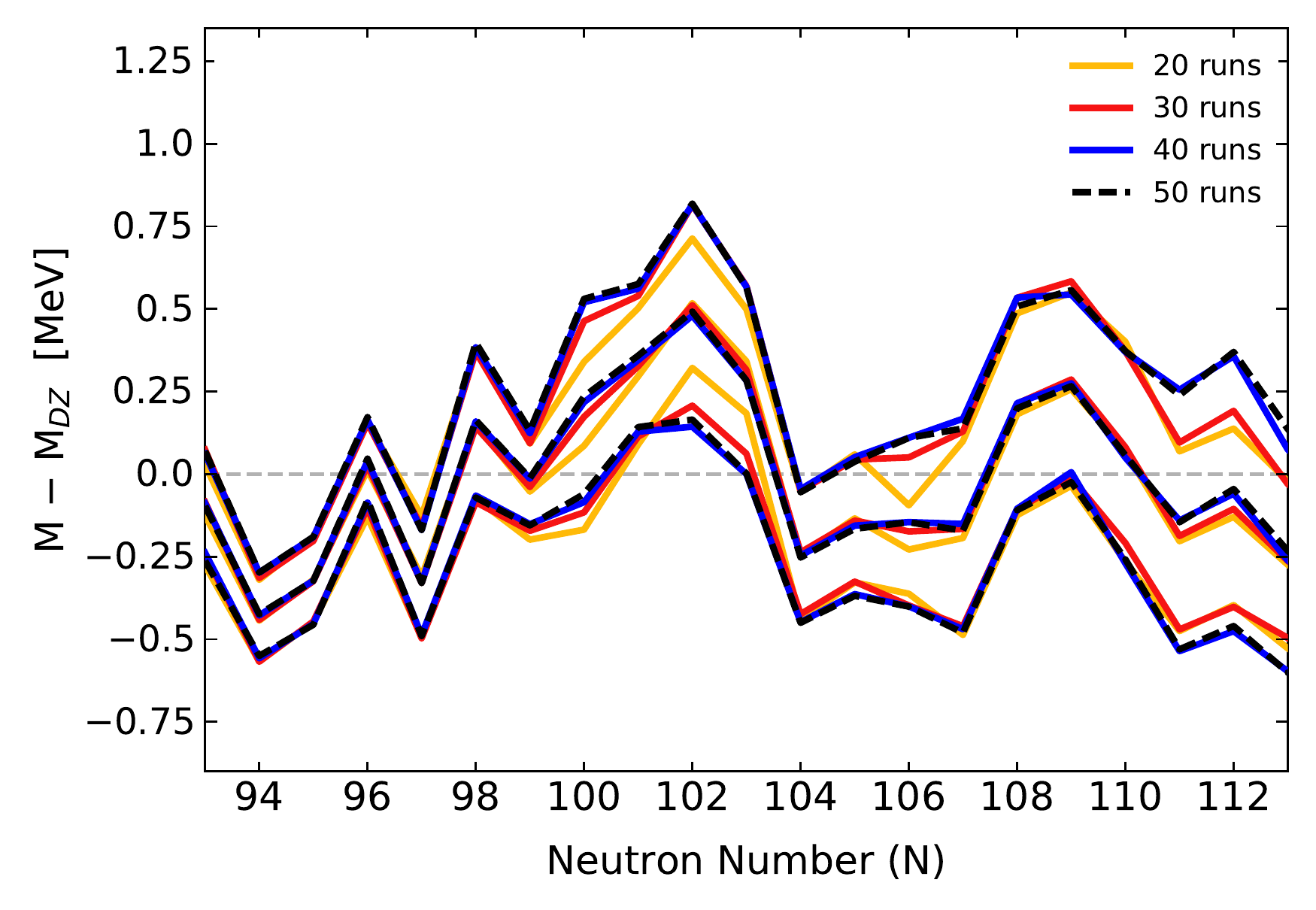}
\end{center}
\caption{The predicted MCMC masses for neodymium ($Z=60$), relative to the DZ mass model, for the astrophysical outflow condition considered in \cite{OrfordVassh2018}. The yellow lines denote the average and standard deviation determined from 20 parallel, independent runs, red lines consider 30 runs, blue lines consider 40 runs, and the black dotted lines show that the solution is converged upon in going from 40 to 50 runs.}
\label{fig:buildstats}
\end{figure}

Since the nuclear rates depend on several MC mass values, and the rare-earth abundances used to calculate the $\chi^2$ are determined by a convolution of nuclear reactions, our parameters become highly correlated. This causes our MCMC procedure to have a long integrated autocorrelation time and slow convergence. Since it is difficult to ensure that an individual run explores the full parameter space, we make use of the parallel chains method of MCMC, which determines the final solution by taking the average and standard deviation of several parallel, independent runs \citep{MCMCHandbook}. This method also has the advantage of providing well-defined errors since uncertainty estimates are not directly dependent on the correlations between MC parameters. Our statistics are therefore determined from the average and standard deviation of the configurations with the lowest $\chi^2$ (best steps) of $50$ independent MCMC runs. Figure~\ref{fig:paramspace} demonstrates that with each run taking a distinct path through parameter space on its way toward its solution with the lowest $\chi^2$, the parallel chains method helps to ensure that the full parameter space is explored. 

Given the use of the parallel chains method of MCMC, we determine convergence to a solution by considering how the average and standard deviation evolve as runs are collected. Figure~\ref{fig:buildstats} demonstrates how the average and standard deviation determined from a set of 20 runs compare to the final full set of 50 runs. The average continues to evolve as statistics are built when considering 20 to 30 runs, but only adjusts slightly after the addition of 10 more runs when moving from a result determined from 30 runs to one determined from 40. However, there exists no significant difference between a 40 run result and 50 run result, implying that the addition of more runs would provide no new information because it would only reinforce what has already been determined from the 50 run search (for a comparison of the full 50 run band to the results of individual runs see Appendix B). Additionally the convergence of the solution found by the parallel chains method of MCMC can be evaluated by comparing the regions of low $\chi^2$ found by the parameter space search of an individual run to the regions of low $\chi^2$ given the full set of all parallel runs (see Figure~\ref{fig:paramspace}). Such an analysis can be used as a diagnostic as to whether an individual run needs to be resumed in order to continue its search and potentially leave its previously identified local minimum. For a discussion of additional convergence diagnostics see Appendix C.

Since we aim to ensure that our mass predictions give results that are consistent with established nuclear properties, we rein in the broad parameter space search by imposing physical constraints. First we consider a comparison with measured mass data by requiring that $\sigma_{rms}(M,M_{AME12}) \leq \sigma_{rms}(M_{DZ},M_{AME12})$ when considering all nuclei with $140\leq A \leq 190$. That is, we require that the root-mean-square (rms) deviation between our mass predictions and AME2012 data \citep{AME2012} is smaller than the deviation between the DZ baseline masses and AME2012. In practice, external checks being applied to the MC parameters in order to veto unphysical solutions are implemented with a modified likelihood function, as is done when considering `truncated' or `censored' data \citep{CensTruncDataBook,EfronCoxLikelihood,ZengCensData}, by including a step function. In our consideration of AME2012 data, the step function has the form
\begin{eqnarray}
\theta(\sigma_{rms}(M,M_{AME12})) = \nonumber \\
\begin{cases}
0, \,\, \text{if}\,\, \sigma_{rms}(M,M_{AME12}) > \sigma_{rms}(M_{DZ},M_{AME12})\\
1, \,\, \text{if}\,\, \sigma_{rms}(M,M_{AME12}) \leq \sigma_{rms}(M_{DZ},M_{AME12}).\\
\end{cases}
\end{eqnarray}

\noindent This condition is checked prior to passing the MC mass values to calculations of reaction and decay rates. If the requirement is not satisfied, new MC mass values are generated.

In addition to the $\sigma_{rms}(M,M_{AME12}))$ requirement, we also consider the odd-even staggering behavior of our MC mass values. This was implemented after initial runs located a number of solutions with an inversion in the odd-even behavior of the one-neutron separation energies (see Appendix D). Since we considered this to be an unphysical mechanism of rare-earth peak formation, we introduced a check regarding the odd-even behavior of our mass solutions using the neutron pairing metric
\begin{equation}
D_n (Z,N) = (-1)^{N+1}(S_n (Z,N+1) - S_n (Z,N)).
\end{equation} 

\begin{figure}
\begin{center}
\includegraphics[scale=0.415]{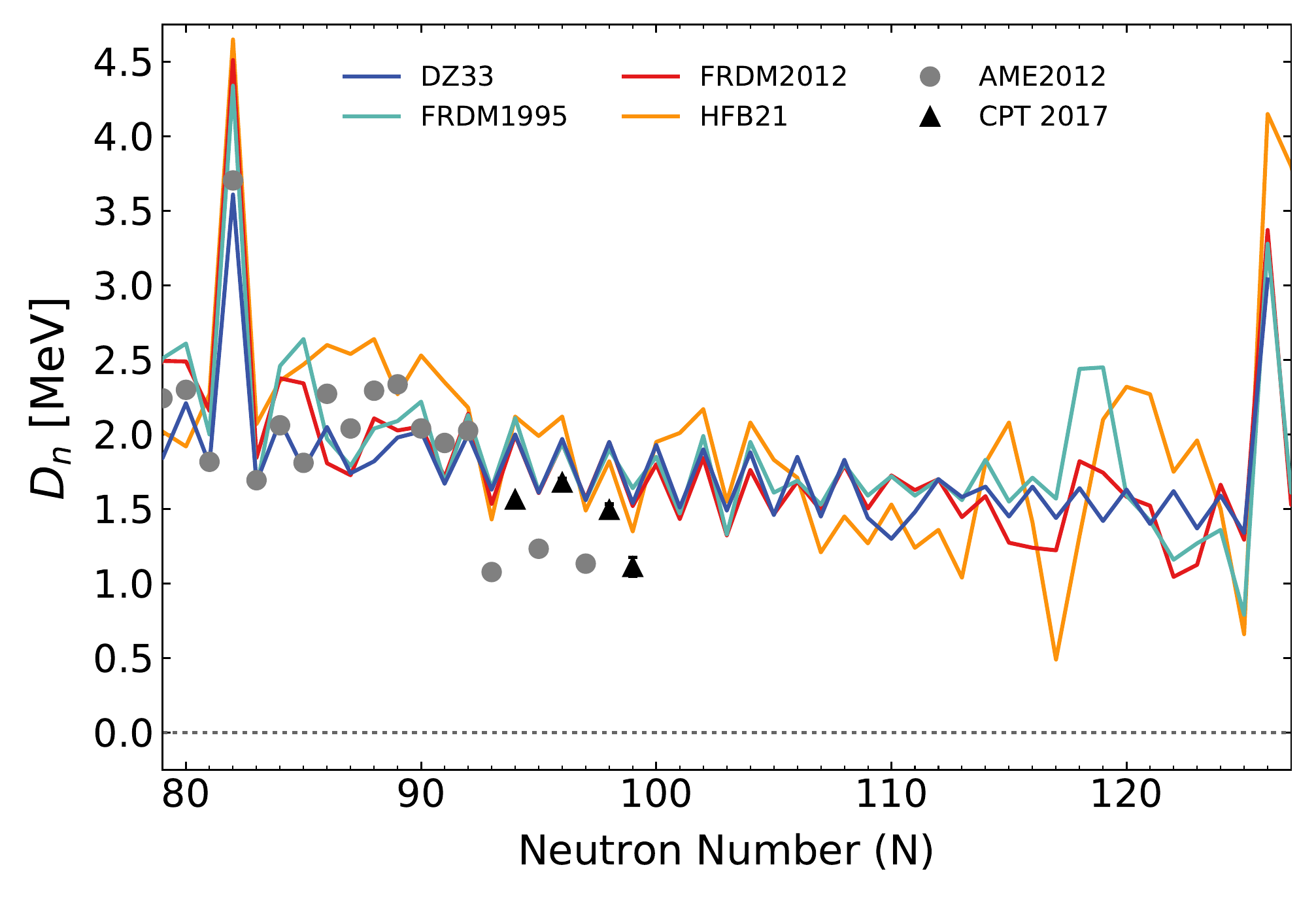}
\end{center}
\caption{The one-neutron pairing metric, $D_n$, for the neodymium chain ($Z=60$) predicted by the models considered in Figure~\ref{fig:massmodelsab} as compared to data from AME2012 \citep{AME2012} and CPT at CARIBU \citep{OrfordVassh2018} data.}
\label{fig:massmodelsDN}
\end{figure}

\noindent As can be seen from Figure~\ref{fig:massmodelsDN}, this metric reveals nuclear structure via sharp transitions between nearby nuclei where local nuclear properties suggest a region to be especially stable, as is clear from predictions at the shell closures $N=82$ and $N=126$. The $D_n$ metric can also hint at nuclear structure effects from pairing or collective effects such as deformation via its features between shell closures. Examining our baseline mass model of DZ demonstrates a case where there is no enhanced stability of rare-earth masses, as evidenced by the purely odd-even behavior of the $D_n$ metric at and around $N=103$. After calculating $D_n$, our algorithm vetoes mass surfaces with an odd-even reversal in their separation energies via modifying the likelihood function to include the step function
\begin{eqnarray}
\theta_1(D_{n}(Z,A)) = \nonumber \\
\begin{cases}
0, \,\, \text{if}\,\, D_{n}(Z,A) \leq 0\\
1, \,\, \text{if}\,\, D_{n}(Z,A) > 0\\
\end{cases}
\end{eqnarray}

\noindent since, as can be seen from Figure~\ref{fig:massmodelsDN}, this metric is predicted to be positive and has never been experimentally observed to have negative values. Additionally, relative to the method described in \cite{OrfordVassh2018}, we include an update to the MCMC procedure to check that along an isotopic chain the value of the $D_n$ metric does not exceed that of the $N=82$ and $N=126$ shell closures (i.e. the height of the largest peaks in Figure~\ref{fig:massmodelsDN}). That is, our modified likelihood function also includes the step function
\begin{eqnarray}
\theta_2(D_{n}(Z,A)) = \nonumber \\
\begin{cases}
0, \,\, \text{if}\,\, D_{n}(Z,A) \geq D_{n, AME12}(Z,Z+82) \\
0, \,\, \text{if}\,\, D_{n}(Z,A) \geq D_{n, DZ}(Z,Z+126) \\
1, \,\, \text{otherwise}. \\
\end{cases}
\end{eqnarray}

\noindent The impact of these $D_n$ metric checks is further discussed in Appendix D. The complete modified likelihood function, which restricts the search to physically meaningful parameters, is then
\begin{equation}
\mathcal{L}' =\mathcal{L}\theta(\sigma_{rms}(M,M_{AME12}))\theta_1(D_{n}(Z,A))\theta_2(D_{n}(Z,A)). 
\end{equation}
 
 \noindent Note that since we use the $\sigma_{rms}$ check against AME2012 data along with the $D_n$ metric checks to reject some combinations of parameters outright before a step is taken, we effectively explore even more of the parameter space than would be implied from examining the steps taken in Figure~\ref{fig:paramspace}.

\section{Distinct Astrophysical Outflows}\label{sec:explaincond}

The nuclear physics feature that our mass adjustments can introduce, such as a sub-shell closure, produces a pile-up of material in order to form the peak. The location where the algorithm finds such a feature to be needed depends on which $r$-process nuclei are dominantly populated when the neutron flux becomes exhausted (freeze-out) and decays to stability begin to take over. Therefore peak formation is determined by two aspects: (1) the initial location of the $r$-process path, i.e. the nuclei most populated along an isotopic chain prior to freeze-out, and (2) the dynamics that govern how the $r$ process proceeds after freeze-out. We therefore considered outflow conditions with distinct behavior: `hot' scenarios in which the path prior to freeze-out is the equilibrium path determined by (n,$\gamma$)$\rightleftarrows$($\gamma$,n) equilibrium (i.e. the Saha equation), and for which photodissociation continues to play a role after freeze-out, and `cold' scenarios in which (n,$\gamma$)$\rightleftarrows$($\gamma$,n) equilibrium fails before the path populates the rare-earth region; we therefore see nuclei closer to the dripline more strongly populated prior to freeze-out, and find little to no influence from photodissociation after freeze-out. We consider such hot and cold scenarios for parameterized outflows that are moderately neutron-rich and low in entropy that will undergo heavy element nucleosynthesis. We emphasize that although considering the heating introduced by nuclear reactions can sometimes make cold dynamics differ from their behavior when such reheating is neglected, this is not the case with all cold scenarios. In fact we find that several scenarios can retain their cold behavior after including the reheating during the nucleosynthesis calculation, and thus cold dynamics remain a physically realizable possibility in astrophysical environments. The outflow conditions considered in this work are all examples that find nuclear reheating to have little to no influence on the expansion dynamics.

Guided by merger simulations, we adopt three distinct types of outflows \citep{Surman+08,Metzger+2008,Perego+14,Fernandez+15,Just+15,Radice18}, which could take place in both accretion disk and dynamical ejecta: (1) a hot outflow with an entropy ($s$) of $30\,k_B$/baryon and a dynamical timescale ($\tau$) of $70$ ms, (2) a cold outflow with $s =10\,k_B$/baryon and $\tau = 3$ ms, and (3) a `hot/cold' outflow with $s = 20\,k_B$/baryon and $\tau = 10$ ms. Here we call this a `hot/cold' outflow since it starts out characterized by hot $r$-process dynamics, and therefore the $r$-process path prior to freeze-out is the equilibrium path, but behaves similar to a cold outflow after freeze-out. All outflows considered here are moderately neutron-rich with an electron fraction ($Y_e$) of 0.20. These outflow parameters are summarized in Table \ref{tab:outflows}. We note that in \cite{OrfordVassh2018} we investigated whether our MCMC result given outflow (1) was a viable solution in cases with similar outflow properties by considering slight adjustments to the entropy and expansion timescale. We found that indeed similar expansion dynamics would require similar mass predictions in order to form a rare-earth peak comparable to the solar data. Therefore, since similar outflows require similar masses, the differences in required masses given distinct outflow conditions such as those in Table \ref{tab:outflows} can be used to discern the type of outflows capable of accommodating both peak formation and the latest mass measurements.

\begin{table}[!t]
\caption{Ejecta Outflow Parameters.}
\label{tab:outflows}
\centering
\begin{ruledtabular}
\begin{tabular}{lcccccc}
Outflow Type & Entropy (s/$k_B$) & Timescale (ms) & $Y_e$ \\ 
\hline
Hot &  30  & 70 & 0.2 \\ 
Hot/Cold & 20 & 10 & 0.2  \\ 
Cold & 10 & 3 & 0.2 \\ 
%\botrule
\end{tabular}
\end{ruledtabular}
\end{table}

All conditions considered here are such that a similar amount of material, that is, a comparable summed mass fraction, reaches the third peak at $N=126$ and beyond. This summed mass fraction is low relative to that suggested by solar abundances since the conditions adopted here were chosen due to their high lanthanide mass fractions. We find that considering the same parameterized conditions with a slight increase in neutron-richness ($Y_e\sim0.17-0.19$) permits the summed mass fractions for the third peak and beyond to be comparable to solar ratios. Importantly we note that the mass solutions found to form the rare-earth peak at $Y_e=0.2$ were still found to support peak formation at these lower $Y_e$ values of $0.17-0.19$ (as is expected from abundance results when outflow parameters were slightly varied in  \cite{OrfordVassh2018}). However, with lower $Y_e$ shifting material to the higher mass numbers at the third peak and beyond, of course the height of the second peak drops. Therefore the specific conditions chosen here are not meant to be able to accommodate the full pattern of $r$-process abundances, but rather represent cases with high lanthanide production that will dominate the total abundance in the rare-earth region. We emphasize that although site ejection will actually be a mass-weighted mixture of outflow conditions, the aim of exploring individual trajectories here is to examine the dynamics that are dominant, since similar dynamics will require similar mass solutions.

\begin{figure}%
    \centering
     \includegraphics[width=8.4cm]{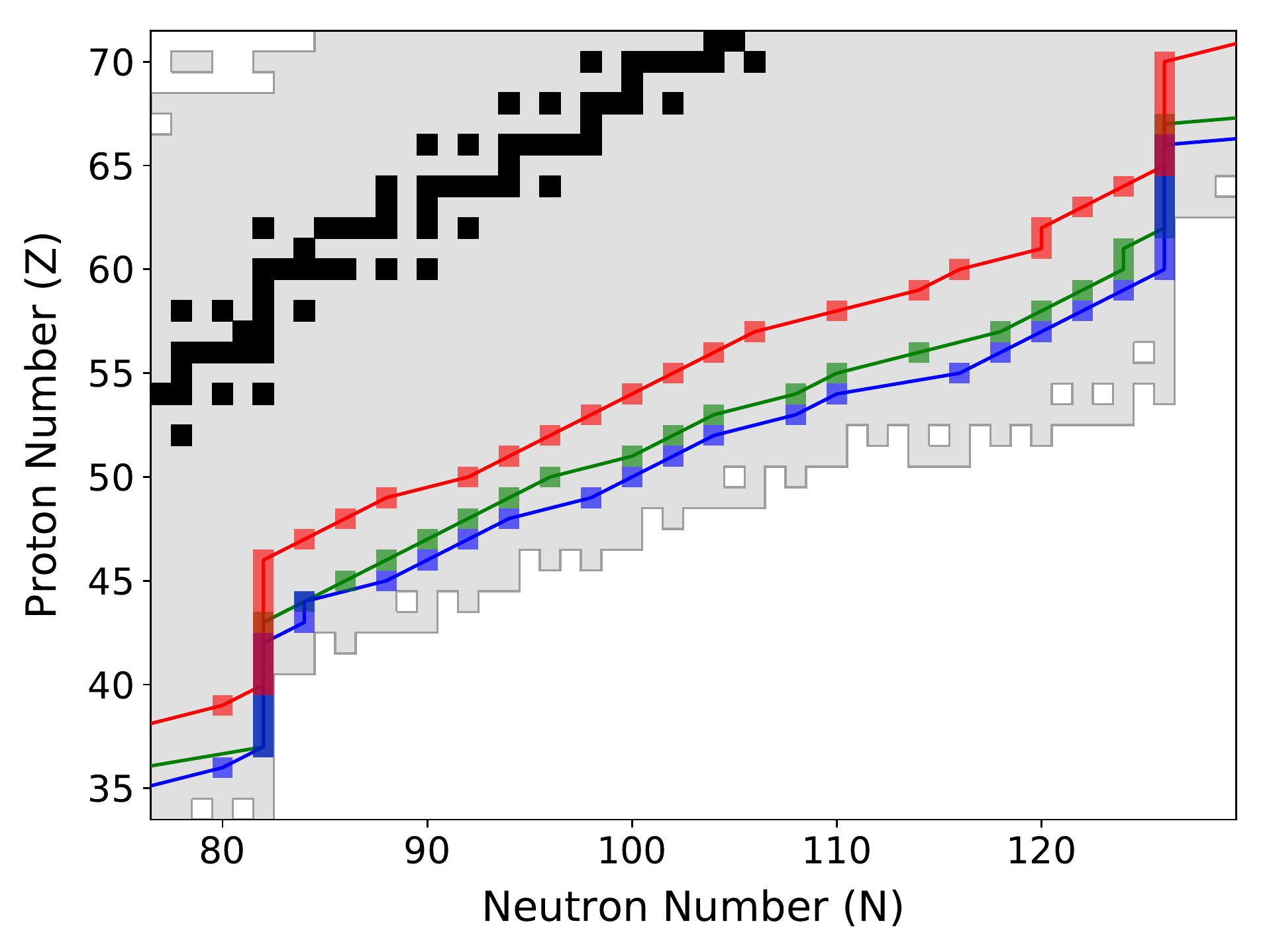}%
     \hspace{0.5cm}
    \includegraphics[width=8.65cm]{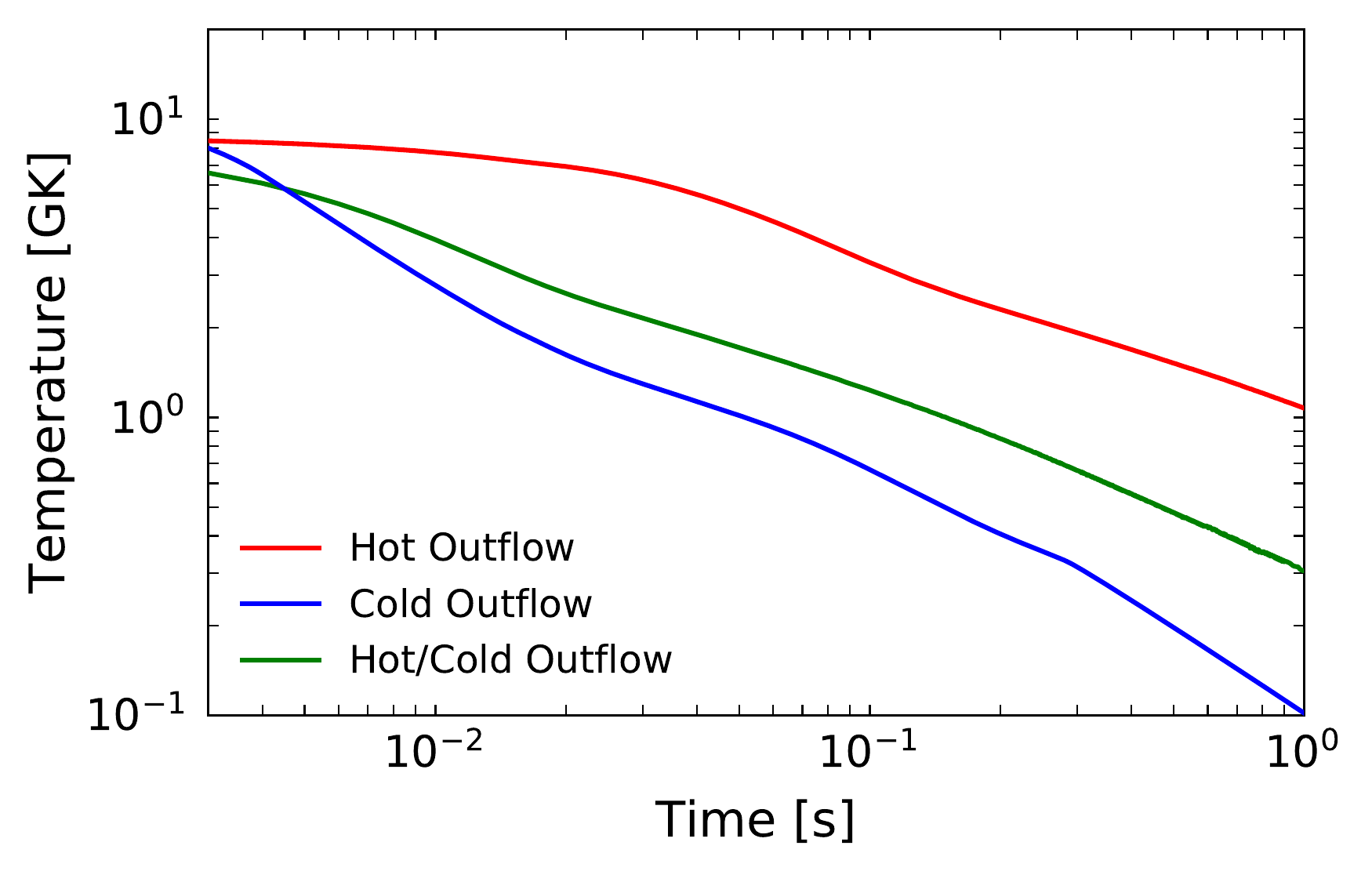}%
   \caption{(Top) A snapshot of the $r$-process path just after reaching the third peak for the `hot' (red), `cold' (blue), and `hot/cold' (green) outflows considered here. The grey region denotes the Duflo-Zuker dripline and black squares are stable nuclei. (Bottom) The temperature evolution as a function of time for these astrophysical trajectories.}
\label{fig:conditions}%
\end{figure}

\begin{figure*}
\begin{center}
\includegraphics[scale=0.54]{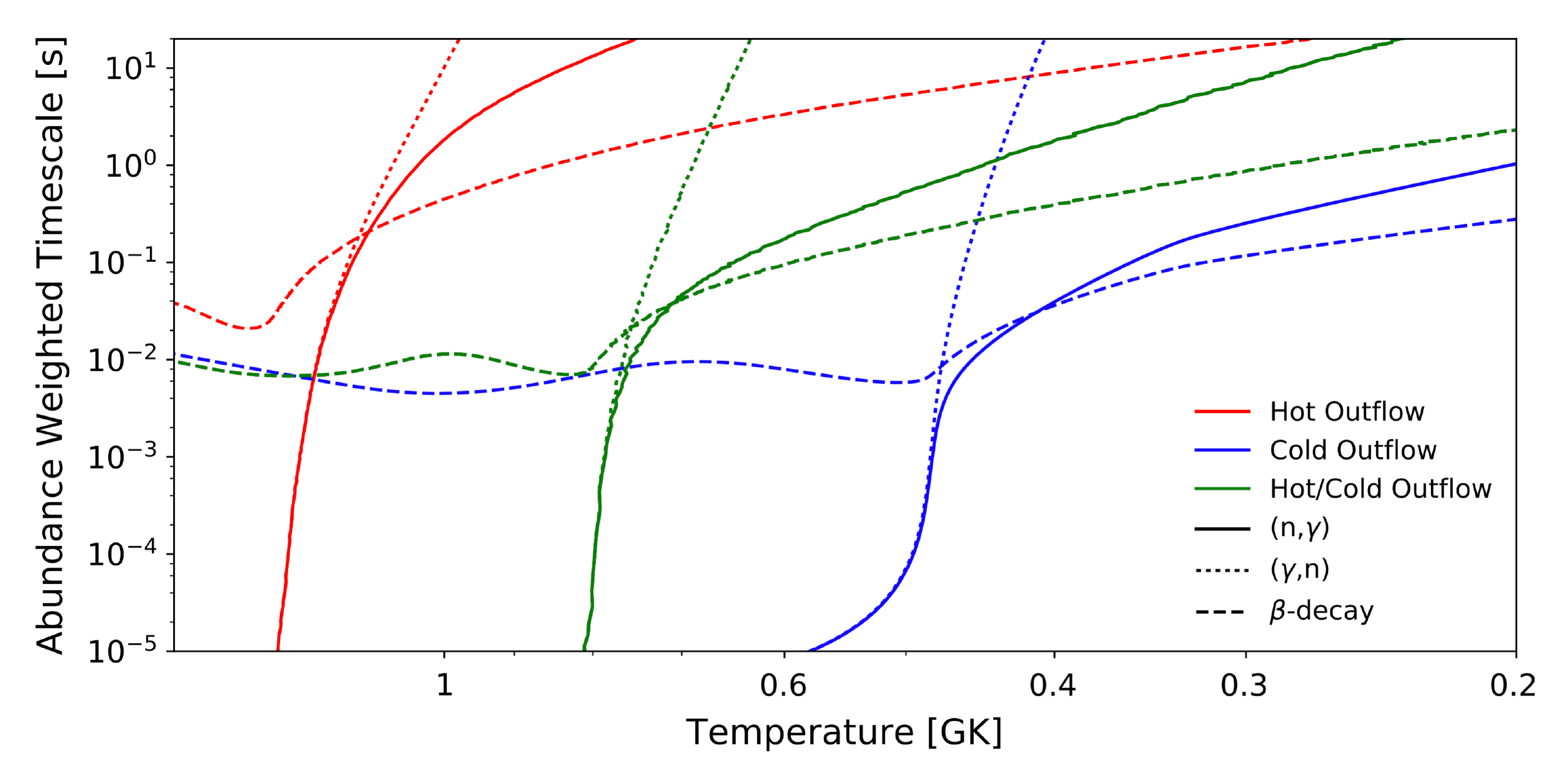}
\end{center}
\caption{The abundance-weighted timescales for neutron capture (solid lines), photodissociation (dotted lines), and $\beta$-decay (dashed lines) as a function of temperature for all three outflow scenarios considered.}
\label{fig:windtimescales}
\end{figure*}  

The temperature evolution, as well as a snapshot of the $r$-process path just as material begins to populate the third peak region at $N=126$, is shown in Figure~\ref{fig:conditions} for these three cases. In the hot outflow, (n,$\gamma$)$\rightleftarrows$($\gamma$,n) equilibrium persists throughout the $r$ process, corresponding to a path closer to stability than in the cold outflow. In contrast, the cold outflow sees photodissociation fall out of equilibrium early and relies almost entirely on $\beta$-decay to compete with neutron capture, making the path lie closer to the dripline at the time when material begins to populate the third peak region. In this case the most populated nuclei in the rare-earth region are not well represented by the equilibrium path even at early times. For the hot/cold outflow, although the $r$-process path prior to freeze-out is the equilibrium path determined by (n,$\gamma$)$\rightleftarrows$($\gamma$,n) equilibrium, the nuclei populated at the time when material begins to populate the third peak region lie closer to the dripline since the hot/cold outflow has a lower initial entropy (i.e. the entropy reported in Table \ref{tab:outflows} which is the value prior to breakout from nuclear statistical equilibrium) than the hot outflow. This implies that the hot/cold case is more dense than the hot case at a given temperature, such that the Saha equation sets an equilibrium path further out in neutron number (at nuclei with lower separation energies). 

The distinct nature of the three outflows considered here can be best understood by examining Figure \ref{fig:windtimescales} which shows the abundance-weighted reaction timescales ($\sim$1/rate) for the reaction and decay channels of importance in these cases. This shows that (n,$\gamma$)$\rightleftarrows$($\gamma$,n) equilibrium dominates early-time dynamics, with abundances at a time prior to freeze-out being well represented by the equilibrium path. However, this equilibrium fails in the cold case very early, before the production of $A\sim 195$ nuclei. These three outflows also find themselves at very different temperatures when freeze-out begins. For instance, the hot/cold outflow has a comparatively low temperature at freeze-out ($\sim$0.8 GK) relative to that of the hot outflow ($\sim$1.2 GK). 
Additionally, the distinct post-freeze-out dynamics shown in Figure~\ref{fig:windtimescales} will vary the influence of late-time neutron capture, photodissociation, and $\beta$-decay in shaping the final rare-earth peak abundances. 
Therefore, since both the population of nuclei along the $r$-process path prior to freeze-out and the dynamics after freeze-out vary among the three outflows described here, we expect these cases to require distinct MCMC mass solutions in order to form the rare-earth peak.
\\
\\
\section{Results}\label{sec:results}

For each of the three outflow conditions discussed in the last section, we obtain $50$ independent, parallel MCMC runs and perform extensive testing and analysis on the MCMC solutions found. In each case, all $50$ runs are compared to evaluate how low of a $\chi^2$ is attainable for the specific outflow being examined. Runs that did not attain a $\chi^2$ around this value are resumed for more time steps to further explore the parameter space. An additional handful of runs, typically ones that found the solutions with the lowest $\chi^2$ of all runs but also had good movement through parameter space, are also resumed for roughly twice as long as the standard run in order to gain further confidence that the solution found was in fact the global rather than local solution. The results from this procedure are presented in this section along with detailed descriptions of the physical mechanism by which the solution for each of the three distinct outflows operates. A summary and comparison of these solutions is presented in the next section. 

\subsection{Hot Outflows}\label{sec:hotcond}

\begin{figure}[h!]%
    \centering
    \includegraphics[width=8.65cm]{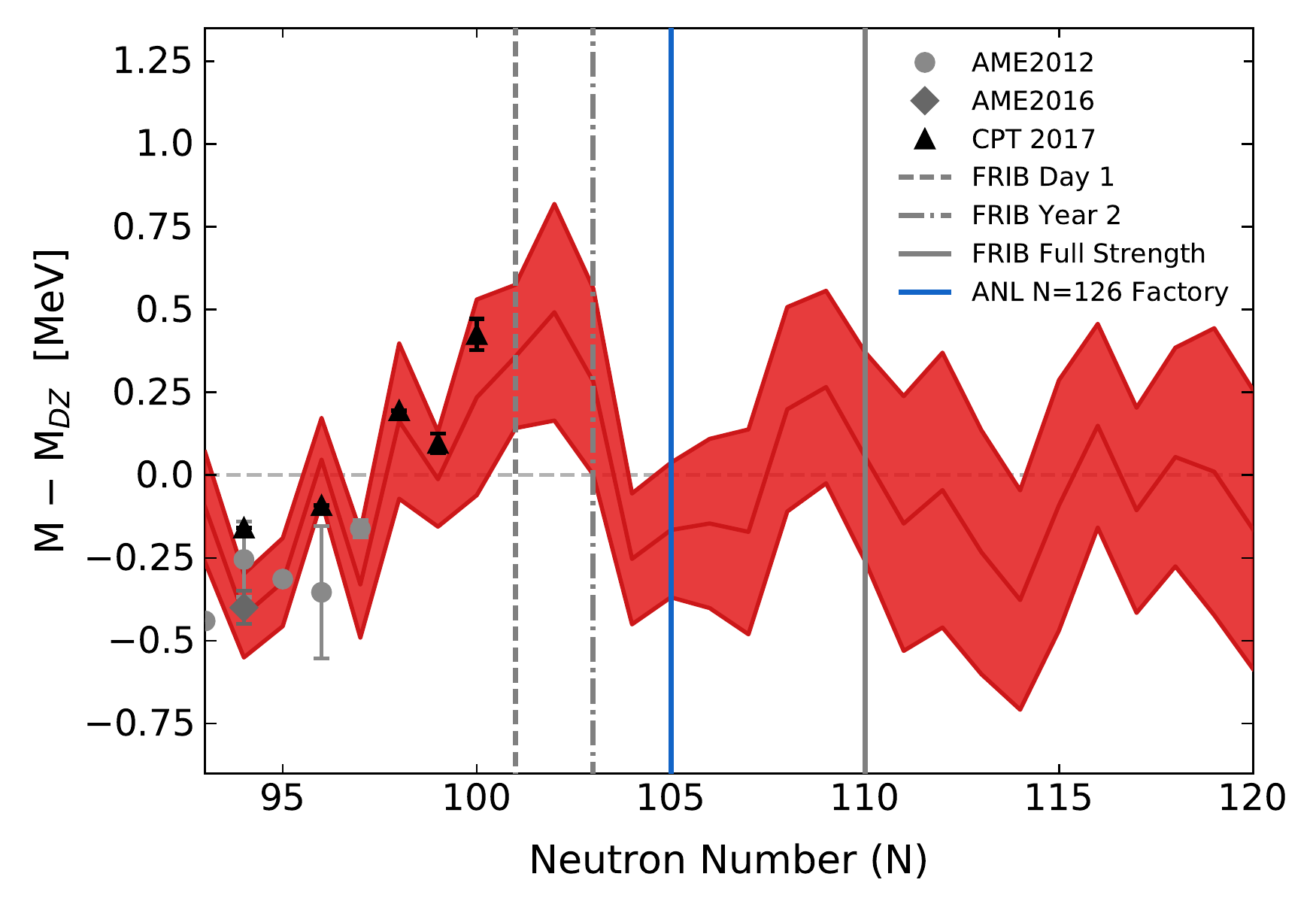}%
    \hspace{0.5cm}
    \includegraphics[width=8.2cm]{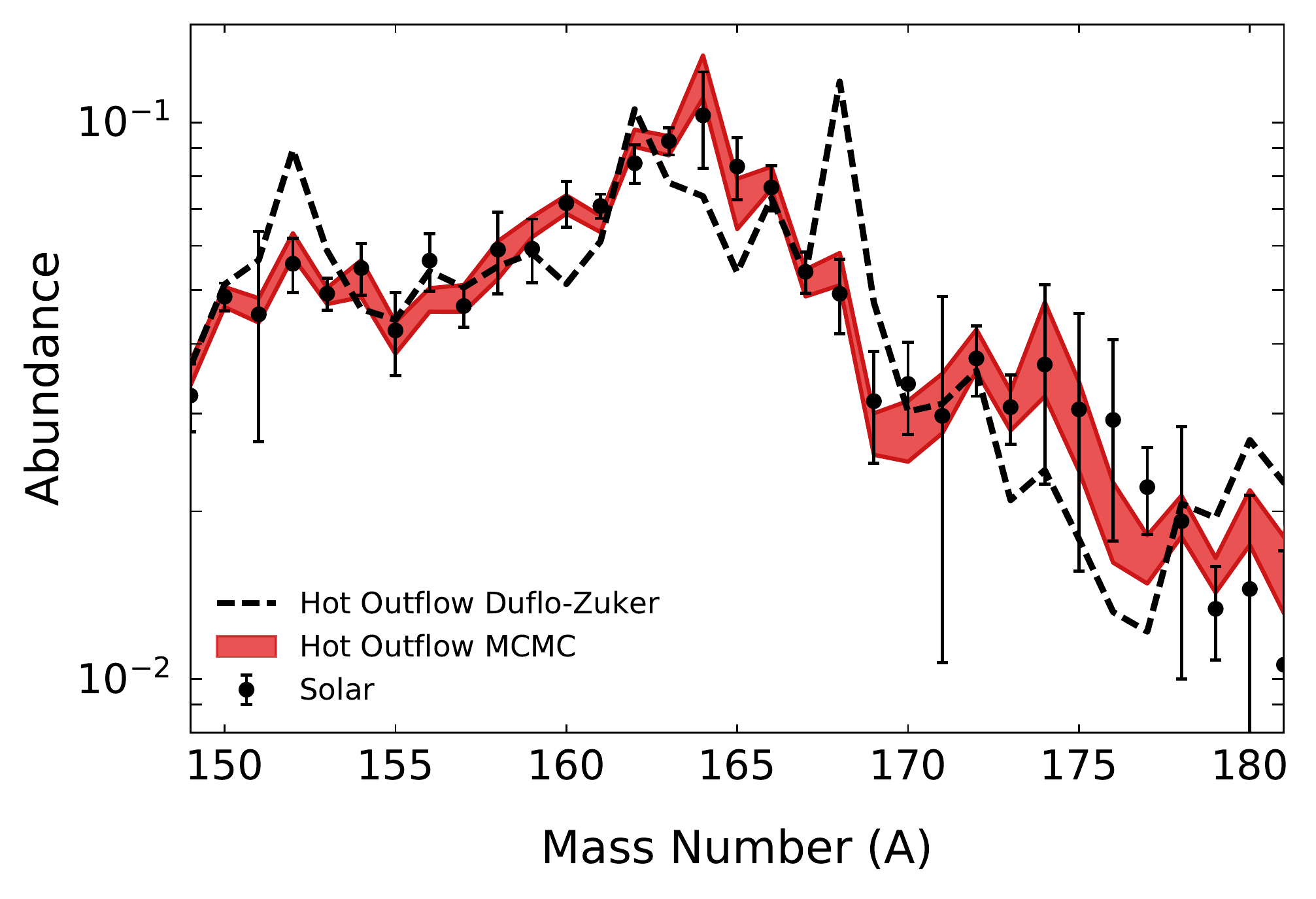}%
   \caption{(Top) The predicted MCMC masses for neodymium ($Z=60$), relative to the Duflo-Zuker mass model, for the hot outflow (red band). The band is produced from the average and standard deviation of results from 50 parallel, independent runs. For comparison the AME2012 data used to guide the calculation are shown, along with more recent data from AME2016 \citep{AME2016} and the CPT at CARIBU \citep{OrfordVassh2018} of which the calculation was not informed. Potential future experimental reaches are shown as vertical lines. (Bottom) The standard deviation of the abundance patterns given by our 50 MCMC runs (red band) as compared to the baseline prediction using DZ (dashed black line).}
\label{fig:masssurfhot1}%
\end{figure}

\begin{figure}
\begin{center}
\includegraphics[scale=0.415]{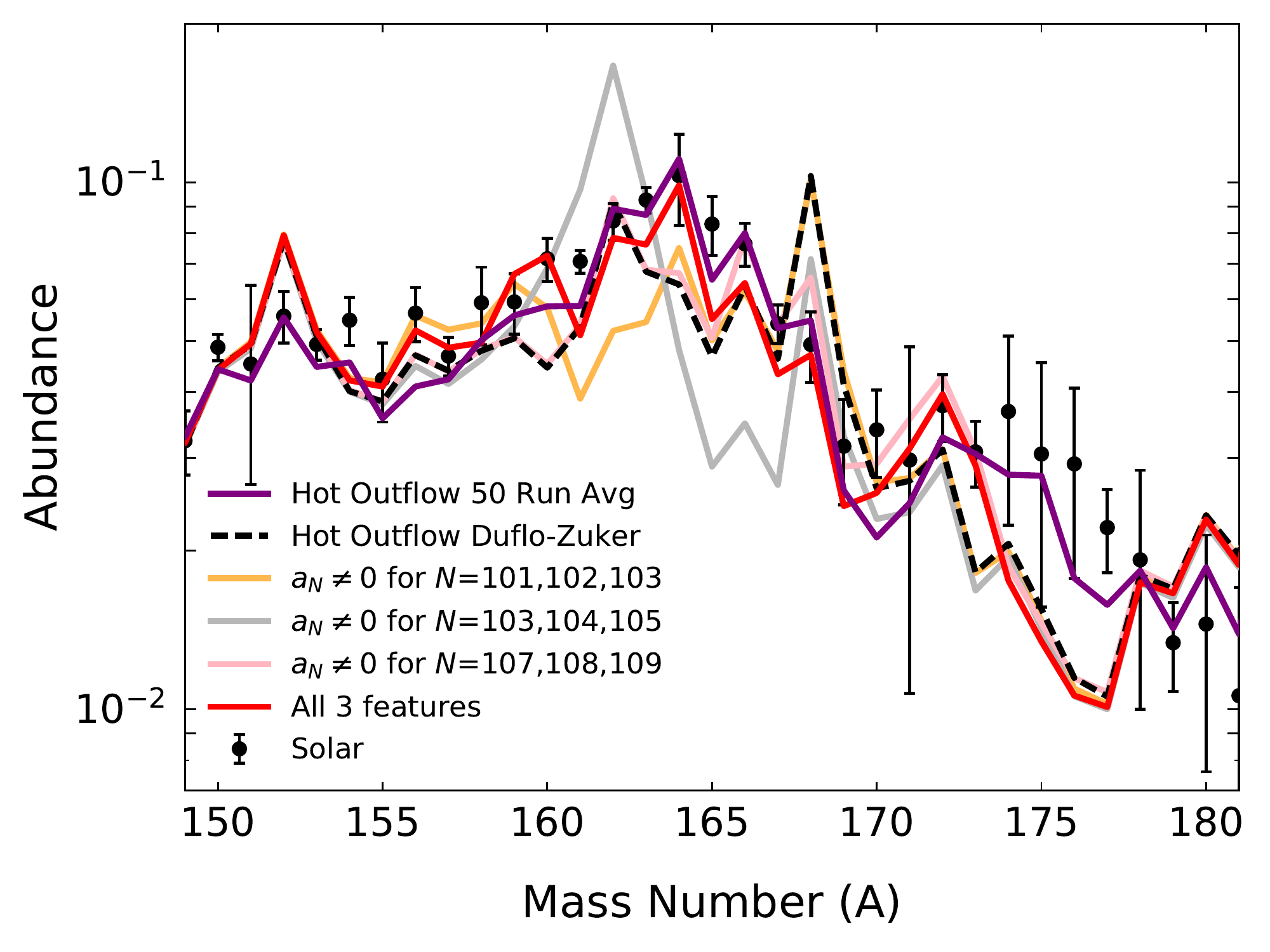}
\end{center}
\caption{The abundance pattern using the average mass values from Figure~\ref{fig:masssurfhot1} as compared to when only the $N=102$ or $N=104$ or $N=108$ key features are applied in an $r$-process calculation. The result when only these three features are combined is also shown.}
\label{fig:hotfeatures}
\end{figure}

Results for mass predictions and abundance patterns for the hot outflow are shown in Figure~\ref{fig:masssurfhot1}. Note that Figure~\ref{fig:masssurfhot1} shows an updated result relative to that in \cite{OrfordVassh2018} since here further $D_n$ metric checks were implemented here. However, this did not significantly affect the MCMC solution for this case (see Appendix D). As described in \cite{OrfordVassh2018}, this solution utilizes a pile-up at $N=104$ in order to form the rare-earth peak. In these hot outflow conditions, pile-up occurs because the updated separation energies produce a kink in the equilibrium path at $N=104$, due to the dip in the mass surface shown by the red band in Figure~\ref{fig:masssurfhot1}. This $N=104$ dip in the mass surface may be accessible by next-generation experiments such as the N=126 Factory and FRIB at full beam strength. The rise in the mass surface at $N=102$ before the dip is also crucial to the solution in order to place the peak center at $A=164$, as can be seen in Figure~\ref{fig:hotfeatures}. The upturn at $N=102$ makes these nuclei less stable than $N=104$ nuclei, promoting further pile-up here at early and late times. With only the dip from $N=103-105$, a peak that is too strong and off-center is produced.

\begin{figure}
\begin{center}
\includegraphics[scale=0.415]{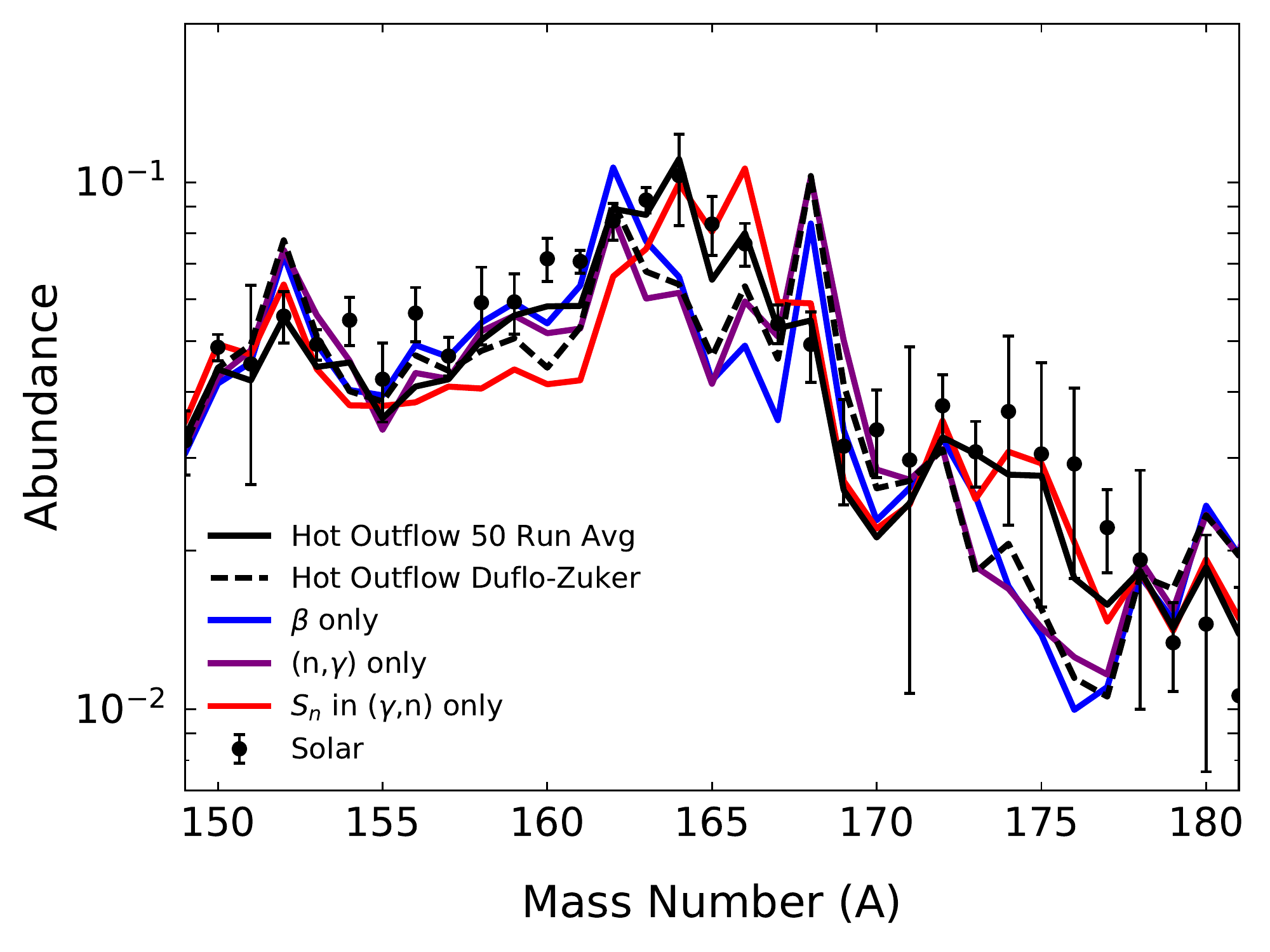}
\end{center}
\caption{The abundance pattern from implementing the updated neutron capture, photodissociation, and $\beta$-decay rates determined from the average mass values in Figure~\ref{fig:masssurfhot1} as compared to when adjustments to only $\beta$-decay or neutron capture or the separation energies in the detailed balance equation for photodissociation are applied.}
\label{fig:hotbetasnncap}
\end{figure}

\begin{figure*}
    \centering
    \includegraphics[scale=0.45]{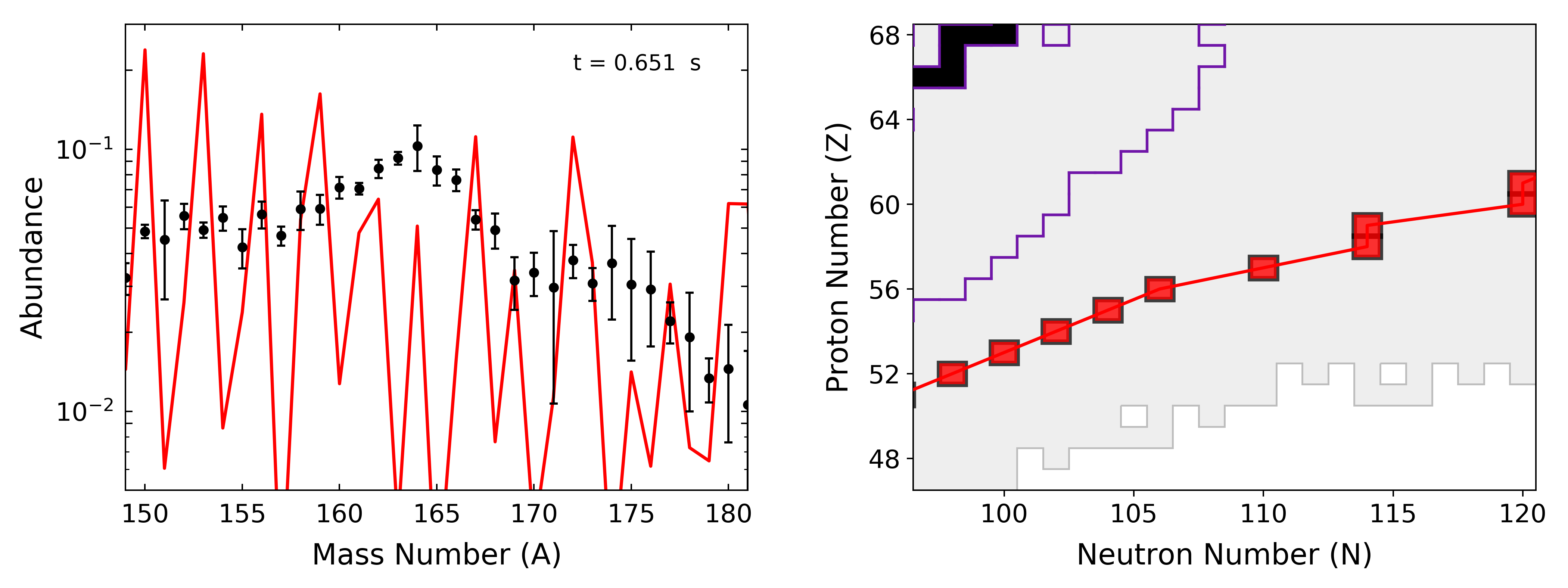}%
    \hspace{0.5cm}
    \includegraphics[scale=0.45]{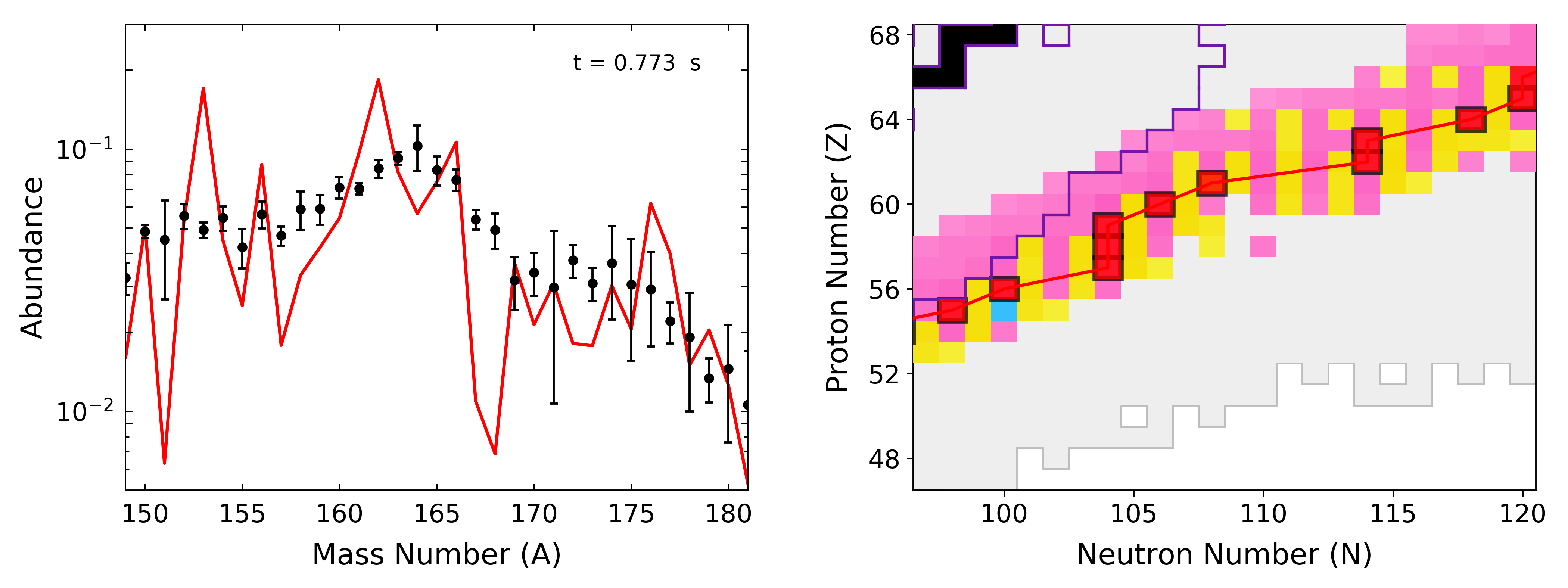}%
    \hspace{0.5cm}
    \includegraphics[scale=0.45]{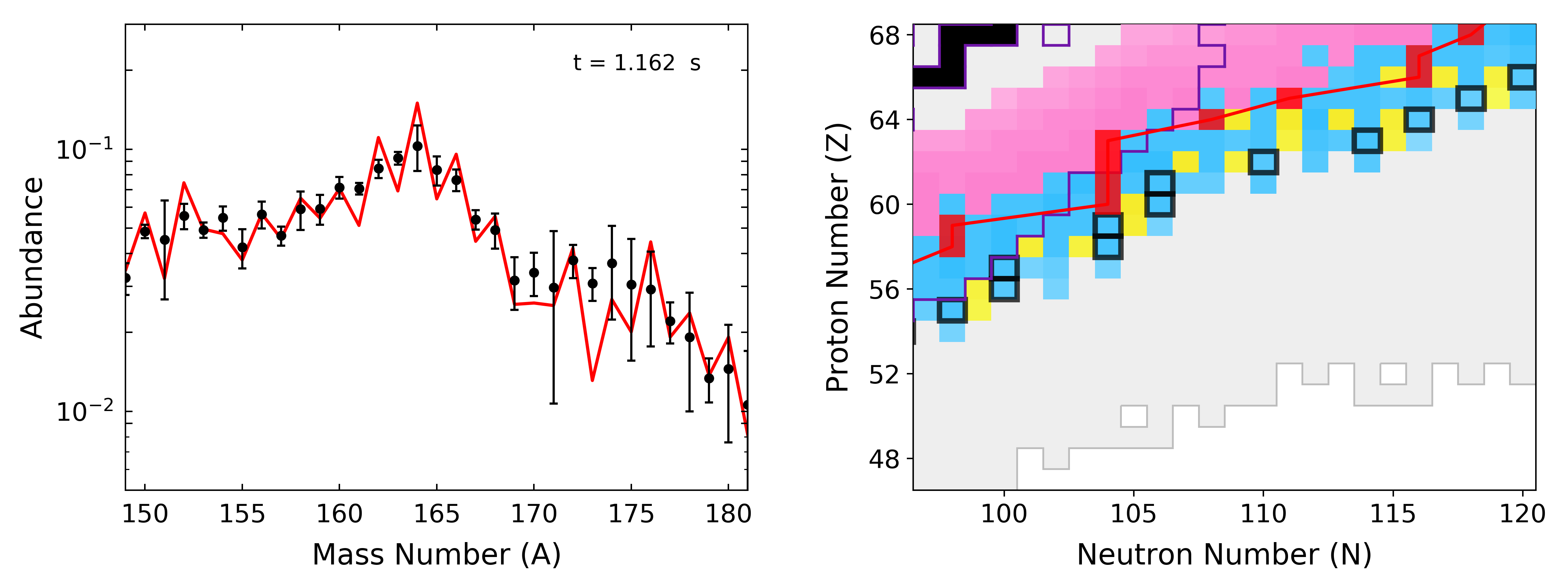}%
   \caption{The time evolution (top to bottom) for the formation of the rare-earth peak when the MCMC mass adjustments determined for the hot outflow (from the 50 run average shown in Figure~\ref{fig:masssurfhot1}) are implemented in the neutron capture, photodissociation, and $\beta$-decay rates. Left panels show the summed abundance as a function of mass number (red line) as compared to solar data. Right panels show the $r$-process path (red) of the most abundant nuclei as compared to the equilibrium path (black boxes) determined by the Saha equation. Additionally populated nuclei are color-coded by the reaction or decay channel that a given species is most influenced by at the time considered: pink when neutron capture dominates, yellow when photodissociation dominates, and blue when $\beta$-decay is dominant. The purple outline shows the range of experimentally established $\beta$-decay half-lives.}
 \label{fig:hot1evo}%
\end{figure*}
  
Since we find that in this hot outflow case the accumulation of peak material ultimately originates from pile-up on the equilibrium path, the adjustments to separation energies are most important and the modifications made by our algorithm to local neutron capture and $\beta$-decay rates play a minor role. This can be seen from Figure~\ref{fig:hotbetasnncap} which compares rare-earth abundances when individual pieces of nuclear data are updated to reflect the mass adjustments shown in Figure~\ref{fig:masssurfhot1}. Recall that photodissociation rates are dependent on both separation energies and neutron capture rates. For the `$S_n$ in ($\gamma$,n) only' case in Figure~\ref{fig:hotbetasnncap} we explore the role of solely the separation energy dependence in the exponential of the photodissociation rate. For the `(n,$\gamma$) only' case in Figure~\ref{fig:hotbetasnncap}, our network instead updates photodissociation to reflect solely the new neutron capture rate and leaves separation energies to be those of DZ. This exercise demonstrates that only adjustments to equilibrium path dynamics, via updating the separation energies considered in the detailed balance equation, are needed to form the peak in such hot outflow conditions. However, the changes introduced in the $\beta$-decay and neutron capture rates from our mass adjustments play a role in shaping the sides of the peak. 

In Figure \ref{fig:hot1evo} we bring together the discussion of key mass surface features and their influence on each reaction and decay channel by showing the evolution of the pile-up that ultimately forms the rare-earth peak in these hot outflow conditions. The figure indicates the dominant reaction or decay that a given species is undergoing at the time considered. Neutron capture dominates when the flow, $F$ (rate times abundance), obeys $\delta_{n,\gamma}=F_{(n,\gamma)} - F_{(\gamma,n)} > 0$  and $|\delta_{n,\gamma}|>F_{\beta}$, photodissociation dominates when $\delta_{n,\gamma} < 0$  and $|\delta_{n,\gamma}|>F_{\beta}$, and $\beta$-decay is dominant when $|\delta_{n,\gamma}|<F_{\beta}$. For the outflow case considered here, the $r$-process path (location of the most abundant nuclei) is well described by the equilibrium path. At early times, there is no preferential pile-up in the rare-earth region and the path is mostly set by odd-even effects. When the onset of freeze-out occurs, the path, still governed by (n,$\gamma$)$\rightleftarrows$($\gamma$,n) equilibrium, begins to move back to stability. After this time the $N=104$ kink in the equilibrium path that causes material to accumulate here emerges and persists throughout the remainder of the process. Thus the material that will populate the center of the peak is found at $N=104$ even at early times. At late times, after the environment falls out of (n,$\gamma$)$\rightleftarrows$($\gamma$,n) equilibrium, the faster $\beta$-decay rates of nuclei at the bottom of the $N=104$ kink in the path, as compared to the slower decay rates at the top of this feature, work as a funneling mechanism that keeps nuclei piled-up in this location. That is, the nuclei near the bottom of the feature will quickly $\beta$-decay to then neutron capture back to $N=104$, keeping the dominant population of nuclei at lower $Z$ aligned with the higher $Z$ nuclei that are decaying more slowly. Therefore, although Figure~\ref{fig:hotbetasnncap} demonstrates that it was only modifications to separation energies that our algorithm needed to exploit in order to form the peak in this hot outflow, and modifications to $\beta$-decay were not influential in peak formation, the late-time funneling produced by standard $\beta$-decay trends is an important aspect of maintaining pile-up. Additionally at the latest times, for which the calculation relies almost entirely on experimentally known decay rates, $\beta$-decay works to smooth the final pattern.

\vspace{0.4cm}
\vspace{0.4cm}

\subsection{Cold Outflows}\label{sec:coldcond}

\begin{figure}%
    \centering
    \includegraphics[width=8.65cm]{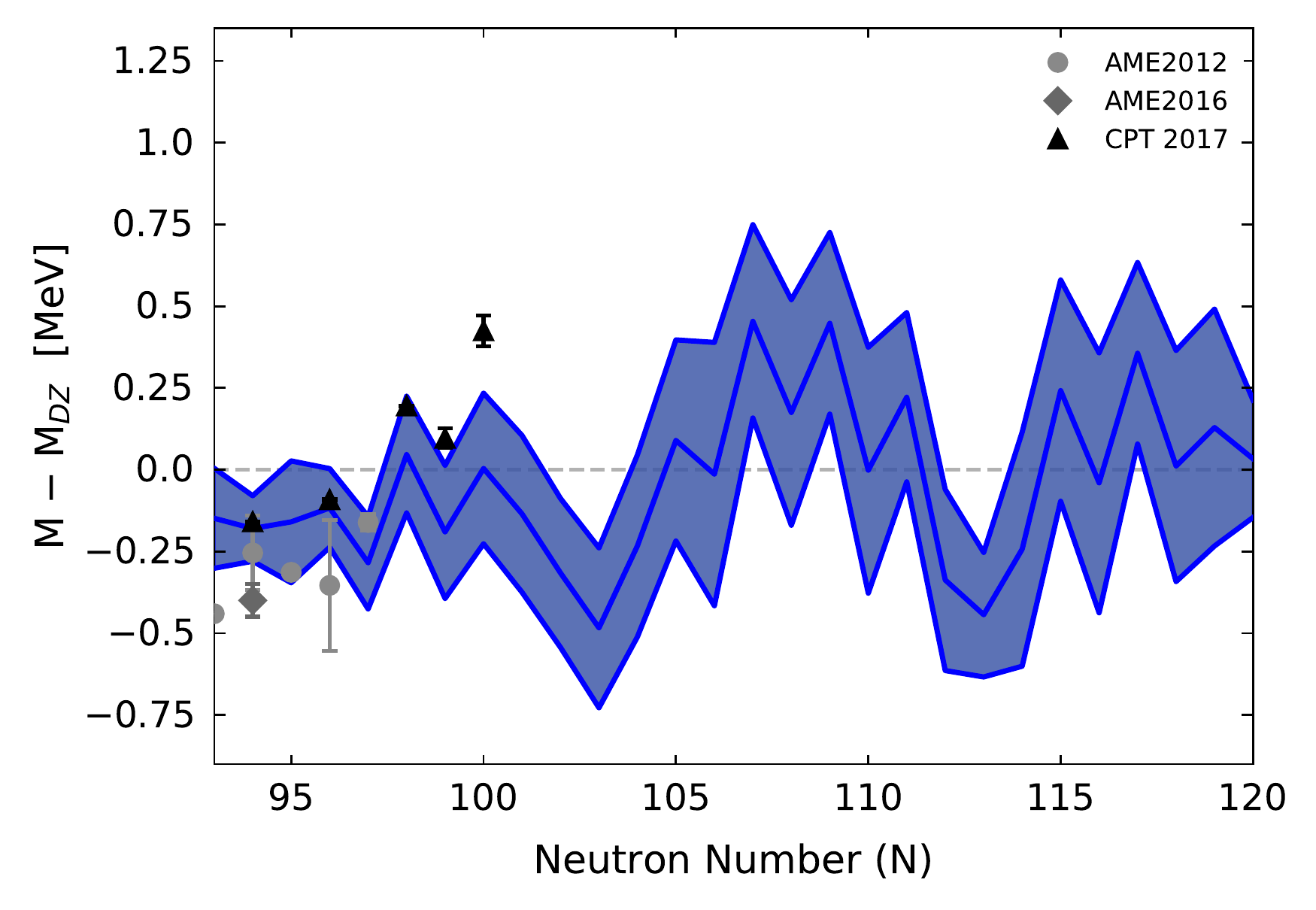}%
    \hspace{0.5cm}
    \includegraphics[width=8.2cm]{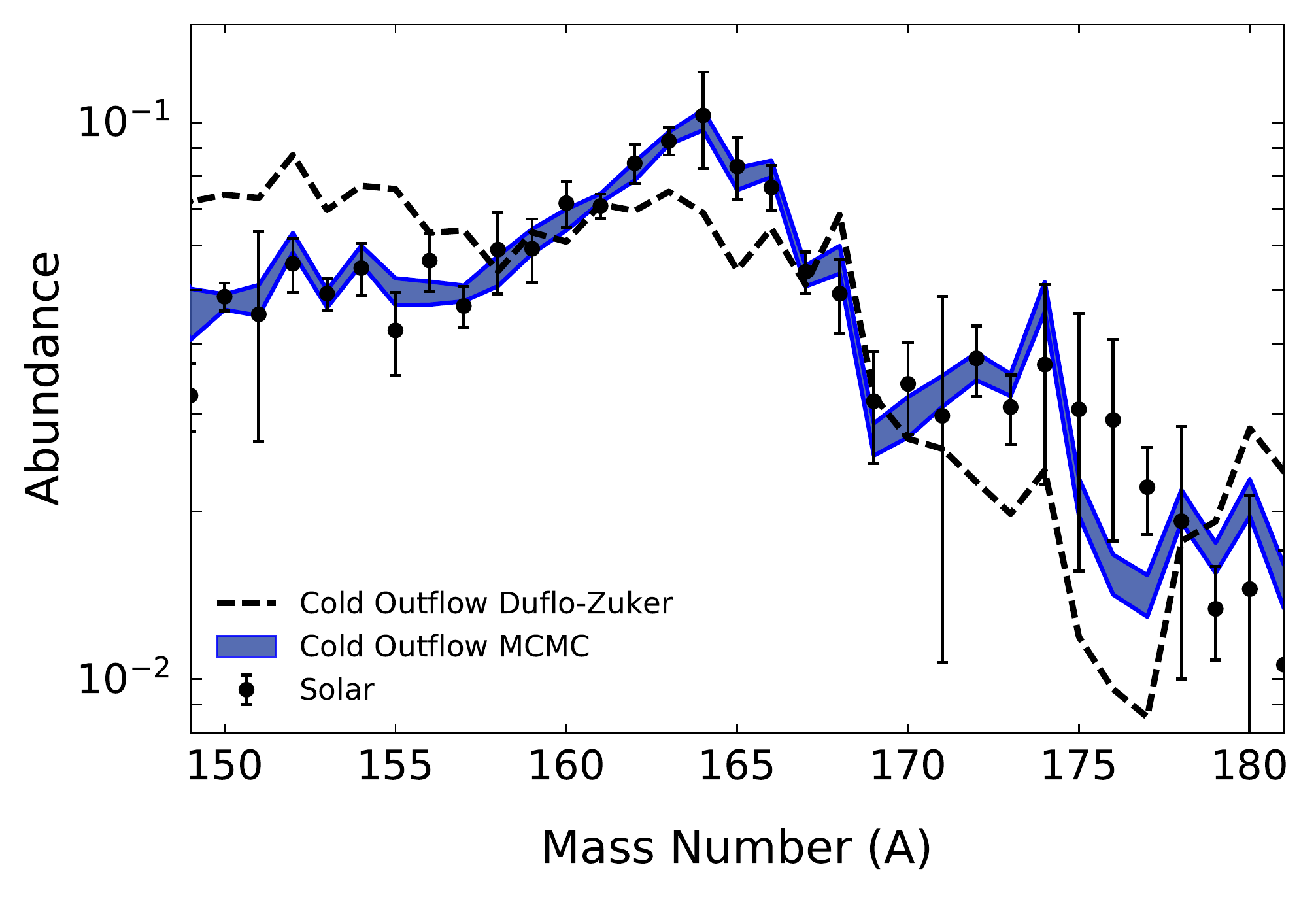}%
   \caption{Same as Figure~\ref{fig:masssurfhot1} but with results for the neodymium mass surface (top) and final abundances (bottom) given the cold outflow.}
\label{fig:masssurfcold}%
\end{figure}

\begin{figure}[!h]
\begin{center}
\includegraphics[scale=0.415]{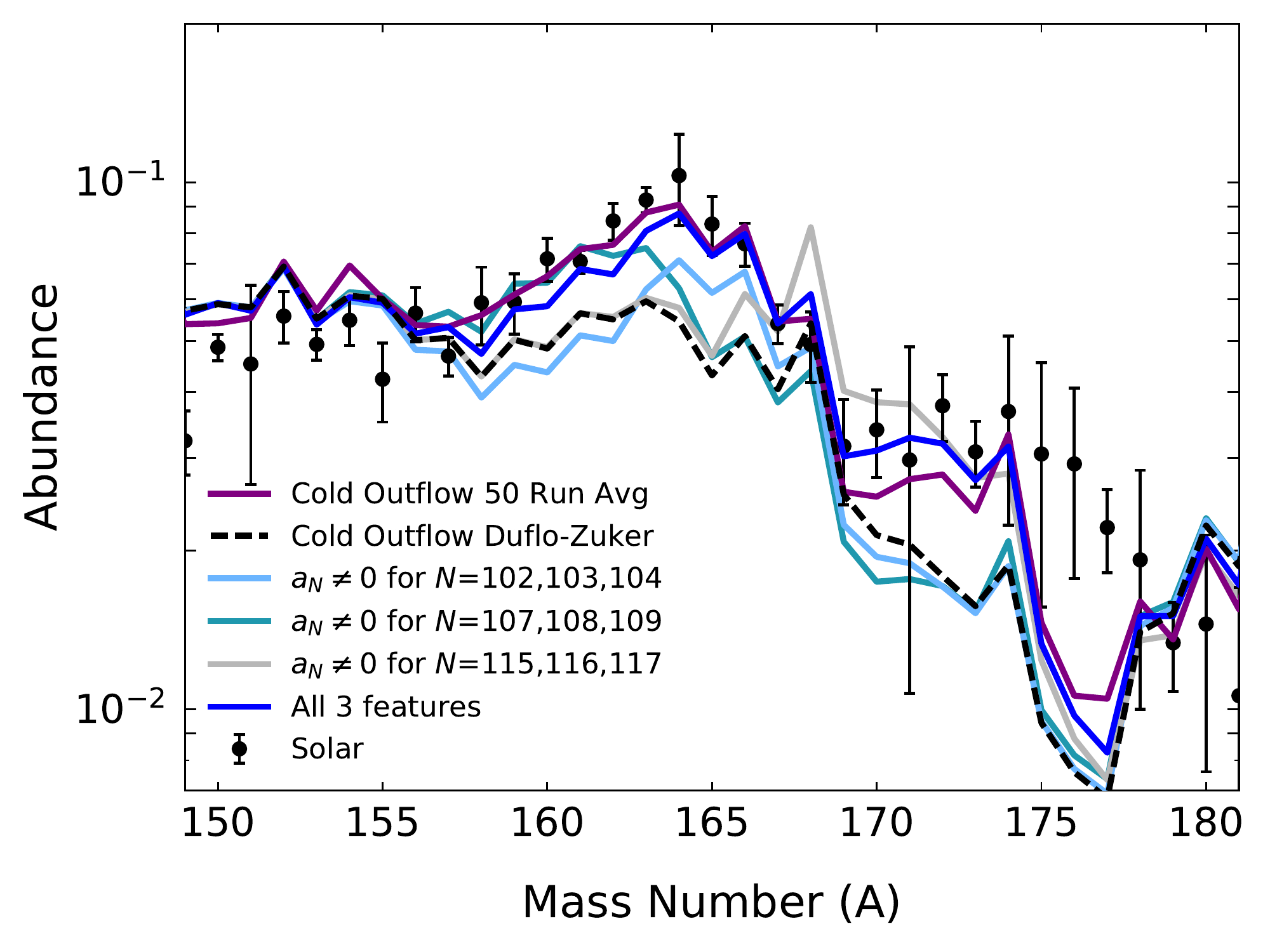}
\end{center}
\caption{The abundance pattern using the average mass values from Figure~\ref{fig:masssurfcold} as compared to when only the $N=103$ or $N=108$ or $N=116$ key features are applied in an $r$-process calculation. The result when only these three features are combined is also shown.}
\label{fig:coldfeatures}
\end{figure}

In the case of cold outflow conditions, at early times the $r$-process path is in the most neutron-rich regions near the dripline. Therefore, unlike the hot case, which was centered at $C=60$, here calculations are centered at $C=58$ since initial tests with both $C=58$ and $C=60$ showed $C=58$ to be able to find the solutions with the lowest $\chi^2$ given these astrophysical outflow conditions. We keep $f=10$ in order to prevent significant changes in the dripline and keep the mass surface effects more localized. Figure~\ref{fig:masssurfcold} shows the results of 50 such parallel, independent MCMC runs. This solution also makes use in part of a pile-up at $N=104$ in order to form the rare-earth peak. However, this is achieved via a dip in the mass surface at odd-N nuclei with $N=103$, a slightly different location than the dip at $N=104$ observed for the hot case. However, as is shown in Figure~\ref{fig:coldfeatures}, this $N=103$ feature is insufficient to form the peak alone. With only the dip from $N=102-104$, a peak that is too weak and off-center is produced. The $r$-process path lying closer to the dripline requires a feature at higher neutron number in order to properly redirect material to the peak region, in this case at $N=108$.  With only the feature at $N=107-109$, a small peak can be produced but is again off-center. Thus it is the $N=103$ and $N=108$ features together that fill in the right and left sides of the peak respectively. 

\begin{figure}[!h]
\begin{center}
\includegraphics[scale=0.415]{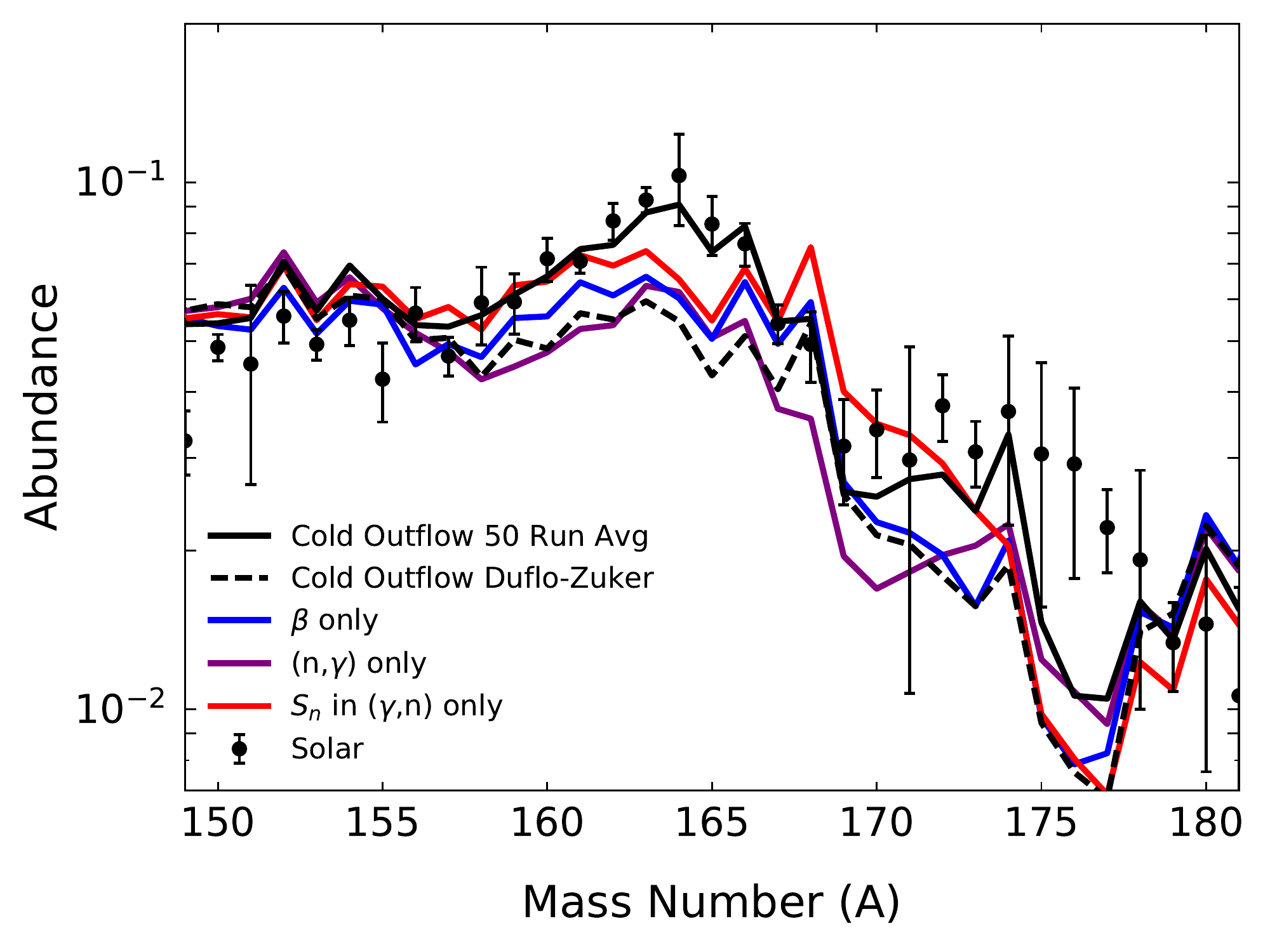}
\end{center}
\caption{The abundance pattern from implementing the updated neutron capture, photodissociation, and $\beta$-decay rates determined from the average mass values in Figure~\ref{fig:masssurfcold} as compared to when adjustments to only $\beta$-decay or neutron capture or the separation energies in the detailed balance equation for photodissociation are applied.}
\label{fig:coldbetasnncap}
\end{figure}

Here the peak formation process occurs so far outside the region of experimentally established $\beta$-decay rates that the adjustments made by the algorithm to the theoretical rates have a stronger influence than in the hot outflow, as can be seen in Figure~\ref{fig:coldbetasnncap}. When only adjustments to the separation energies considered in the detailed balance equation are considered, more material accumulates in the rare-earth region than in the baseline case, but no real peak structure is seen. However, when only the neutron capture adjustments are applied, a clear peak structure is seen even though the height is lacking due to a need for early-time accumulation of material. Therefore in such cold outflows the dynamics are a complex interplay between neutron capture, photodissociation, and $\beta$-decay.

\begin{figure*}
    \centering
    \includegraphics[scale=0.45]{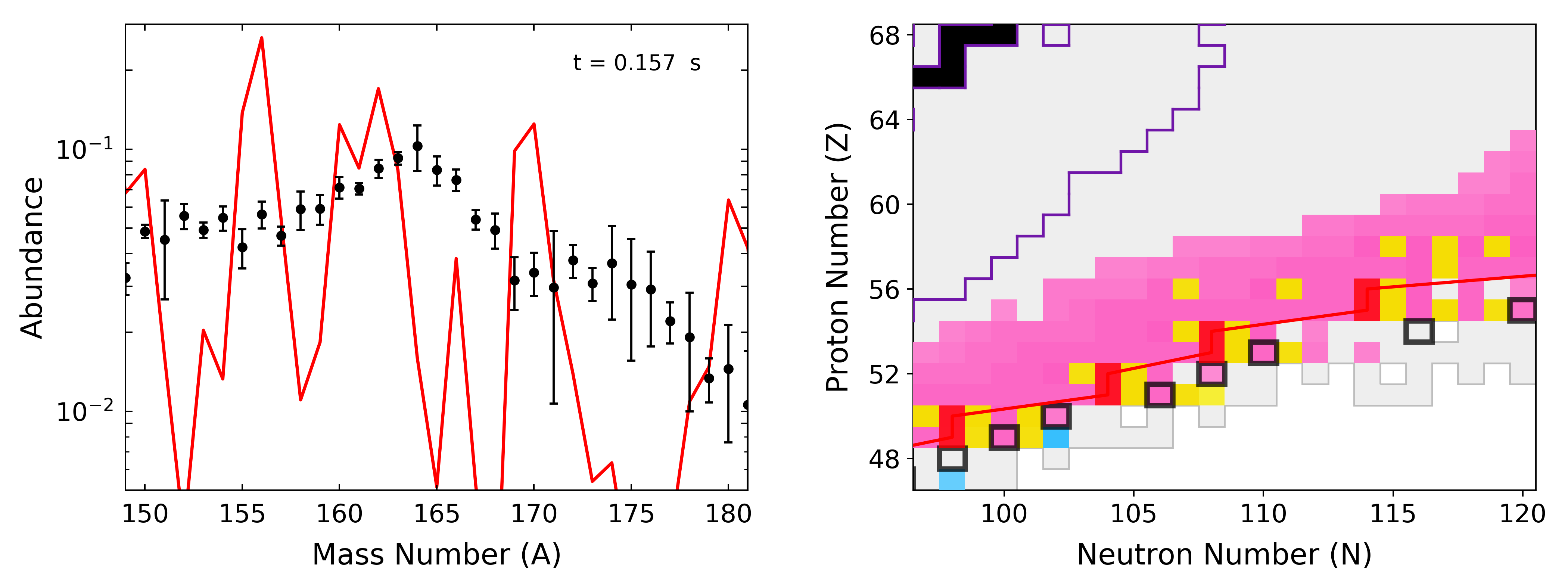}%
    \hspace{0.5cm}
    \includegraphics[scale=0.45]{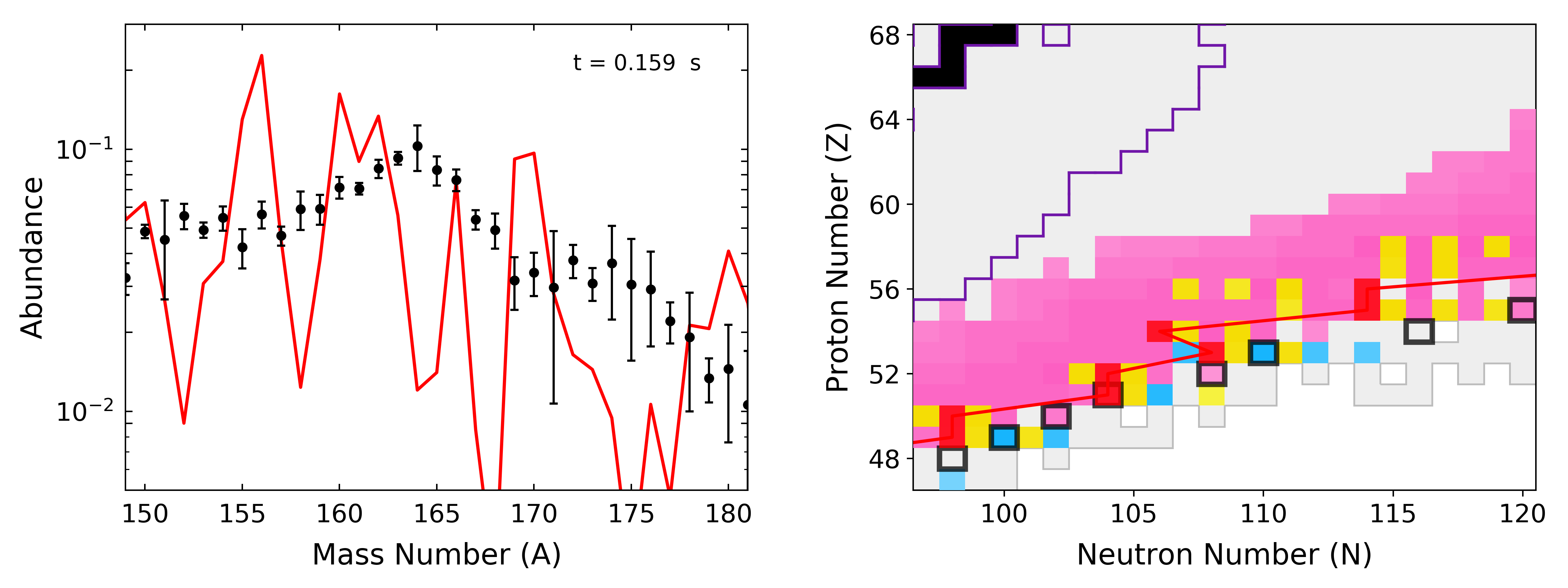}%
    \hspace{0.5cm}
    \includegraphics[scale=0.45]{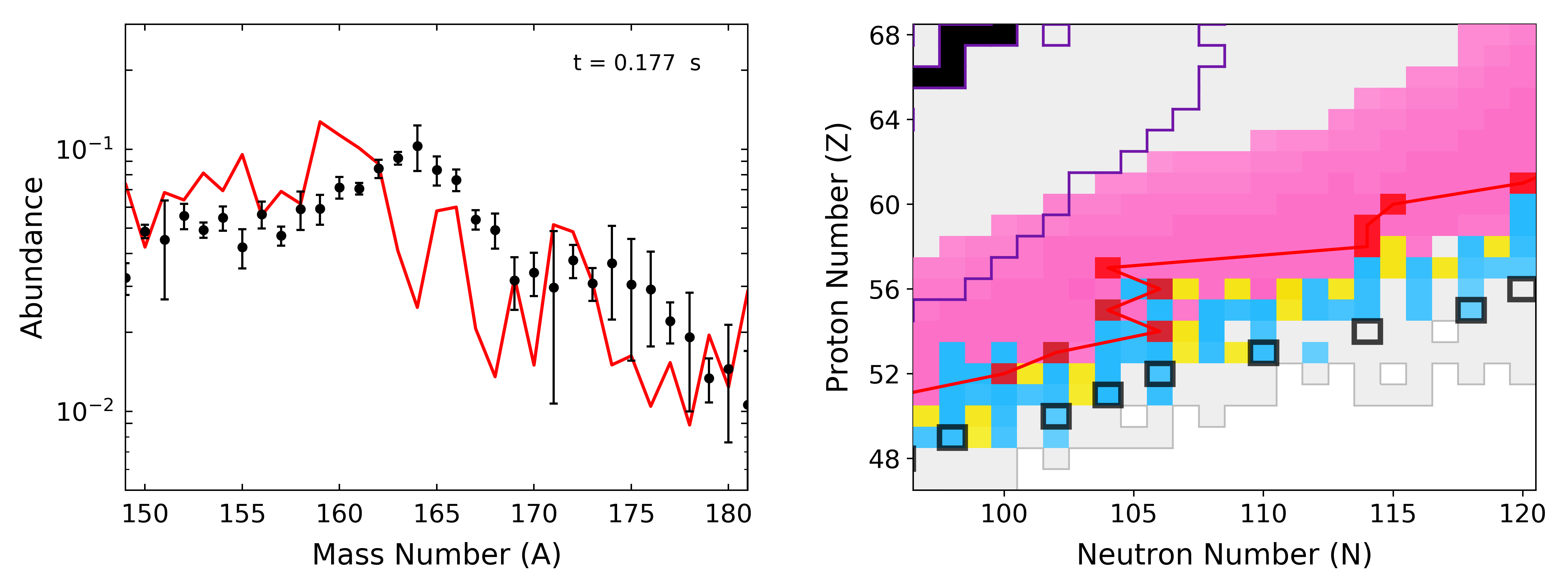}%
   \hspace{0.5cm}
    \includegraphics[scale=0.45]{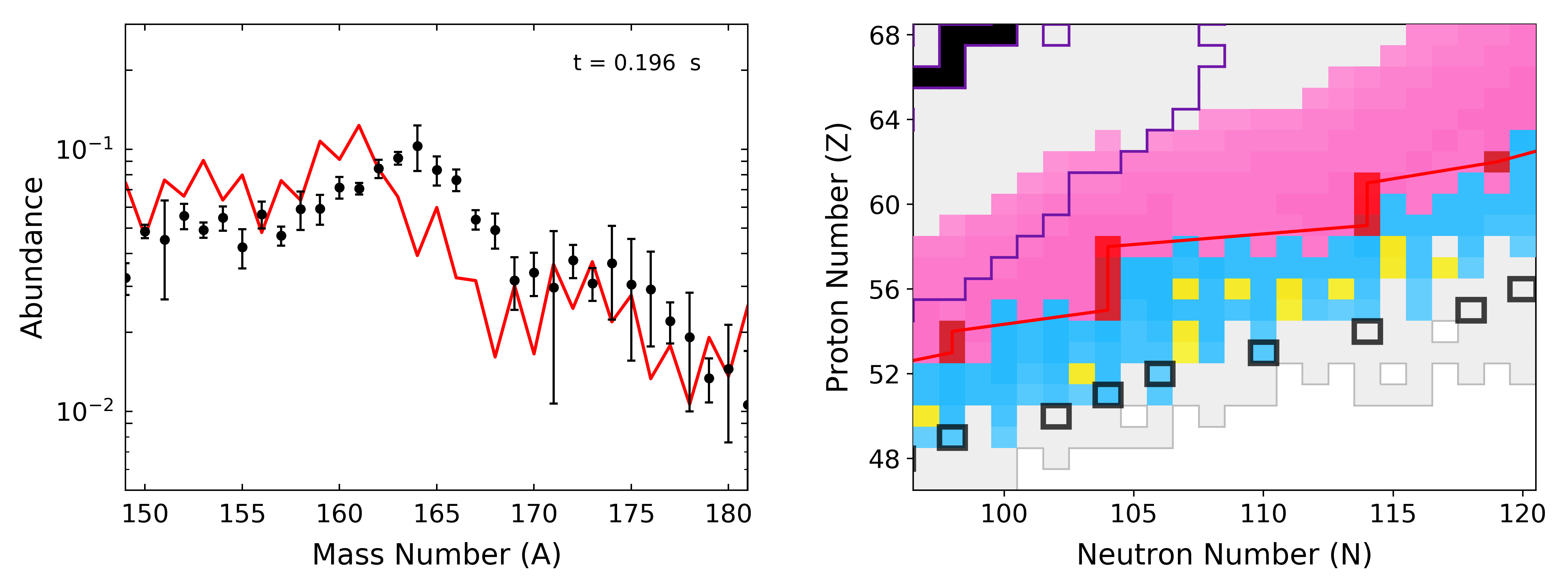}%
   \caption{Same as for Figure~\ref{fig:hot1evo} but showing the time evolution given our MCMC mass and rate adjustments determined for the cold outflow condition (from the 50 run average shown in Figure~\ref{fig:masssurfcold}).}
\label{fig:coldevo}%
\end{figure*}

The evolution of the formation of the peak in this cold outflow case is shown in Figure~\ref{fig:coldevo} . Since such outflows fall out of (n,$\gamma$)$\rightleftarrows$($\gamma$,n) equilibrium early, the equilibrium path does not define the $r$-process path even at early times. Early local competition between neutron capture and photodissociation initiates the pile-ups at $N=104$ and $N=108$ that we find to be responsible for peak formation. The influence of $\beta$-decay can also be seen early, with $N=108$ material being transferred to $N=106$. Throughout the rest of the calculation, it is the competition between $\beta$-decay and neutron capture that will determine the structure of local pile-up features. The material found at $N=106$, ultimately originating from $N=108$, is later transferred to $N=104$, where a strong pile-up persists to late times. The peak is found off-center to the left throughout the majority of the calculation and is eventually moved into place by late-time neutron capture.

\subsection{Hot/Cold Outflows}\label{sec:hccond}

Lastly we explore outflow conditions that fall out of equilibrium in a manner not well represented by solely hot or cold criterion, a hot/cold outflow, as was described in Section~\ref{sec:explaincond}. In such outflows, the path at early times is in more neutron-rich regions than the hot case but does not push all the way to the neutron dripline as does the cold case. Therefore, since preliminary calculations centered at $C=58$ and $C=60$ were both able to find solutions with a low $\chi^2$ abundance pattern, we gather statistics using runs centered at $C=60$ while again keeping $f=10$. 

\begin{figure}%
    \centering
    \includegraphics[width=8.65cm]{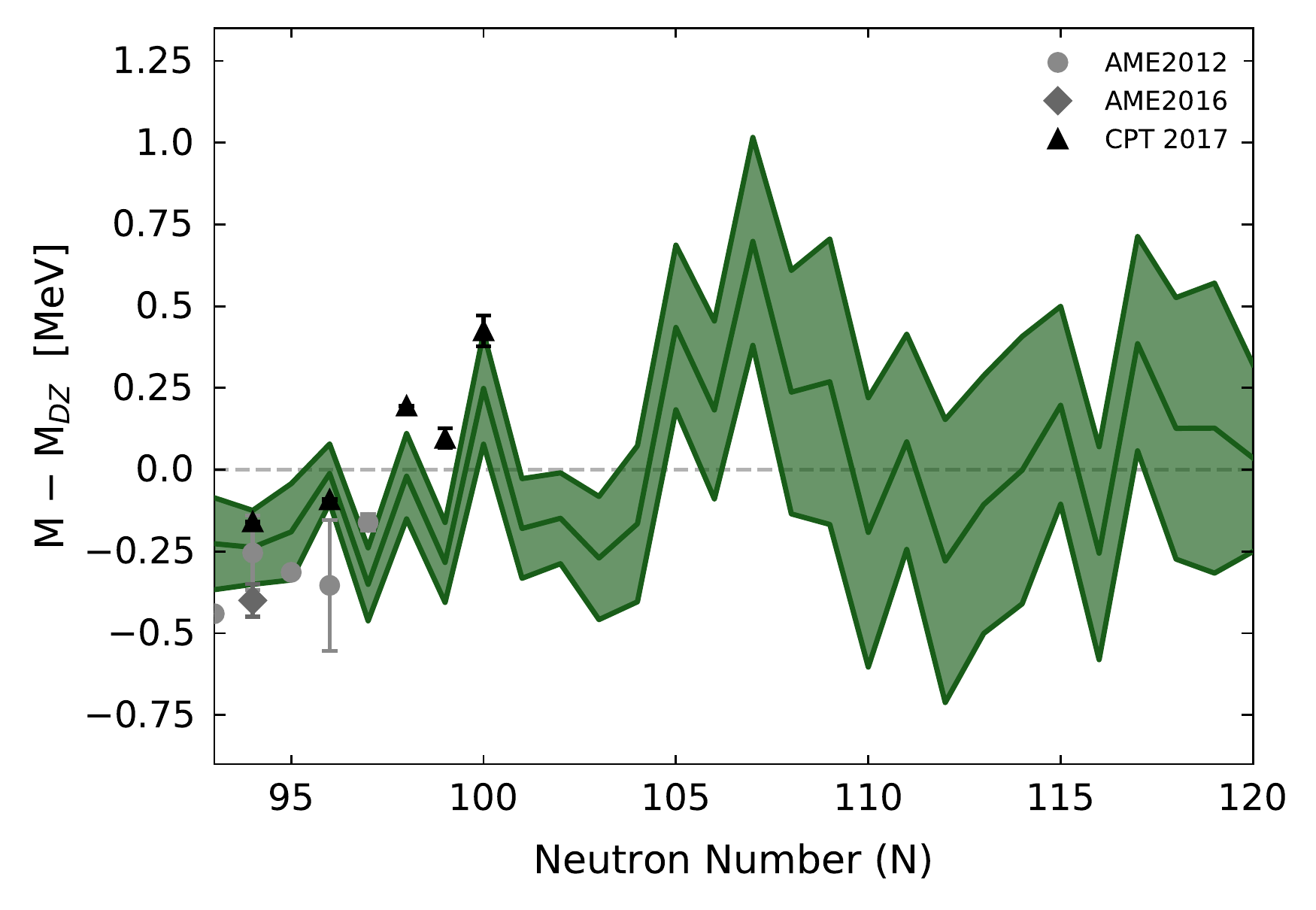}%
    \hspace{0.5cm}
    \includegraphics[width=8.2cm]{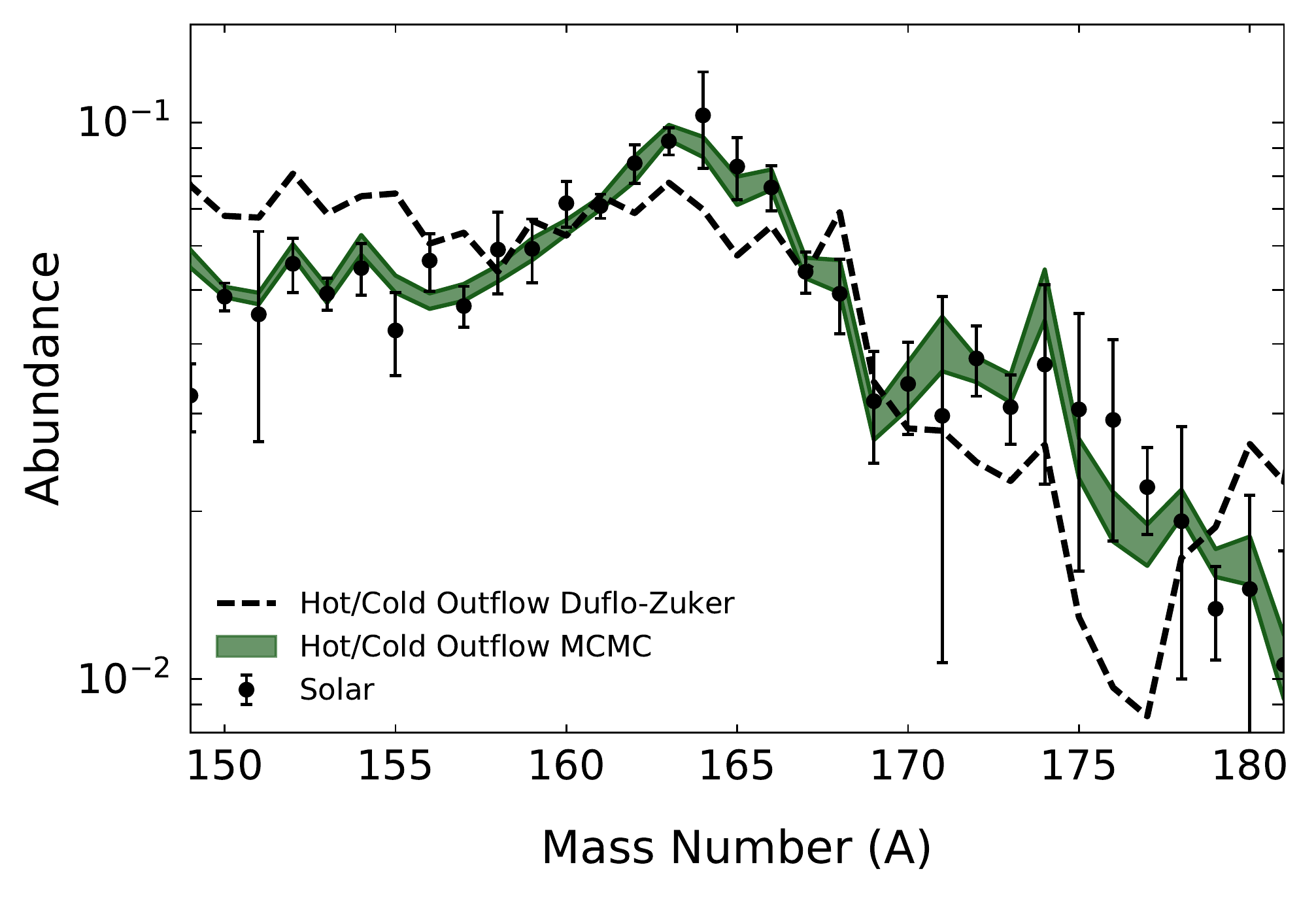}%
   \caption{Same as Figure~\ref{fig:masssurfhot1} but with results for the neodymium mass surface (top) and final abundances (bottom) given the hot/cold outflow.}
\label{fig:masssurfbetw}%
\end{figure}

\begin{figure}
\begin{center}
\includegraphics[scale=0.415]{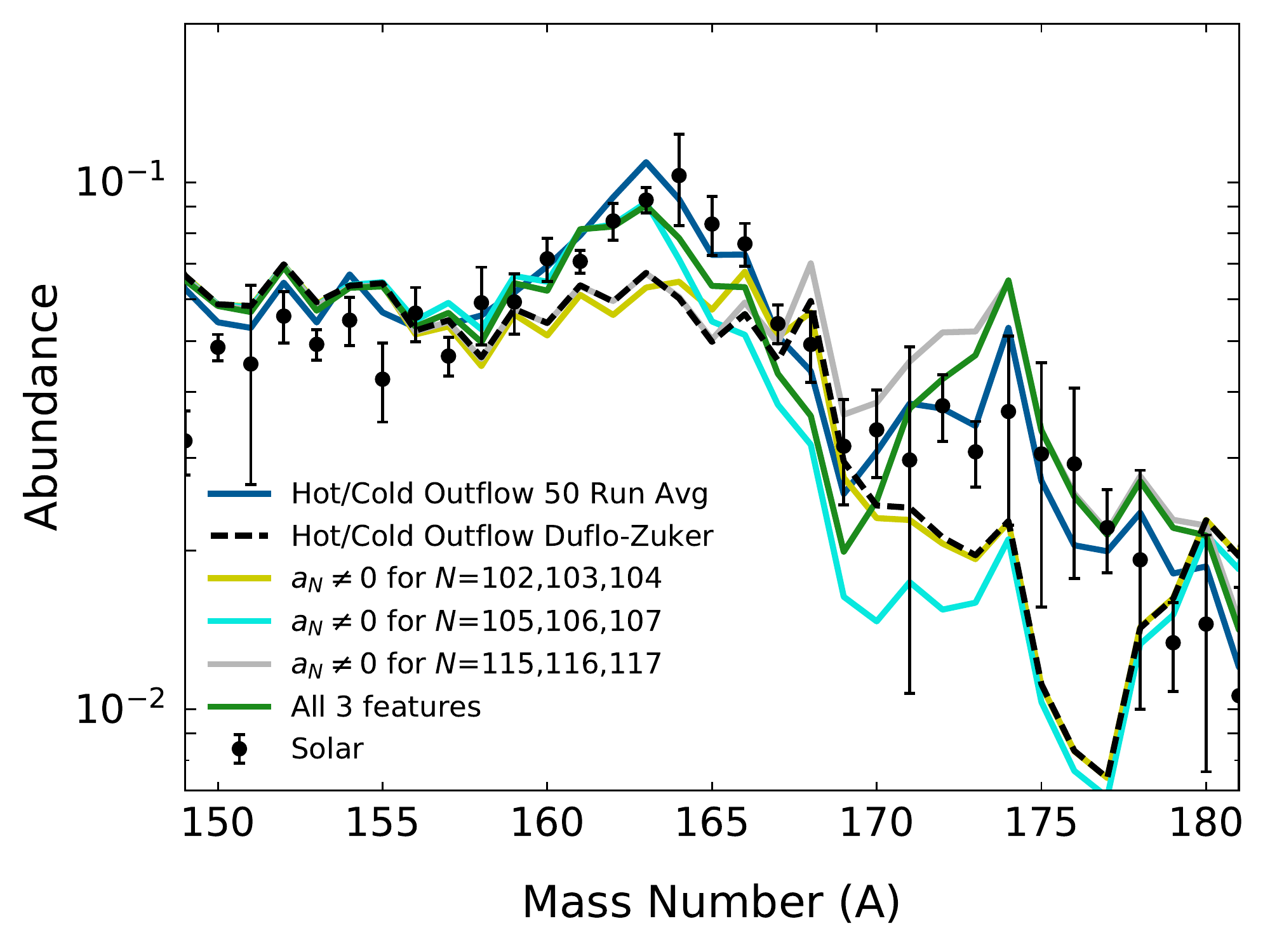}
\end{center}
\caption{The abundance pattern using the average mass values from Figure~\ref{fig:masssurfbetw} as compared to when only the $N=103$ or $N=106$ or $N=116$ key features are applied in an $r$-process calculation. The result when only these three features are combined is also shown.}
\label{fig:betwfeatures}
\end{figure}

Figure~\ref{fig:masssurfbetw} shows the results of 50 such parallel, independent MCMC runs. This solution also makes use of a late-time pile-up at $N=104$ in order to form the rare-earth peak where, as was the case in cold outflow conditions, this is achieved via a dip in the mass surface at odd-N nuclei with $N=103$. However, it should be noted that in this case several of the 50 runs were able to achieve solutions with a low $\chi^2$ with the dip in the mass surface found at $N=104$ rather than $N=103$. As is shown in Figure~\ref{fig:betwfeatures}, this $N=103$ feature is not the dominant source of pile-up and mostly works to fill in the right edge of the peak. Rather, as in the cold case, with the $r$-process path lying in more neutron-rich regions than in the hot case, a feature at a neutron number higher than $N=104$, in this case at $N=106$, is required. With only the feature at $N=105-107$, a peak can be produced in the proper location, with other features having more subtle effects such as assisting in building peak height.

\begin{figure}
\begin{center}
\includegraphics[scale=0.415]{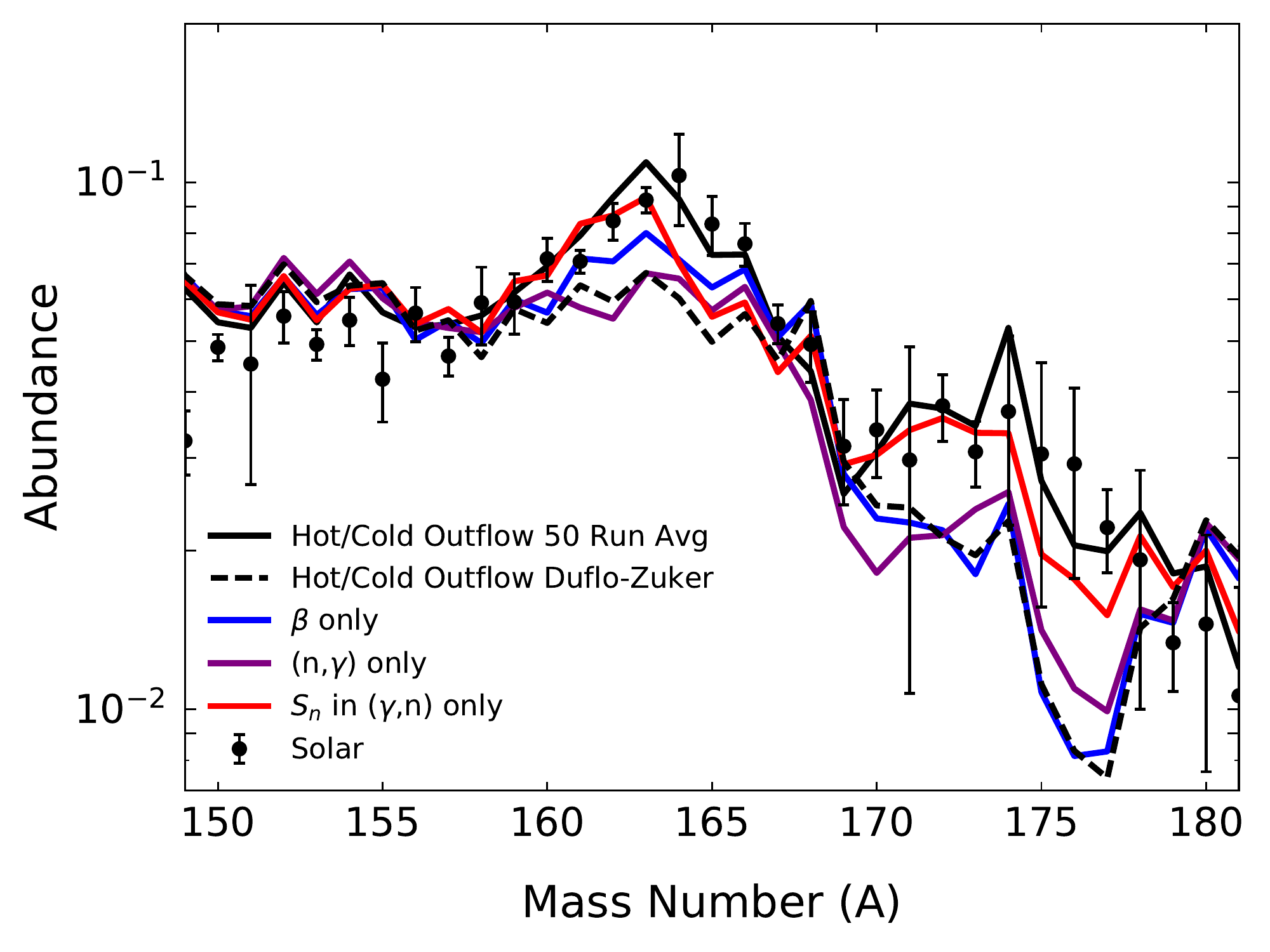}
\end{center}
\caption{The abundance pattern from implementing the updated neutron capture, photodissociation, and $\beta$-decay rates determined from the average mass values in Figure~\ref{fig:masssurfbetw} as compared to when adjustments to only $\beta$-decay or neutron capture or the separation energies in the detailed balance equation for photodissociation are applied.}
\label{fig:betwbetasnncap}
\end{figure}

In this case, as in the case of cold outflows, peak formation occurs outside the region of experimentally established $\beta$-decay rates. However, as can be seen from Figure~\ref{fig:betwbetasnncap}, since here $\beta$-decay is most influential at late times, the peak cannot be produced via changes to $\beta$-decay alone. Similarly, neutron capture adjustments alone are insufficient to form the peak, and in fact produce abundances very near the baseline result. When only adjustments to the separation energies considered in detailed balance are applied, a peak structure is clearly produced and centered in the proper location, but it is however insufficient to fill in the right side of the peak. Therefore, in such outflow conditions, it is primarily adjustments to (n,$\gamma$)$\rightleftarrows$($\gamma$,n) equilibrium via changes to the separation energies followed by the late-time shaping of the peak from $\beta$-decay that are responsible for peak structure.

\begin{figure*}
    \centering
    \includegraphics[scale=0.45]{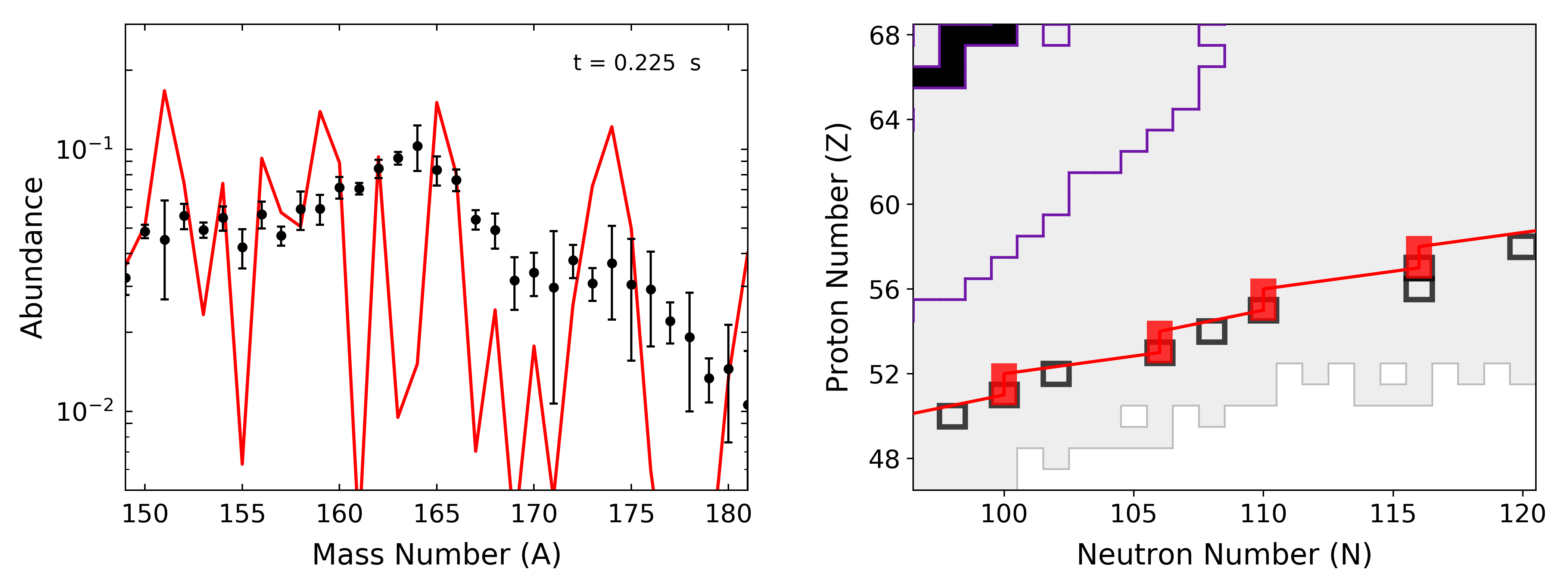}%
    \hspace{0.5cm}
    \includegraphics[scale=0.45]{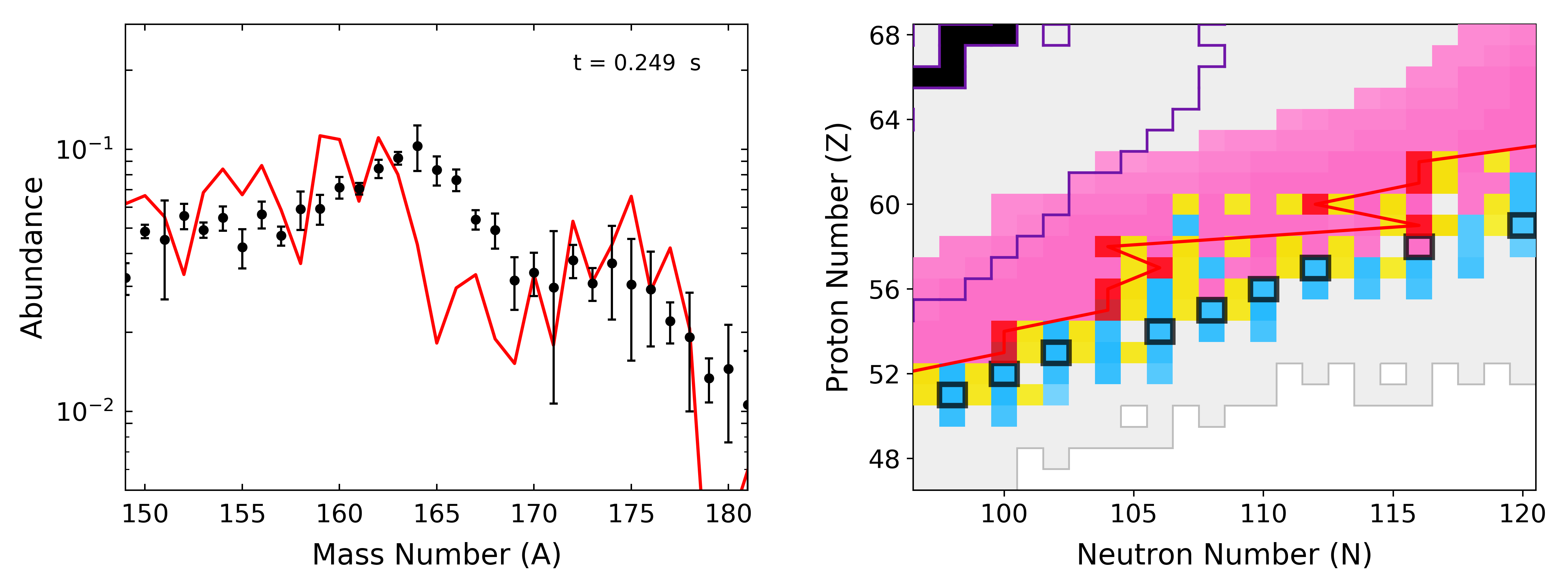}%
    \hspace{0.5cm}
    \includegraphics[scale=0.45]{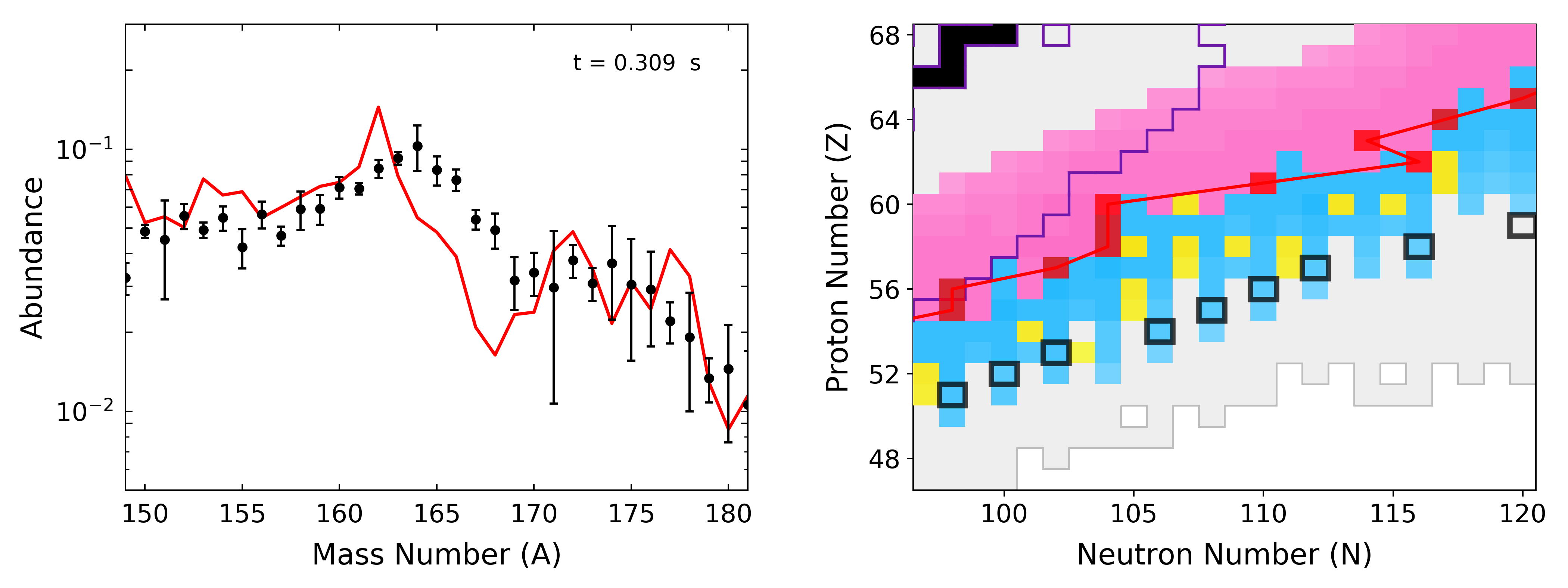}%
   \caption{Same as for Figure~\ref{fig:hot1evo} but showing the time evolution given our MCMC mass and rate adjustments determined for the hot/cold outflow condition (from the 50 run average shown in Figure~\ref{fig:masssurfbetw}).}
\label{fig:betwevo}%
\end{figure*}

The evolution of the formation of the peak for this hot/cold outflow is shown in Figure~\ref{fig:betwevo}. At early times (n,$\gamma$)$\rightleftarrows$($\gamma$,n) equilibrium is obeyed, and the equilibrium path is correlated with the $r$-process path. It is through changes in the separation energies of the detailed balance equation that early-time adjustments to the equilibrium path produce a pile-up of material at $N=106$. However, here, unlike in the hot case, equilibrium is not obeyed into late-times, and local competition between neutron capture, photodissociation, and $\beta$-decay transfers the early pile-up of material from $N=106$ to $N=104$. Later $\beta$-decay works to smooth the sharply formed peak while late-time neutron capture shifts peak material from $A=162$ to $A=163$.

\section{Comparing the Distinct Solutions found for Different Astrophysical Outflows}\label{sec:compareresults}

In the last section, we presented the mass surface solutions found by our MCMC method for each of the three outflow conditions explored and outlined the peak formation mechanism in each case. Here we comment on the statistical diagnostics of each case and compare these solutions directly. Table \ref{tab:MCMCResults1} provides a summary of parameterization values and diagnostic criterion, as well as the key features introduced by our mass adjustments that we find most influential to the formation of the main peak region (from $A\sim158-168$). Although the baseline $\chi^2$ for the abundances predicted by the Duflo-Zuker mass model differ among the three cases, the $\chi^2$ that can be achieved by our MCMC procedure is similar, with the hot/cold outflow runs able to achieve fits to the rare-earth peak with the lowest $\chi^2$. The number of steps taken by our Monte Carlo runs and the acceptance rate of runs are similar in these three cases since we work to ensure that the parameter space is being properly explored. 

\begin{table*}
\caption{MCMC Results Given Three Distinct Outflows.}
\label{tab:MCMCResults1}
\centering
\begin{ruledtabular}
\begin{tabular}{lccccccccc}
Outflow Type & $C$ & $f$ & Baseline $\chi^2$ & Avg. $\chi^2$ & Avg. $\#$ of Steps & Acceptance Rate Range & Most Influential for Main Peak  \\ 
\hline
Hot & 60 & 10 & 200.07    & 22.69 & 16,800 & 18.31$\%$ - 51.82$\%$ & $N=102,104$   \\ 
Hot/Cold & 60 & 10 &  217.18 & 17.70 & 15,155 & 23.12$\%$ - 65.62$\%$ & $N=103,106$  \\ 
Cold &  58 & 10 & 285.71 & 21.57 & 17,095 & 13.80$\%$ - 62.14$\%$ & $N=103,108$  \\ 
\end{tabular}
\end{ruledtabular}
\end{table*}

\begin{figure}[h!]%
    \centering
    \includegraphics[width=8.3cm]{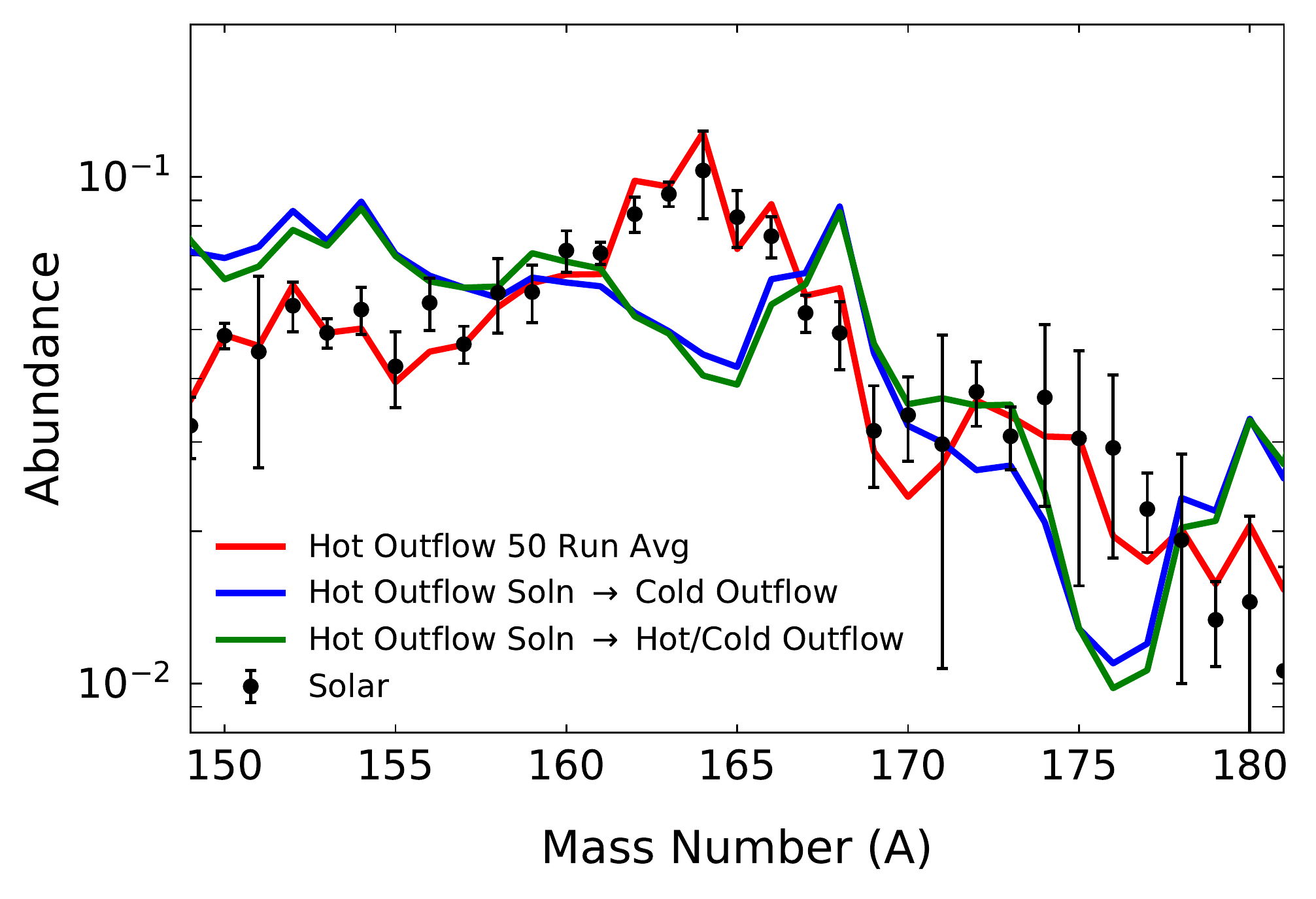}%
    \hspace{0.5cm}
    \includegraphics[width=8.3cm]{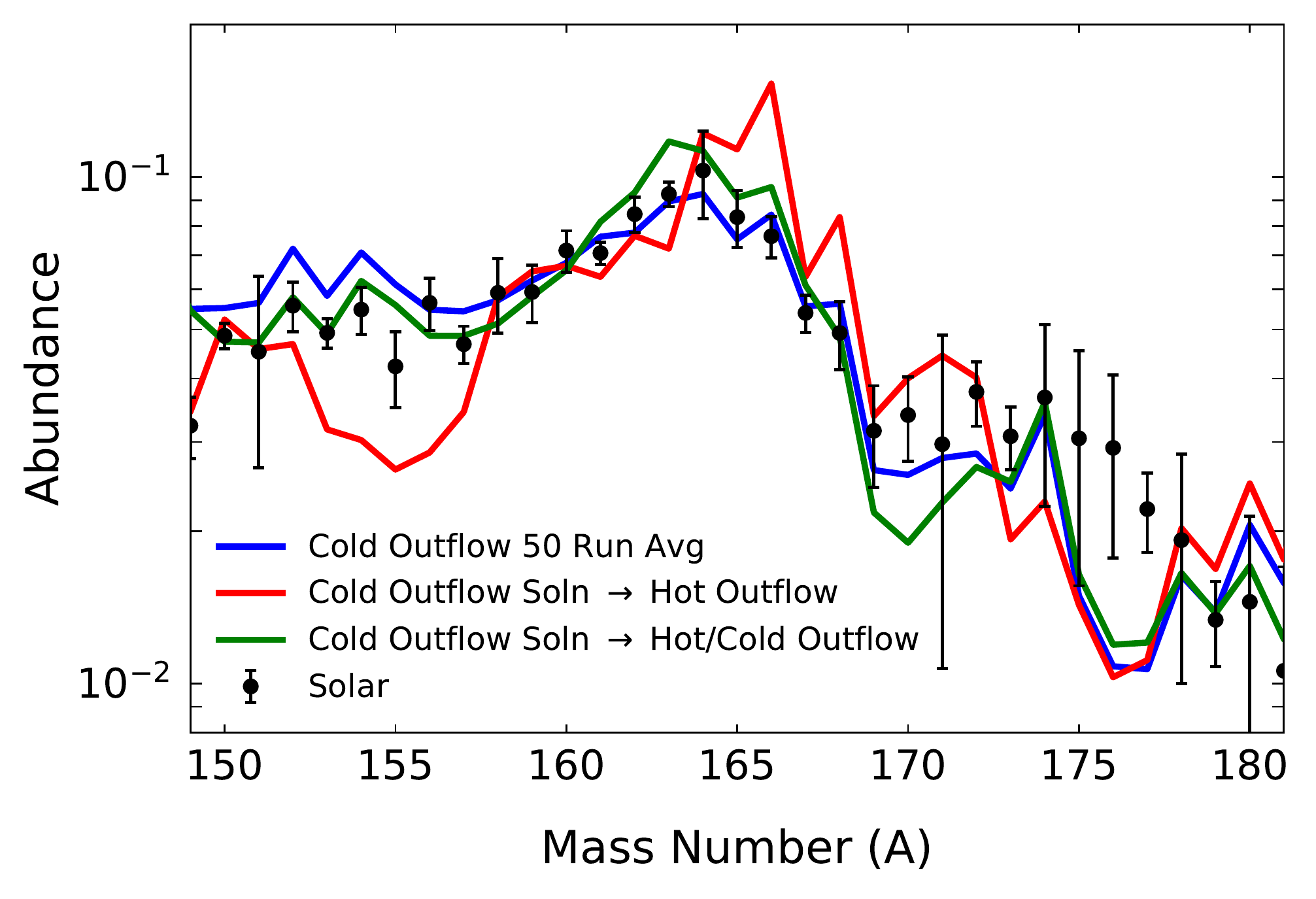}%
    \hspace{0.5cm}
     \includegraphics[width=8.3cm]{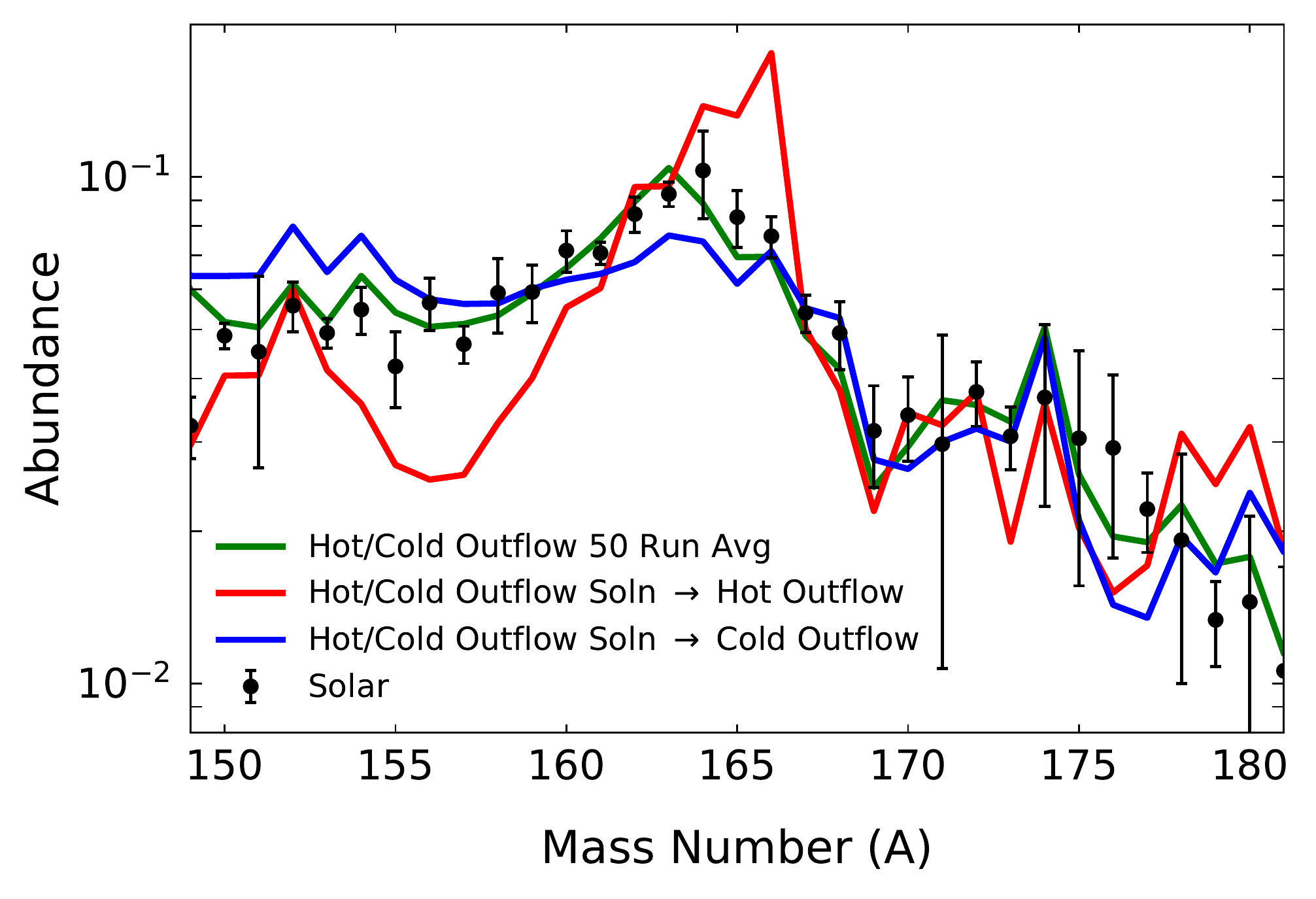}%
   \caption{Rare-earth peak abundance predictions when the masses from Figure~\ref{fig:masssurfhot1} are applied in nucleosynthesis calculations for all three outflows conditions (top), as compared to when the masses from Figures~\ref{fig:masssurfcold} (middle) and Figure \ref{fig:masssurfbetw} (bottom) are applied in all outflows considered here.}
\label{fig:abundallsolns}%
\end{figure}

Given the common occurrence of pile-up at $N=104$, it may appear that the three mass solutions are all equally capable of peak formation in any type of astrophysical outflow. However, it is important to recognize that each solution is tied to the distinct dynamics of the trajectory explored. This is represented in Figure~\ref{fig:abundallsolns} where we consider the formation of the rare-earth peak with the solutions presented in Figures~\ref{fig:masssurfhot1}, \ref{fig:masssurfcold}, and \ref{fig:masssurfbetw} in all three outflow conditions. The solution found to well form the peak in hot outflows produces a peak that is shifted greatly in mass number in cold and hot/cold outflows. The $N=108$ feature crucial to peak formation in the cold case is sufficient to produce a peak with outflows that are less cold, but produces too strong a peak that is off-center in both hot and hot/cold outflow cases. The $N=106$ pile-up produced by the mass surface solution in the hot/cold case gives an off-center peak at $A\sim166$ in hot outflows, while in cold outflows a peak that is, on average, flat is produced with this solution.       

\begin{figure*}
\begin{center}
\includegraphics[scale=0.7]{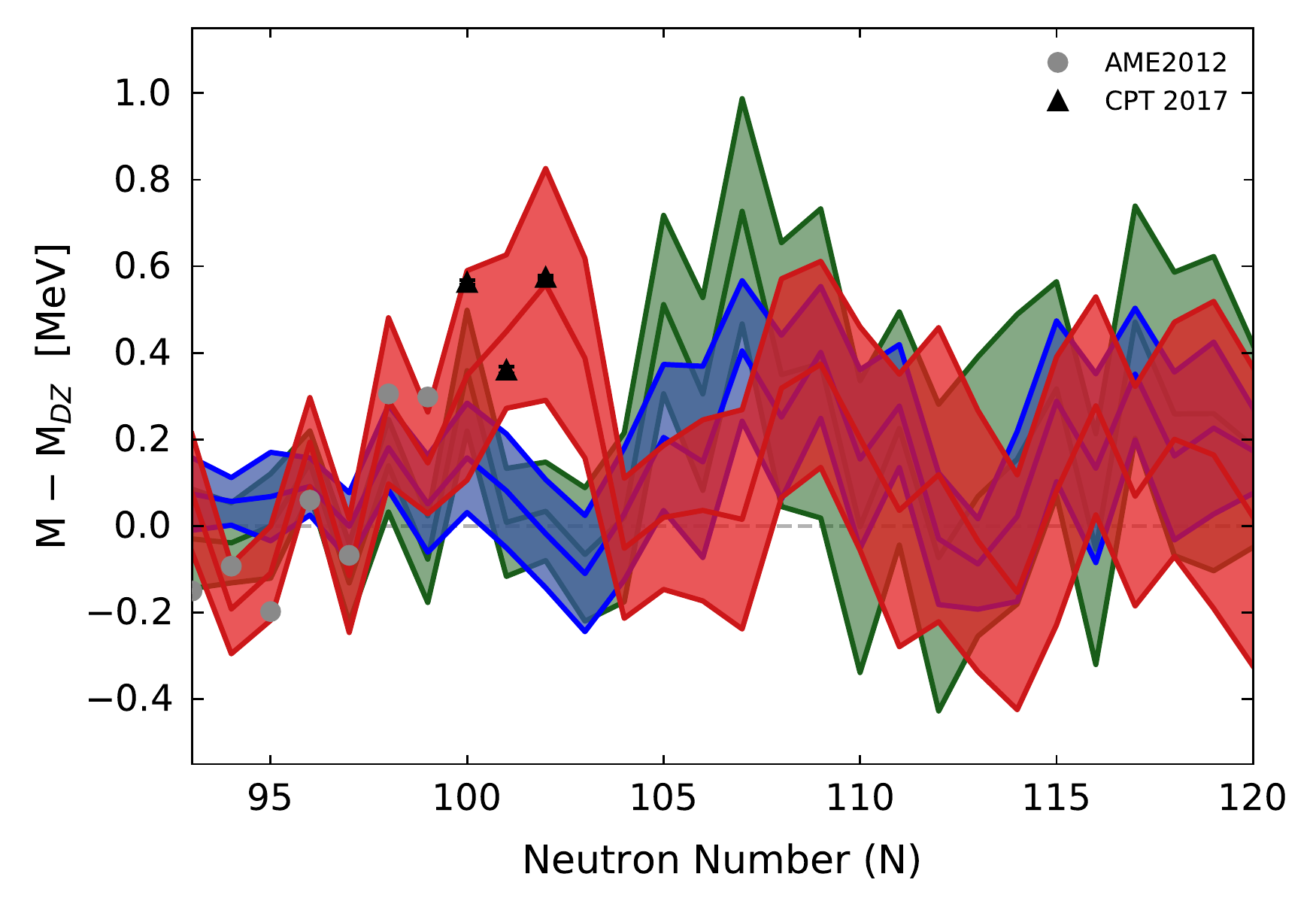}
\end{center}
\caption{The mass predictions for the samarium isotopic chain for all outflow conditions considered as compared to the most neutron-rich mass data available from the CPT at CARIBU \citep{OrfordVassh2018}.}
\label{fig:allwindSm}
\end{figure*}

In Section~\ref{sec:results} we showed the MCMC mass predictions for all outflow cases for the neodymium isotopic chain. Here experiments have probed neutron-rich isotopes up to $N=100$. Since key features for our predictions in each case begin to be distinguishable just after this neutron number, neodymium comparisons would suggest that all cases are currently consistent with experimental masses (except for perhaps the cold case since it does not well reproduce the exact height of the $N=100$ rise seen in data). Luckily, mass values are available for more neutron-rich isotopes at higher $Z$. In Figure~\ref{fig:allwindSm} we compare our solutions directly with samarium ($Z=62$) experimental mass data, which reach up to $N=102$. It is clear that this comparison suggests hot outflows to be most consistent with the latest measurements. However, recall that peak formation in  the cold case is most influenced by nuclei near $Z=58$, so measurements at lower $Z$ would be best to evaluate whether this remains a viable peak formation mechanism. With the hot/cold case solution centered at $Z=60$, samarium masses bear more weight on peak formation in such outflows. However, recall from Section~\ref{sec:hccond} that $N=103$ was a supporting feature in this case, playing a relatively minor role in comparison with the need for pile-up at $N=106$. Therefore measurements at higher $N$ are needed. 

Additionally, in this work we intentionally explored outflows that do not host fission because fission fragments can greatly influence abundances in very neutron-rich ejecta. We note that very neutron-rich dynamical ejecta can also exhibit both `hot' and `cold' outflow conditions, so the path and freeze-out arguments laid out for the moderately neutron-rich cases in this work can be examined in such cases as well. The complexities that can potentially be introduced by fission deposition should be considered alongside local deformation of rare-earth elements before more conclusive statements can be made as to whether hot dynamics are indeed the most favorable outflow conditions to form the rare-earth peak.

\section{Conclusions}\label{sec:conc}

The usefulness and broad applicability of statistical methods has led to an increase in the popularity of such approaches in a wide variety of scientific applications. Here we made use of statistical techniques to model the unknown nuclear physics properties that could be responsible for the pile-up forming the $r$-process rare-earth abundance peak. Such statistical methods, which examine the overlap between nuclear physics properties and astrophysical observations, are able to fully exploit the capabilities of next-generation experiments and advancing observational techniques.

We presented how we have evolved our statistical approach to soundly search parameter space and generate well-defined statistical uncertainties by utilizing the parallel chains method of MCMC. Additionally, by implementing a modified likelihood function, we are able to train the algorithm to obey known nuclear physics. Here we instruct the algorithm to take into account known masses of isotopes lighter than those being modeled as well as some general nuclear physics properties by considering the $D_n$ metric. We found that the algorithm successfully meets AME2012 data when it is given this information in a general way by considering an rms deviation summed over many isotopes. We note that using a modified likelihood function approach to train the algorithm could be extended to consider other properties as neutron-rich rare-earth nuclei are further probed by future experiment.

We apply our method to the $r$-process rare-earth abundance peak since it is a window into the astrophysical outflow conditions that dominated the synthesis of lanthanide elements in our solar system. For the three distinct types of outflows explored here, we find that the features generating the pile-up that forms the rare-earth peak correlate with the type of outflow conditions. Outflows that push out toward the dripline, that is, outflows with cold $r$-process dynamics that see photodissociation drop out of competition early, show a need to pile up abundances at a higher neutron number, later transferring this bulk of material to lower neutron numbers. For the cold and hot/cold cases explored here such an early pile-up is found at $N=108$ and $N=106$, respectively. In the case of hot outflows with dynamics governed by (n,$\gamma$)$\rightleftarrows$($\gamma$,n) equilibrium, the $r$-process path finds itself comprised of isotopes that are less neutron-rich than is the case for outflows with cold dynamics. The hot case considered here forms the rare-earth peak via a pile-up at $N=104$ which occurs early and persists during the decay back to stability. This $N=104$ feature is tantalizingly close to the mass measurements of neutron-rich rare-earth nuclei recently reported by the CPT at CARIBU, in particular being just two units in neutron number away from the recent measurement of $^{164}$Sm.

Here we focused on applying our method to outflow conditions that do not significantly populate fissioning nuclei. Outflows such as these are predicted to occur in a variety of astrophysical scenarios from the dynamical ejecta of neutron star mergers to accretion disk winds. Should fission take place in the $r$ process, fission fragments could play a role in shaping lanthanide abundances, introducing a potential connection between late-time fission deposition and the rare-earth peak \citep{GorielySPY}. However, the predicted distributions of fission yields for neutron-rich actinides vary widely \citep{GorielySPY,GorielyGMPGEF,Eichler15,Mendoza-Temis+15,RobertsWahl,Shibagaki,VasshGEF2019,VasshFRLDM2019}, and the fissioning nuclei that most impact abundances are well outside the reach of experimental facilities \citep{VasshGEF2019}. Thus whether fission deposition can influence rare-earth peak abundances is highly uncertain. Therefore progress in understanding the synthesis of these elements is best made with further studies of neutron-rich lanthanide properties, such as the masses, which are within the reach of future facilities such as FRIB and the N=126 Factory. Interestingly, a past study suggested that when even-A abundances are interpolated and compared to odd-A abundances, their ratio approaches one near A = 130, 163, and 195, which supports conclusions that rare-earth abundances are determined by the influence of local nuclear properties during the r-process decay to stability as opposed to fission deposition \citep{MartiSuess}. We will explore very neutron-rich outflow conditions for which fission deposition occurs alongside an enhanced stability of rare-earth nuclei in future MCMC investigations.

For the moderately neutron-rich outflow conditions considered in this work, we find that the most neutron-rich CPT mass data for samarium ($Z=62$) at $N=100-102$ agree best with our MCMC results given the case of a hot $r$ process that does not push out to the dripline. The dip in the predicted mass data at $N=103$, which cold type dynamics use to keep material in place at later times, is not reflected by Sm mass measurements. If the trends suggested in the Sm isotopes also persist through lower $Z$ where data cannot yet reach, the latest mass measurements favor rare-earth peak production to occur in hot scenarios. Trends at lower $Z$ are of importance since the $r$-process path can be found further from stability in cold and hot/cold scenarios as compared to the hot scenario. With peak formation in the cold outflow most influenced by isotopes near the $Z=58$ isotopic chain, and features needed to form the peak in the hot case concentrated near $Z=60$, higher $Z$ elements such as Sm have less influence on rare-earth peak abundances. Other caveats should be noted. The $N=103$ feature found in the cold and hot/cold cases is not the primary feature of peak formation in these outflow conditions since pile-up must first occur at higher neutron numbers. Therefore measurements are needed at higher neutron numbers where predicted features are concentrated. 

Our results demonstrate that the formation of the rare-earth peak is intimately tied to both the outflow conditions and nuclear properties of isotopes that are near current experimental data. Therefore future measurements that push further into the neutron-rich lanthanides can discriminate between different outflow scenarios. Although here we use our approach to explore the role of neutron-rich masses, this method can be responsive to both new experimental information and theoretical developments for individual pieces of nuclear data such as neutron capture or $\beta$-decay rates. It is through such collaborative efforts between theory and experiment that the nature of lanthanide production in astrophysics can be understood.

\acknowledgments
N.V. would like to thank Zolt\'an Toroczkai, Jorge Piekarewicz, Maria Piarulli, Daniel Phillips, Ian Vernon, and Michael Grosskopf for useful discussions. N.V. would also like to acknowledge the Information and Statistics in Nuclear Experiment and Theory (ISNET) conference series for valuable insights. The work of N.V., G.C.M., M.R.M., and R.S. was partly supported by the Fission In R-process Elements (FIRE) topical collaboration in nuclear theory, funded by the U.S. Department of Energy. Additional support was provided by the U.S. Department of Energy through contract numbers DE-FG02-02ER41216 (G.C.M), DE-FG02-95-ER40934 (R.S.), and DE-SC0018232 (SciDAC TEAMS collaboration, R.S.). R.S. and G.C.M also acknowledge support by the National Science Foundation Hub (N3AS) Grant No. PHY-1630782.
M.R.M. was supported by the US Department of Energy through the Los Alamos National Laboratory. Los Alamos National Laboratory is operated by Triad National Security, LLC, for the National Nuclear Security Administration of U.S.\ Department of Energy (Contract No.\ 89233218CNA000001). 
This work was partially enabled by the National Science Foundation under Grant No. PHY-1430152 (JINA Center for the Evolution of the Elements). This work utilized the computational resources of the Laboratory Computing Resource Center at Argonne National Laboratory (ANL LCRC) and the University of Notre Dame Center for Research Computing (ND CRC). We specifically acknowledge the assistance of Stanislav Sergienko (ANL LCRC) and Scott Hampton (ND CRC). This manuscript has been released via Los Alamos National Laboratory report number LA-UR-20-24194.

\appendix

\section{Solar Data with Symmetrized Error Bars}
\begin{figure}[h!]
\begin{center}
\includegraphics[scale=0.42]{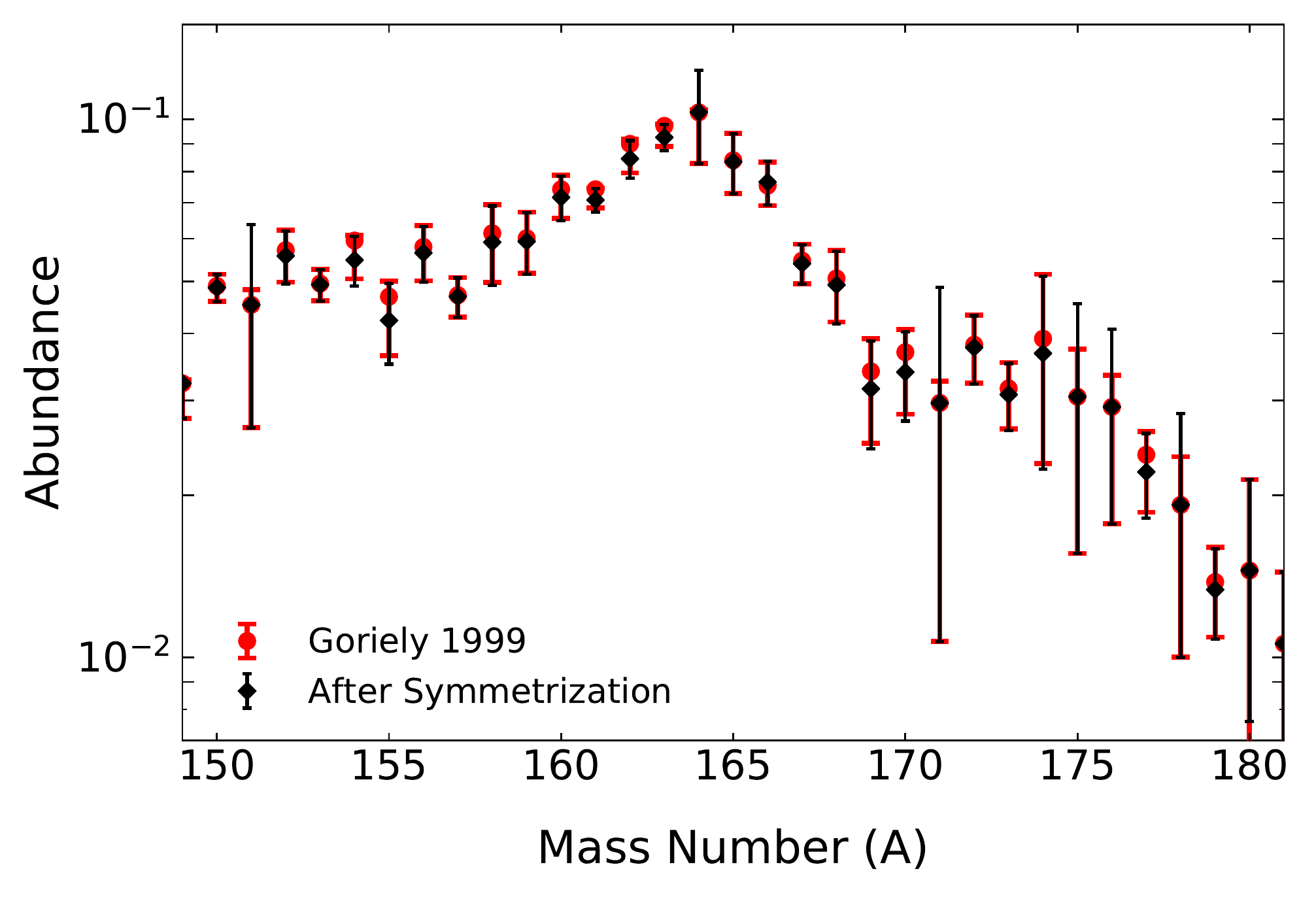}
\end{center}
\caption{Comparison between the $r$-process solar abundance residuals from \cite{goriely99} and the data applied in this work with symmetrized errors.}
\label{fig:SymmEB}
\end{figure}

Evaluations that report uncertainties for the solar $r$-process residuals are rare, and the most widely used, that of \cite{goriely99}, reports asymmetric uncertainties, $Y^{+a}_{-b}$. 
However, likelihood functions involving asymmetric errors are a rarely applied present area of study in Bayesian methods \citep{Barlow2003,Barlow2004,Marazzi2004}. Here we choose to symmetrize the solar uncertainties in order to make use of a likelihood function more consistent with established Bayesian methods. Symmetrization of asymmetric uncertainties is needed in many practical applications, as is the case with AME and NUBASE evaluations of mass and decay data \citep{AME2016,NUBASE2016}. Thus, we adopt the symmetrization approach outlined in Appendix A of the NUBASE2016 data evaluation \citep{NUBASE2016}. Rather than simply taking a midpoint and averaging the uncertainties $a$ and $b$ (as has been done with AME evaluations in the past), this method maps an asymmetric normal distribution into a symmetric normal distribution by finding the median, $m$, that divides the asymmetric distribution into two equal areas:

\begin{equation}
  m =
  \begin{cases}
    Y + \sqrt{2}a\,\text{erf}^{-1}\left(\frac{a-b}{2a}\right), & a > b \\
    Y + \sqrt{2}b\,\text{erf}^{-1}\left(\frac{a-b}{2b}\right), & b > a.
  \end{cases}
\end{equation}

\noindent where we take $\text{erf}^{-1}(z)\approx \sqrt{\pi} z/2$. The variance of this equivalent symmetric distribution centered at $m$ is given by $\sigma^2 = (1-2\pi) (a-b)^2 +ab$. Since large errors in the solar abundance data can lead to some significant differences between $m$ and the original data point $Y$, we do not apply this procedure when the percent difference before and after symmetrization is greater than $10\%$. For these cases we instead take the larger of the two errors to represent the variance and leave the center unchanged. Figure~\ref{fig:SymmEB} shows explicitly how the symmetrized solar data used in this work compare to those reported in the original evaluation.

\section{Comparison of Individual MCMC Runs to the Parallel Chains Band}

\begin{figure}[h!]%
    \centering
     \includegraphics[width=8.75cm]{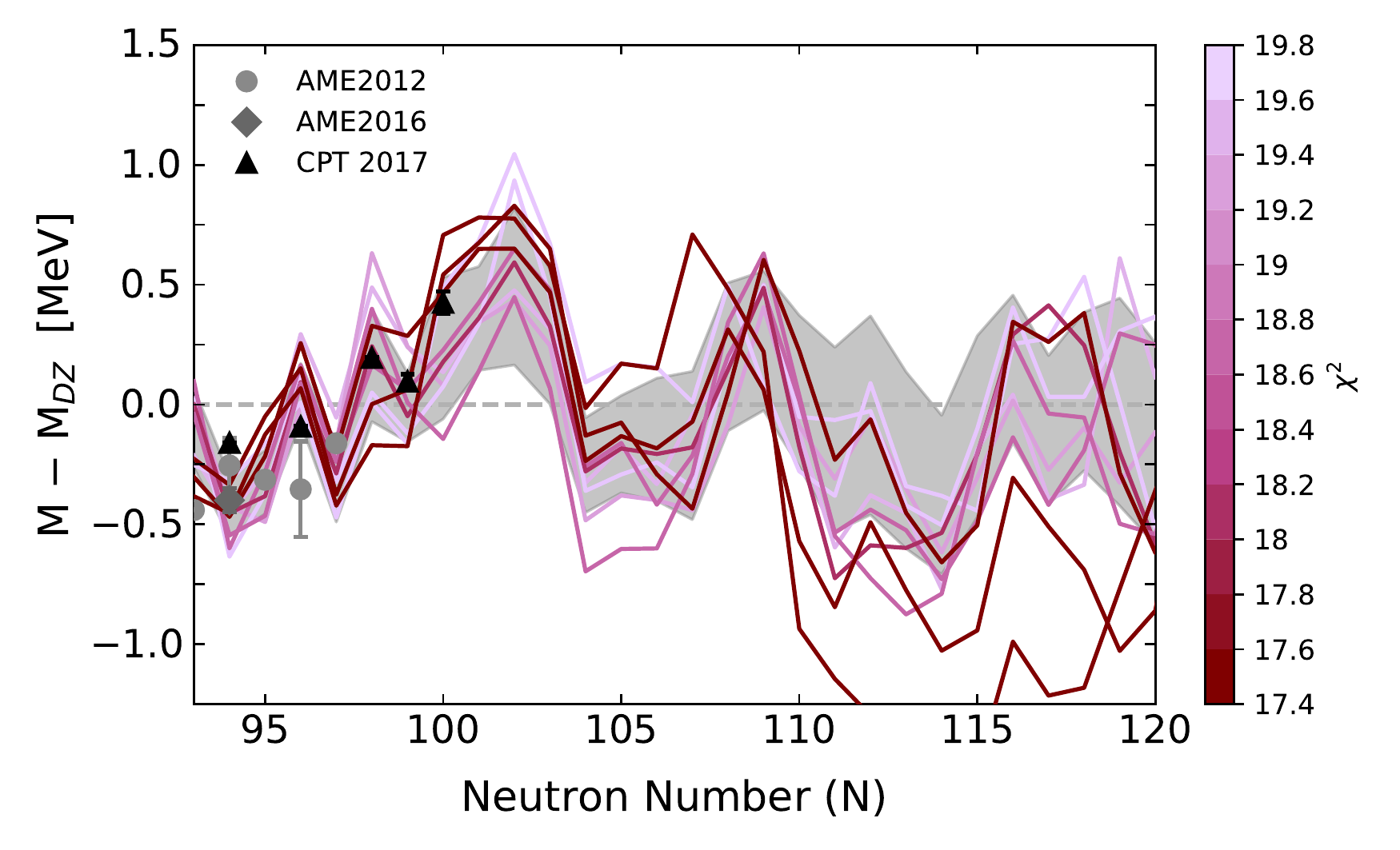} %
     \hspace{0.5cm}
    \includegraphics[width=8.75cm]{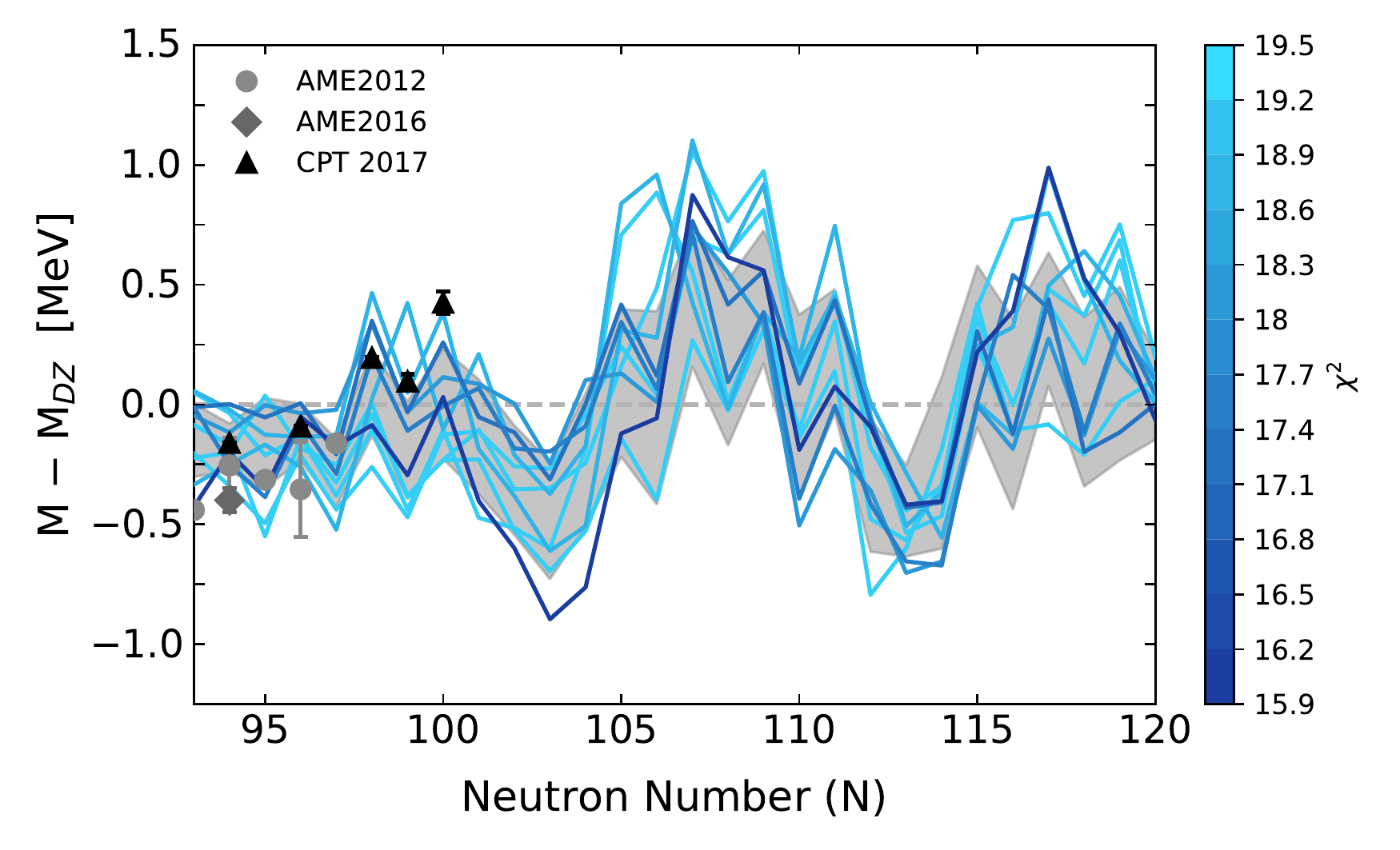} %
    \hspace{0.5cm}
    \includegraphics[width=8.75cm]{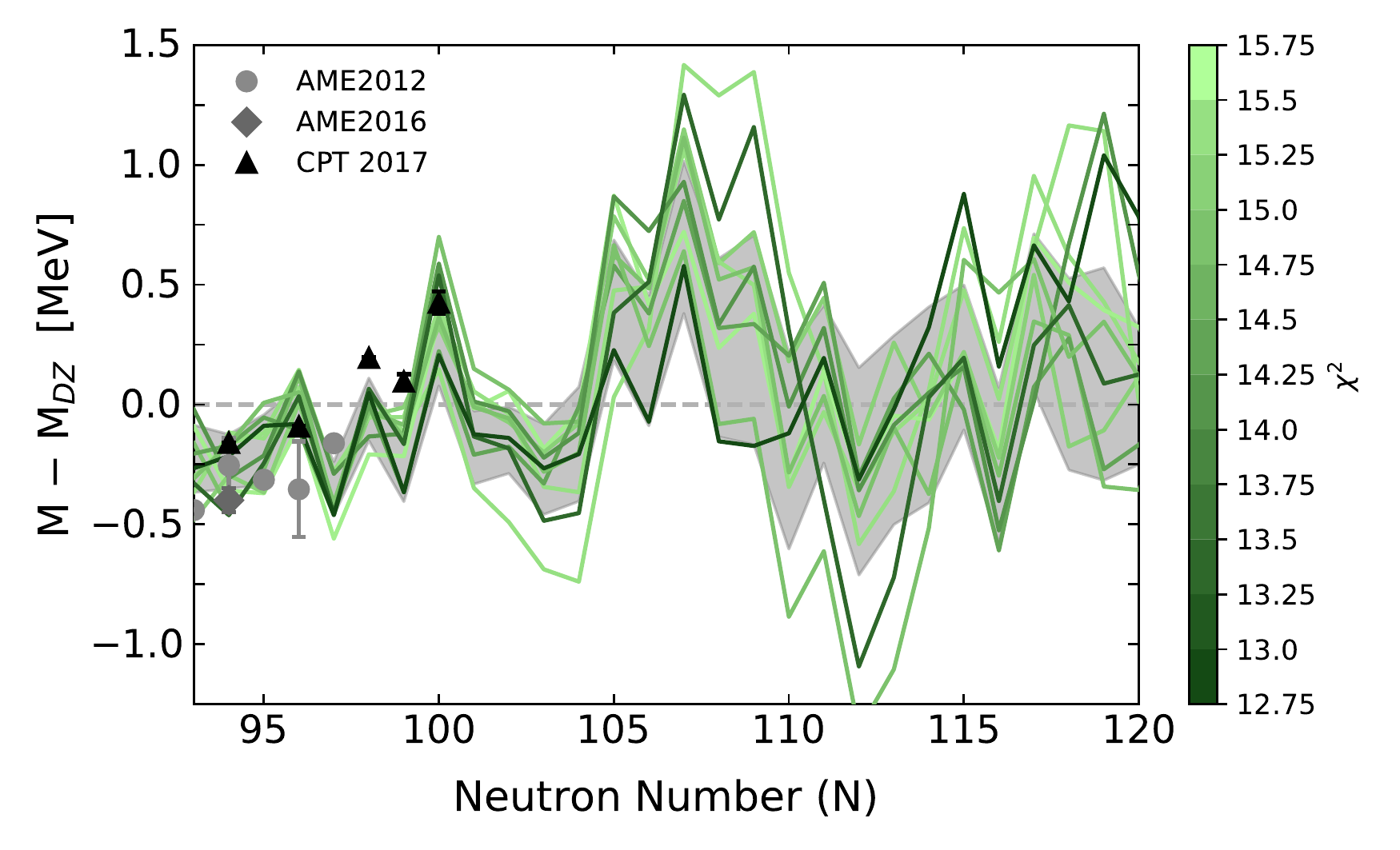} %
   \caption{The reported band given all 50 parallel chains (grey) as compared to results for the 10 runs with lowest $\chi^2$ colored as a function of $\chi^2$ for the hot (top), cold (middle), and hot/cold (bottom) astrophysical outflows considered in Sec.~\ref{sec:hotcond}, \ref{sec:coldcond}, and \ref{sec:hccond} respectively.}
\label{fig:indruns}%
\end{figure}

\begin{figure}[h!]%
    \centering
     \includegraphics[width=8.25cm]{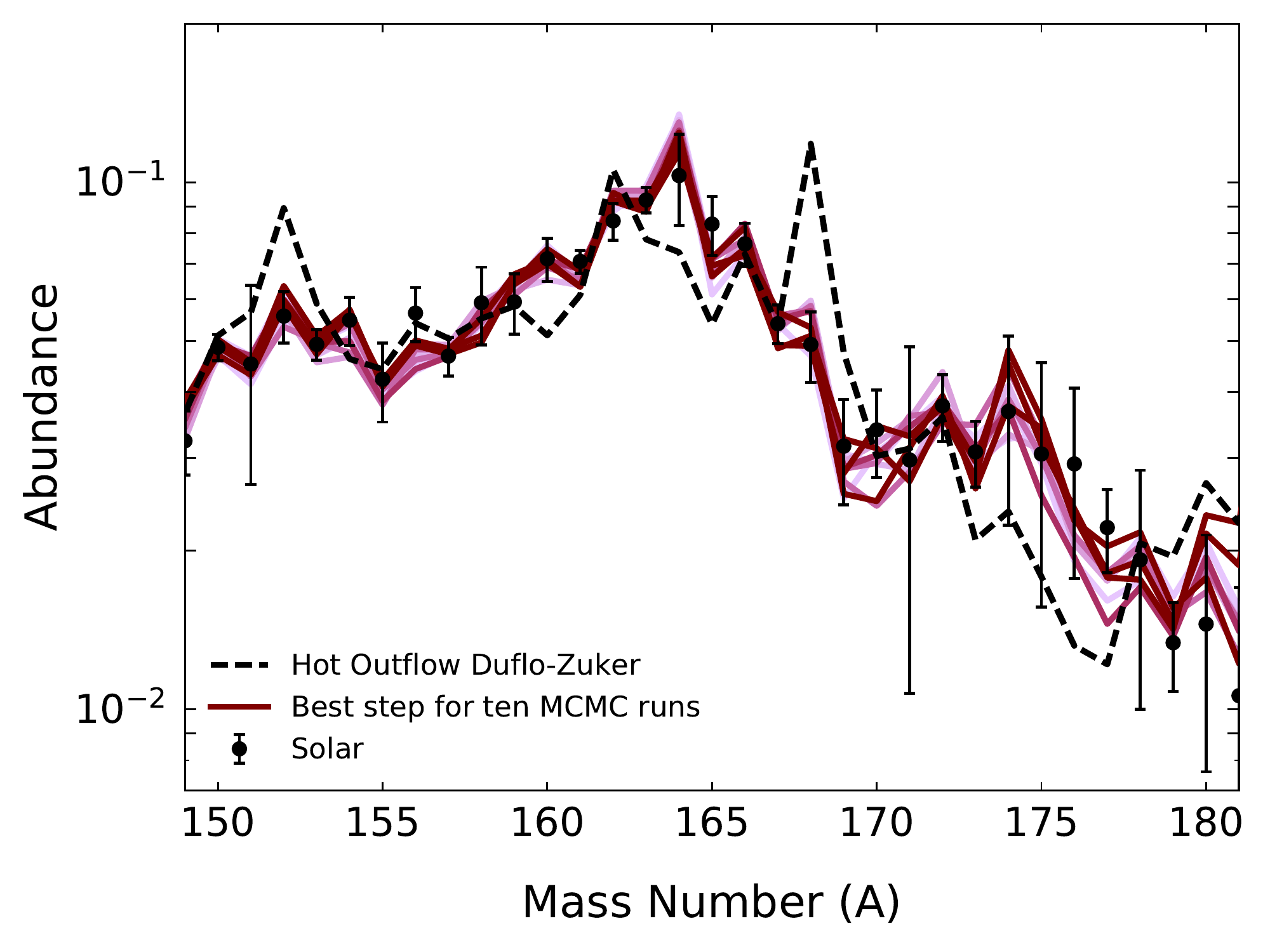} %
     \hspace{0.5cm}
    \includegraphics[width=8.25cm]{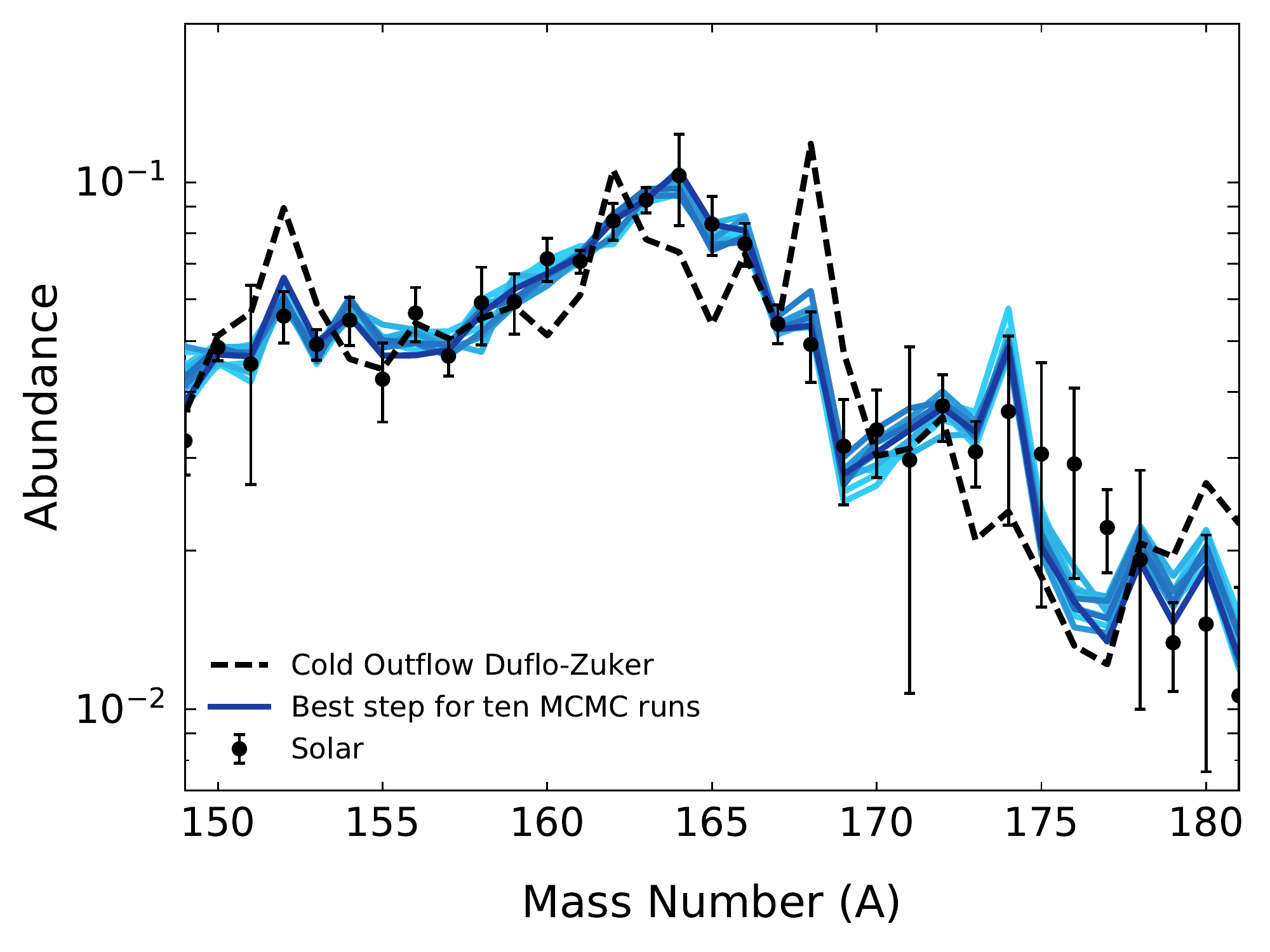} %
    \hspace{0.5cm}
    \includegraphics[width=8.25cm]{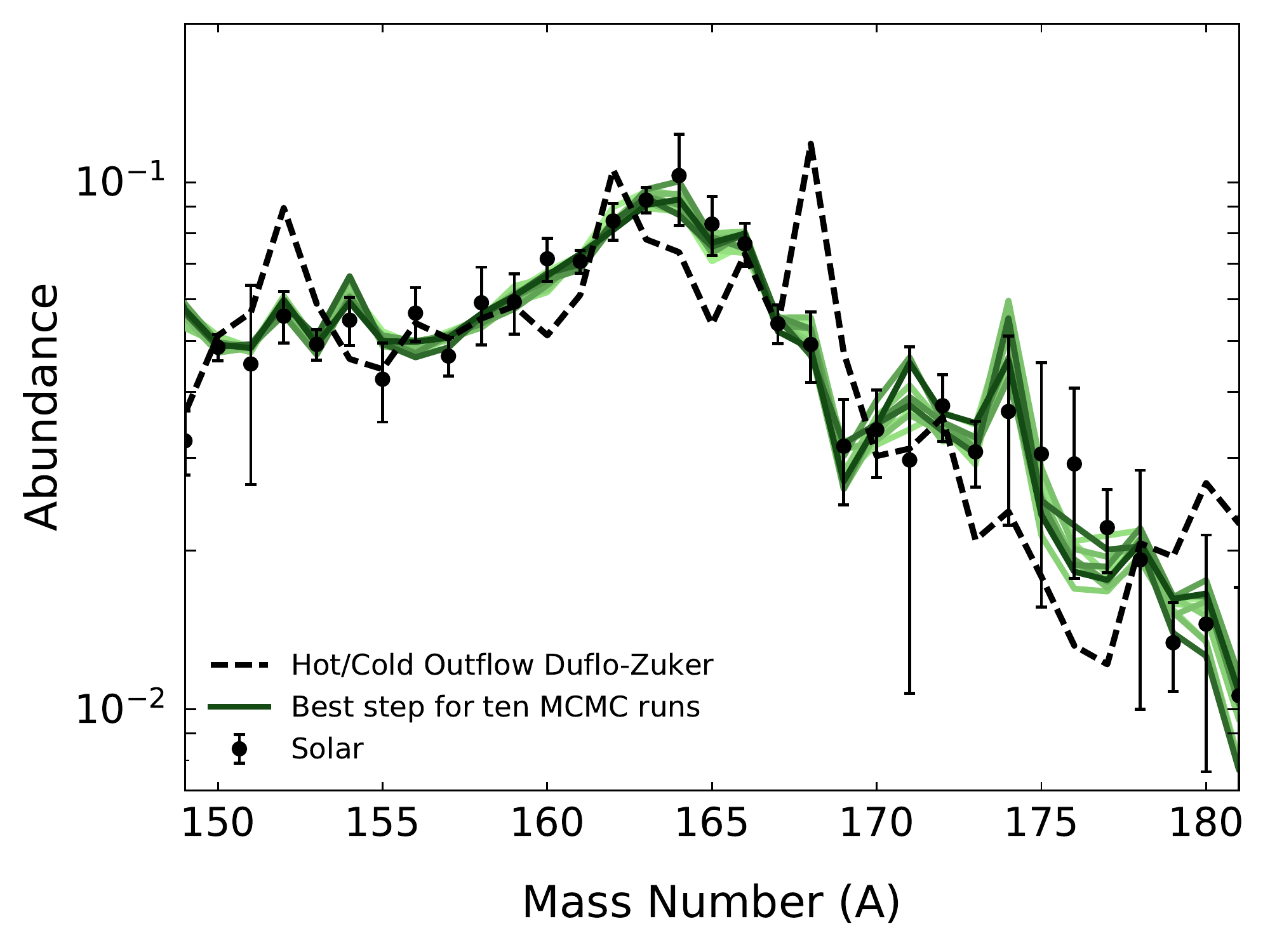} %
   \caption{The corresponding abundance patterns for the 10 MCMC runs shown in Fig.~\ref{fig:indruns} for the hot (top), cold (middle), and hot/cold (bottom) astrophysical outflows. The black dashed line shows the baseline abundances found prior to our mass adjustments when using the Duflo-Zuker mass model.}
\label{fig:abindruns}%
\end{figure}

We compare the results obtained using the parallel chains method of MCMC to the solutions with the lowest $\chi^2$ found by 10 individual runs in Fig.~\ref{fig:indruns} along with the corresponding abundances in the rare-earth region in Fig.~\ref{fig:abindruns}. This demonstrates that the parallel chains method of MCMC is best for our particular problem since there are several runs within the set of 50 with very similar $\chi^2$ values. The large sample size provided by the parallel chains method produces error bars that capture the trends that dominate our solutions. The agreement amongst runs also tends to be best for masses at neutron numbers that either are informed by experimental data or are found to be greatly influential on rare-earth peak formation. Note when considering the variation among mass predictions of individual runs that for comparison nuclear mass models often have rms differences with respect to experimental data that are greater than 0.3 MeV. Additionally we point out that since these plots consider mass corrections to the Duflo-Zuker mass model, we are actually considering a very small correction to the total nuclear binding energy, which is larger than 1 GeV for neutron-rich lanthanides. Therefore small mass differences lead to big changes in the calculated abundances shown in Fig.~\ref{fig:abindruns}. It should also be noted when comparing abundance predictions of the individual runs that the plotting scale is logarithmic.

\section{Convergence of a Parallel Chains Set and the Gelman-Rubin Metric}

\begin{figure}%
    \centering
     \includegraphics[width=8.6cm]{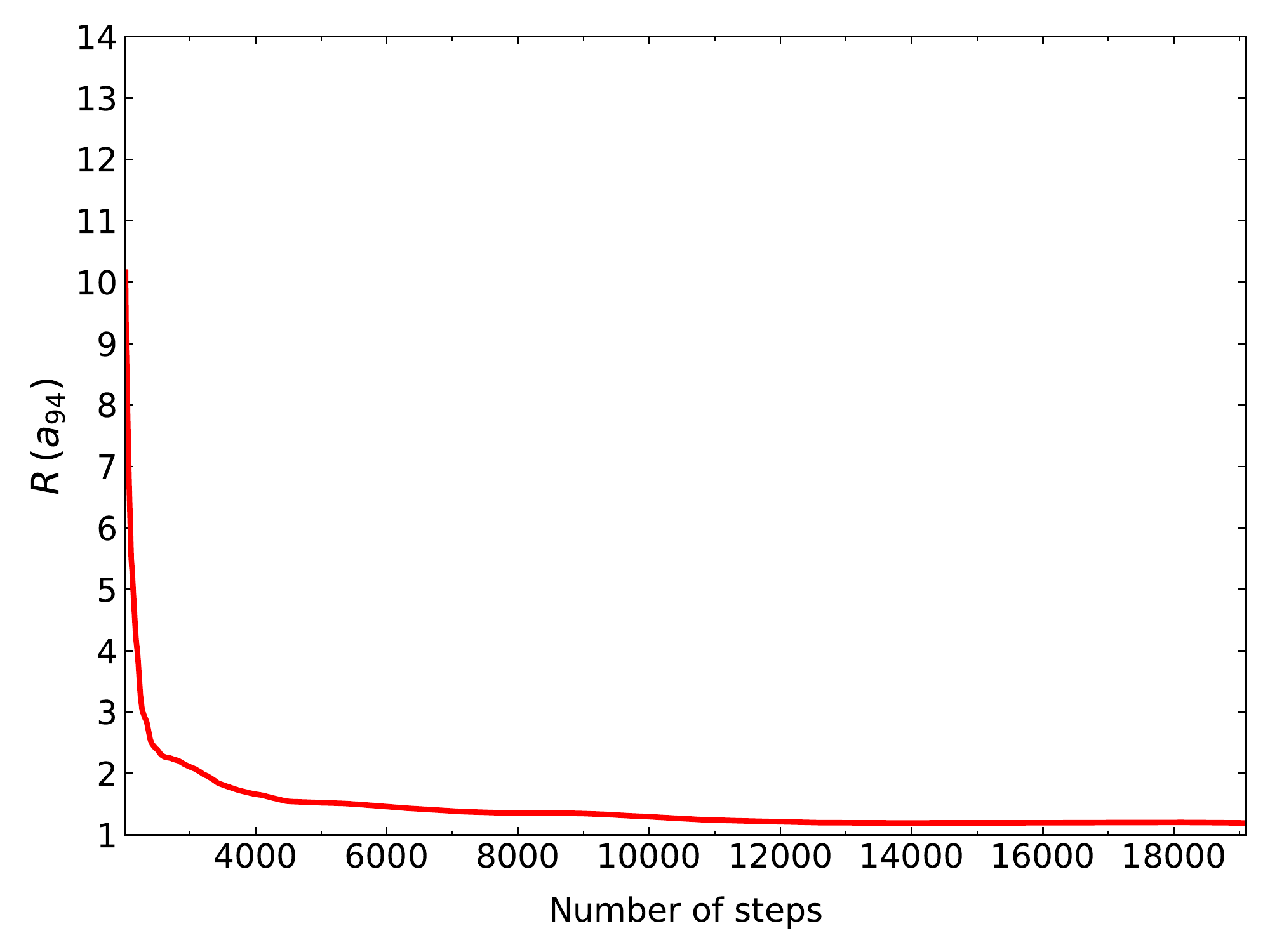}%
     \hspace{0.5cm}
    \includegraphics[width=8.6cm]{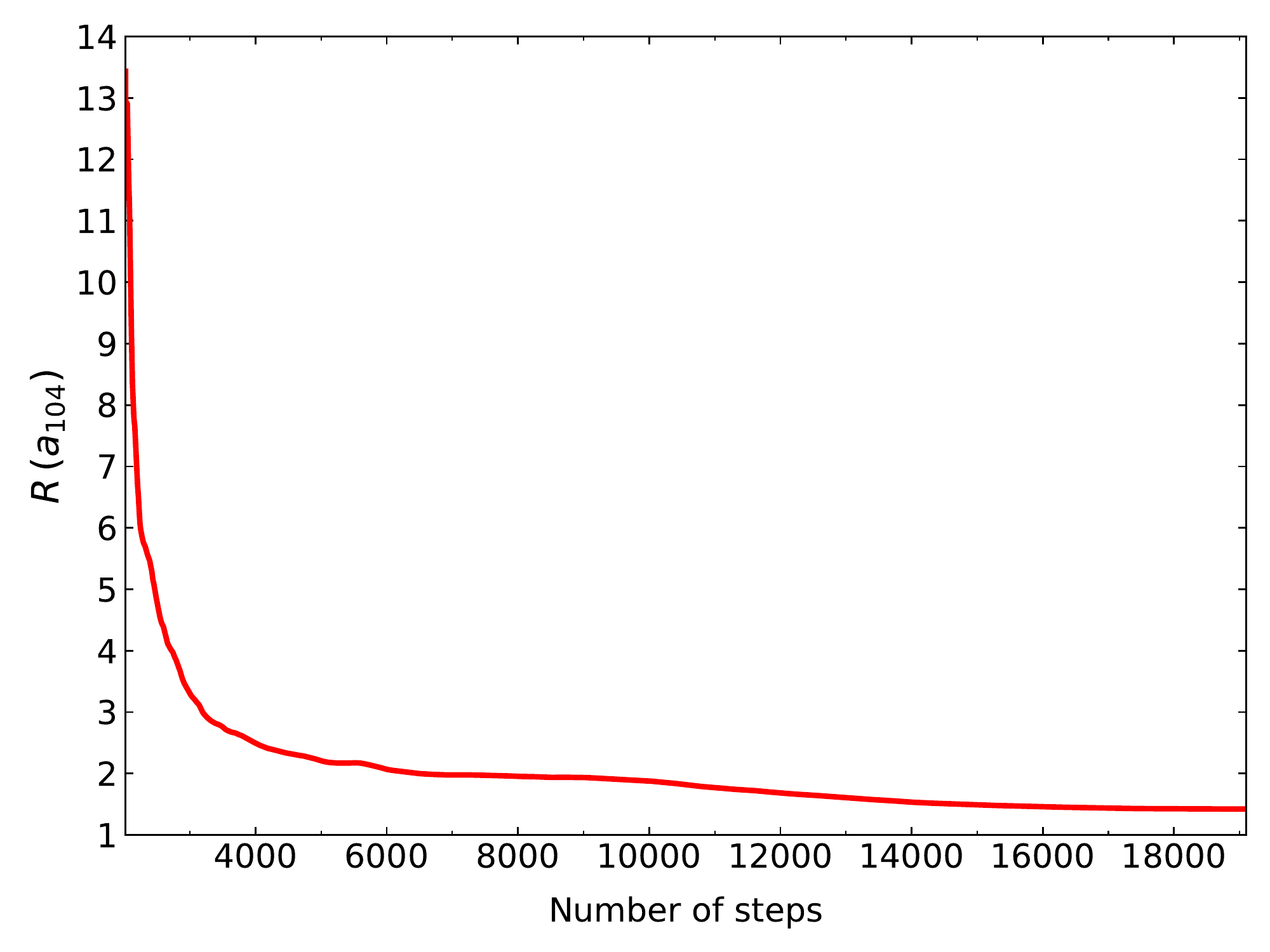}%
\caption{The Gelman-Rubin metric for our $a_{94}$ parameter (top) and $a_{104}$ parameter (bottom) as a function of time step when the 20 longest of our MCMC runs are considered.}
\label{fig:gelmanrubin}%
\end{figure}

We previously demonstrated the convergence of our parallel chains set via visual inspection of the parameter space search shown in Figure~\ref{fig:paramspace}, the convergence of error bands between 40 run and 50 run sets demonstrated in Figure~\ref{fig:buildstats}, as well as the reported acceptance rates, which verify that our chains are moving around parameter space. To consider another convergence metric, here we calculate the Gelman-Rubin measure \citep{GelmanRubin} for the case of the hot astrophysical outflow considered in the manuscript, which was used to discuss our other convergence diagnostics in Section~\ref{sec:meth}. For calculating this metric, we consider the 20 MCMC runs that have the largest number of steps, with all runs taking at least $\sim$20000 steps (we note that we checked that our reported solution with 50 runs is reproduced by this particular subset of 20). Figure ~\ref{fig:gelmanrubin} shows the results for our $a_{94}$ and $a_{104}$ parameters. Our Gelman-Rubin metrics show asymptotic-like behavior (with small downward trends near $\sim$20000 steps) and approach values close to 1. Specifically for the cases shown in Fig.~\ref{fig:gelmanrubin} we find final values of $R\,(a_{94})=1.196$ and $R\,(a_{104})=1.42$ with values of the Gelman-Rubin metric for all parameters being less than $1.95$.

Since for our particular MCMC problem we use a Hauser-Feshbach code to recalculate $\beta$-decay rates given the adjusted mass and then run a nucleosynthesis network, a given step takes $\sim20-60$ seconds. Therefore, to make the most of our computational resources, we selected a subset of 10 MCMC chains to run for longer. We found that these extended runs (all run for an additional $\sim$10000 steps and reaching $\sim$30000 steps in total) did not locate a distinctly different solution with lowest $\chi^2$ and therefore did not change our overall results. We note that if we calculate the Gelman-Rubin metric for this subset of the 10 longest runs we do see the metric further approach 1, with $R\,(a_{94})=1.179$ and $R\,(a_{104})=1.41$ and with metric values for all parameters being less than $1.88$. This small decrease in $R\,(a_{104})$ from 1.42 to 1.41 highlights the computational expense of pushing toward lower values of the Gelman-Rubin metric with our particular MCMC problem.

We note that although the literature tends to aim for values of the Gelman-Rubin metric less than $1.2$, this might not be practical for our application given the complexity of our problem, as well as the computational expense of a step, and given that our choice of parameterization was motivated by physics reasons rather than aiming for model parameters that can achieve a particularly low variance. Our parameterization is set up in a way that does not make any a priori assumptions as to whether a given parameter is influential in peak formation for a given type of astrophysical condition. Parameters such as $a_{94}$ which are indirectly informed by experimental mass data can achieve the lowest Gelman-Rubin measures while parameters such as $a_{104}$ (which control the masses of nuclei with $N=104$ neutrons) have no experimental information guiding them. For this and many other parameters it is solely the the rare-earth abundances from $A=150-180$ that provide guidance and this range of abundances is affected by the properties of many nuclei and therefore many MCMC parameters. Given this, our calculated Gelman-Rubin values are reasonable since they all approach 1.

\section{Impact of $D_n$ metric check}

We use the neutron pairing metric, $D_n (Z,N) = (-1)^{N+1}(S_n (Z,N+1) - S_n (Z,N))$, to provide the algorithm with feedback as to whether mass parameters have entered an unphysical regime. It is a useful diagnostic since it is clearly connected to nuclear structure, being influenced by odd-even effects and being largest at closed shells. Additionally, a negative value for the $D_n$ metric implies a reversal in the odd-even staggering of the one-neutron separation energies and such odd-even behavior is not supported by any nuclear physics models or experimental measurements to date.

\begin{figure}[h!]%
    \centering
     \includegraphics[width=8.4cm]{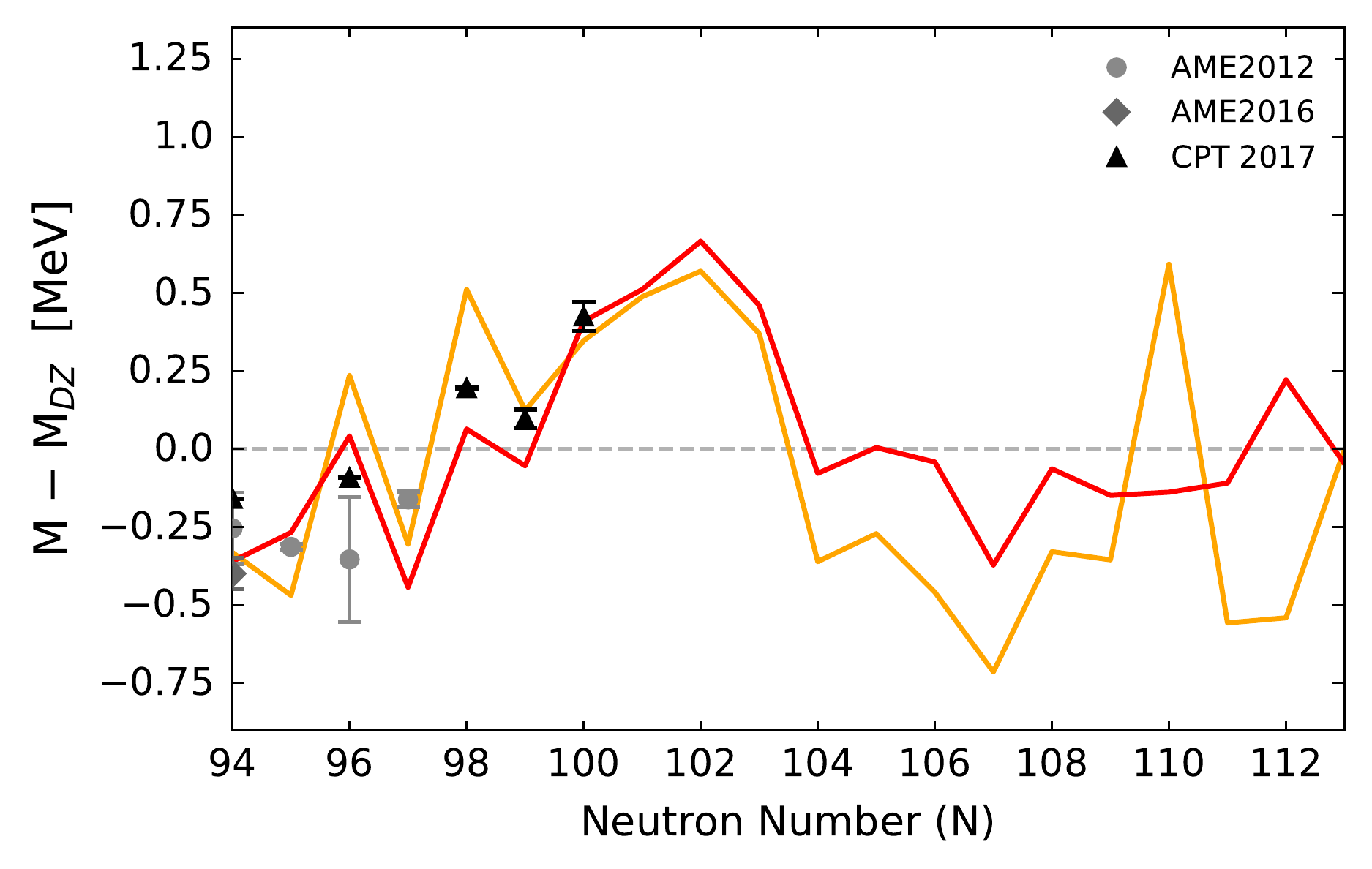} %
     \hspace{0.5cm}
    \includegraphics[width=8.3cm]{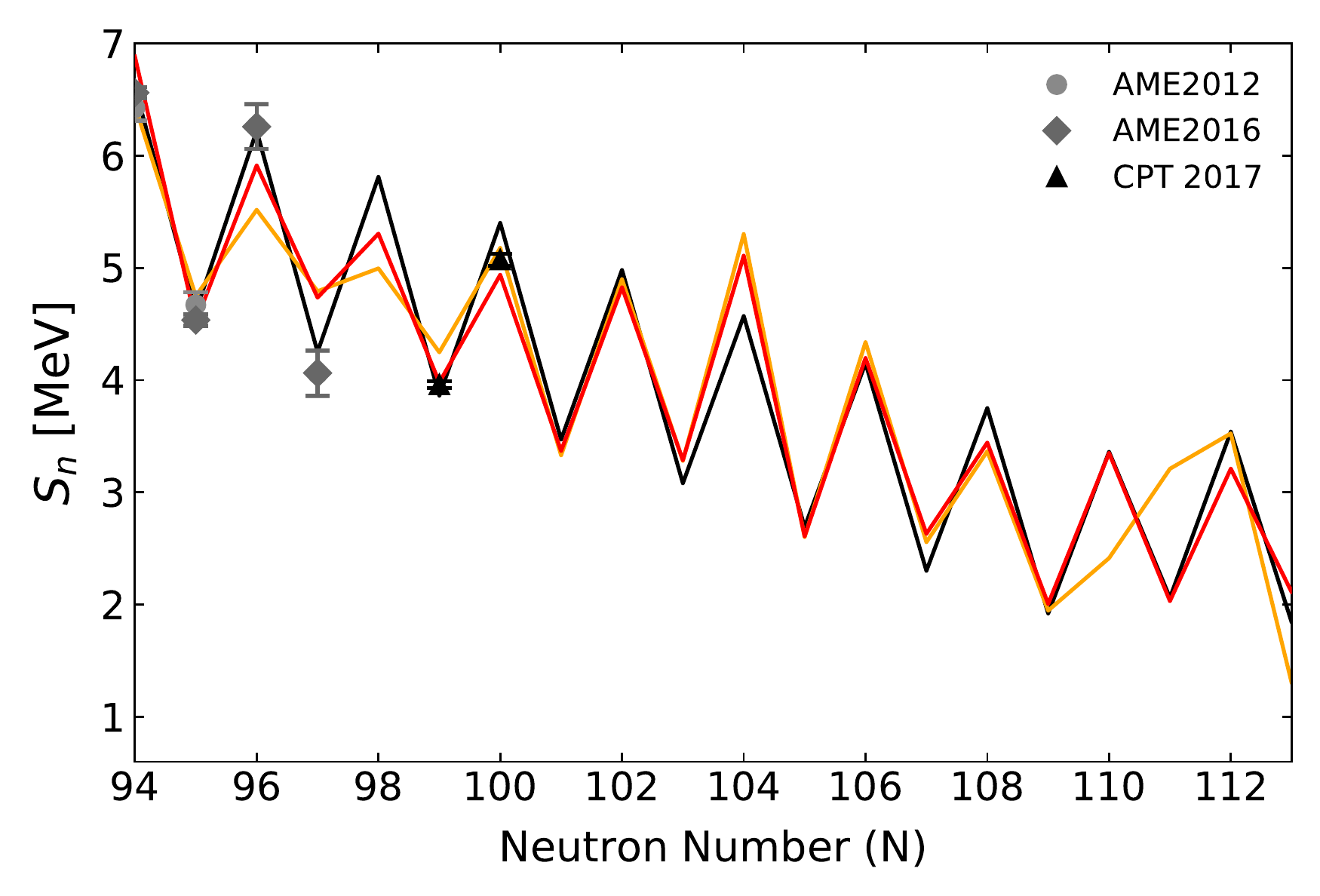} %
    \hspace{0.5cm}
    \includegraphics[width=8.4cm]{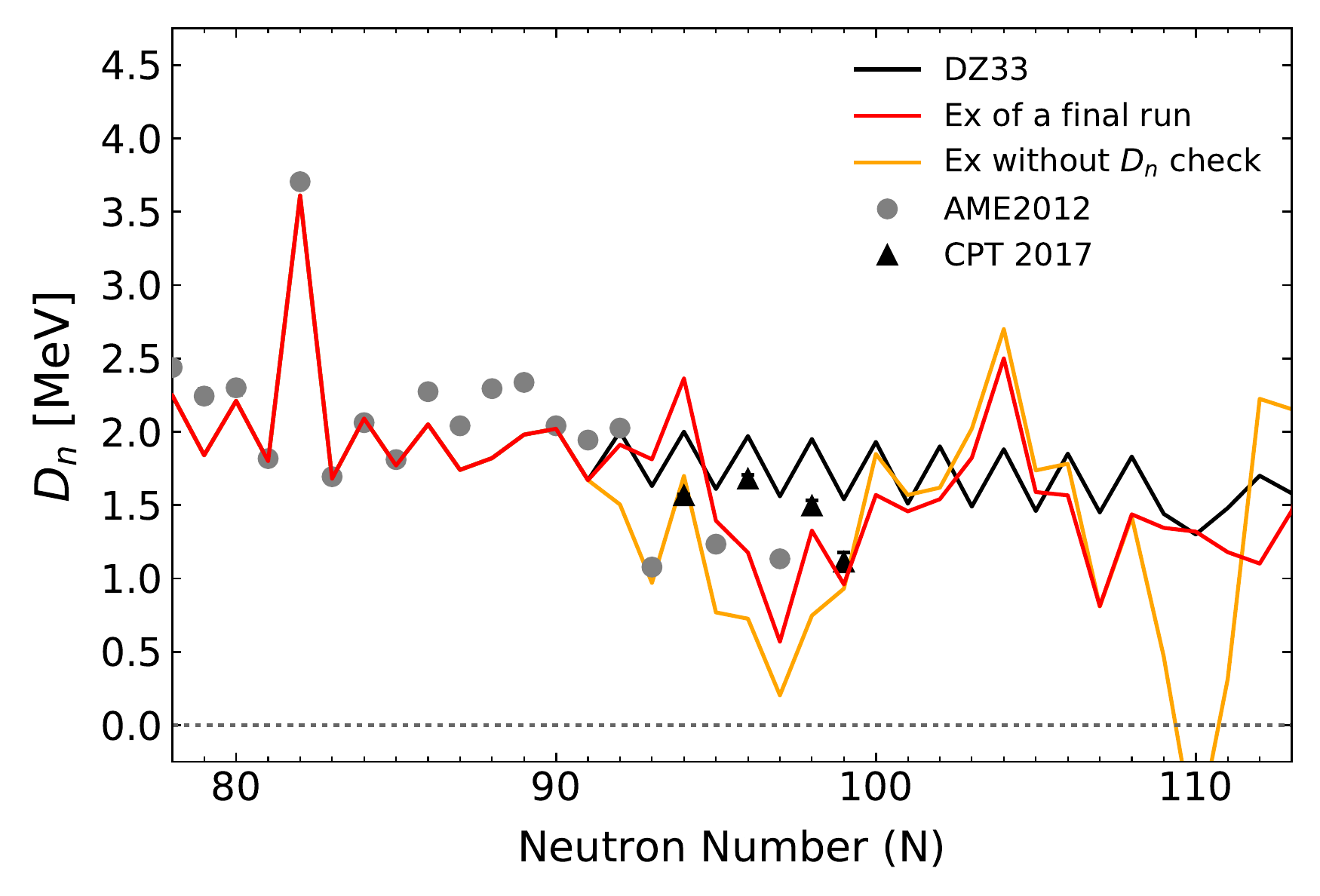} %
   \caption{The predictions for the mass surface (top), one-neutron separation energy (middle), and $D_n$ metric (bottom) prediction for an MCMC run that implemented the requirement $D_n>0$ (red) as compared to an MCMC solution found prior to implementing this check (orange). Sharp transitions in mass as is shown by the orange line produce an odd-even reversal in $S_n$ which we took to be unphysical.}
\label{fig:b4after}%
\end{figure}

In order to ensure that algorithm parameters maintain physical values, our modified likelihood function enforces that at each time step we check that $D_n>0$ before propagating the MC parameters to the nuclear rates and calculating the likelihood ratio. The $D_n$ check thus prevents computational resources from being spent in unphysical regimes. Prior to implementing this check, roughly $40-50\%$ of preliminary runs located solutions that violated the condition that $D_n$ remain positive. One such solution is shown in Figure~\ref{fig:b4after}. Solutions with $D_n <0$ are able to effectively produce pile-up in the abundances via an odd-even reversal in the one-neutron separation energy, as can be seen at $N=110$ in the figure.

In addition to requiring that $D_n$ remain positive, we implement the criterion that at neutron numbers between $N=82$ and $N=126$, the $D_n$ metric cannot be larger than the values at these closed shells. That is, with $D_{82}$ being the value of this metric given by AME2012 mass data and $D_{126}$ being the value of this metric predicted by the Duflo-Zuker mass model, at each step we check $D_n < D_{82}$,  $D_n < D_{126}$, $D_n - D_{N-1} < D_{126} - D_{125}$, and $D_n - D_{N-1} < D_{82} - D_{81}$. Such a check is performed not only for the isotopic chain at which the calculation is centered, i.e. $Z=C$, but also checked for the nearby isotopic chains $Z=C-1$, $Z=C+1$, and $Z=C+2$. These checks on the height of the $D_n$ metric were implemented following several solutions located during preliminary runs with the cold outflow. In such cold outflows where the $r$-process path lies close to the neutron dripline, the algorithm exploited the ability to effectively produce a new shell closure between $N=82$ and $N=126$, which is not supported by experimental measurements. 

\begin{figure}
\begin{center}
\includegraphics[scale=0.485]{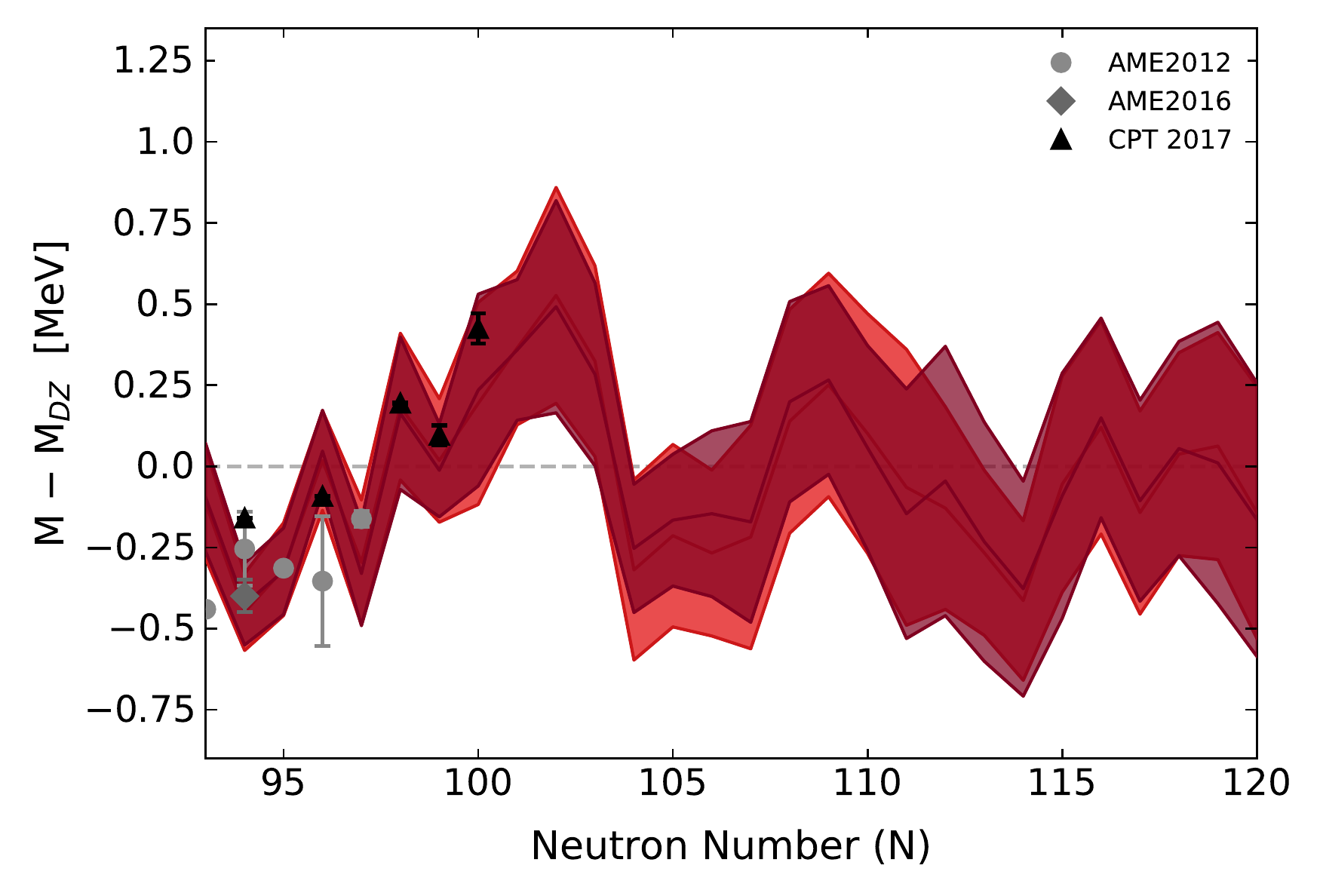}
\end{center}
\caption{Comparison between the updated result with runs that did not violate the $D_n$ height check (maroon) and the published result from \cite{OrfordVassh2018}, which required only $D_n >0$ (light red).}
\label{fig:hotoldnew}
\end{figure} 

By recursively examining our runs for the hot outflow case previously published in \cite{OrfordVassh2018}, we found that 21 of 50 runs violated the $D_n$ height check at some neutron number. However, in this case, such violations were minor and were not key features forming the rare-earth peak. To produce a full 50 run result as is presented in the main text that obeys all $D_n$ metric conditions obeyed by the cold and hot/cold cases, we collected 21 new MCMC runs with such checks in place. A comparison of our previously published result to the new result presented in this work is shown in Figure~\ref{fig:hotoldnew} which confirms that the $D_n$ height check did not significantly affect the MCMC solution we find for this hot outflow condition.

\section{Adjusting the Plotting Scale of Each Isotopic Chain}\label{sec:datatrends}

\begin{figure*}
\begin{center}
\includegraphics[scale=0.6]{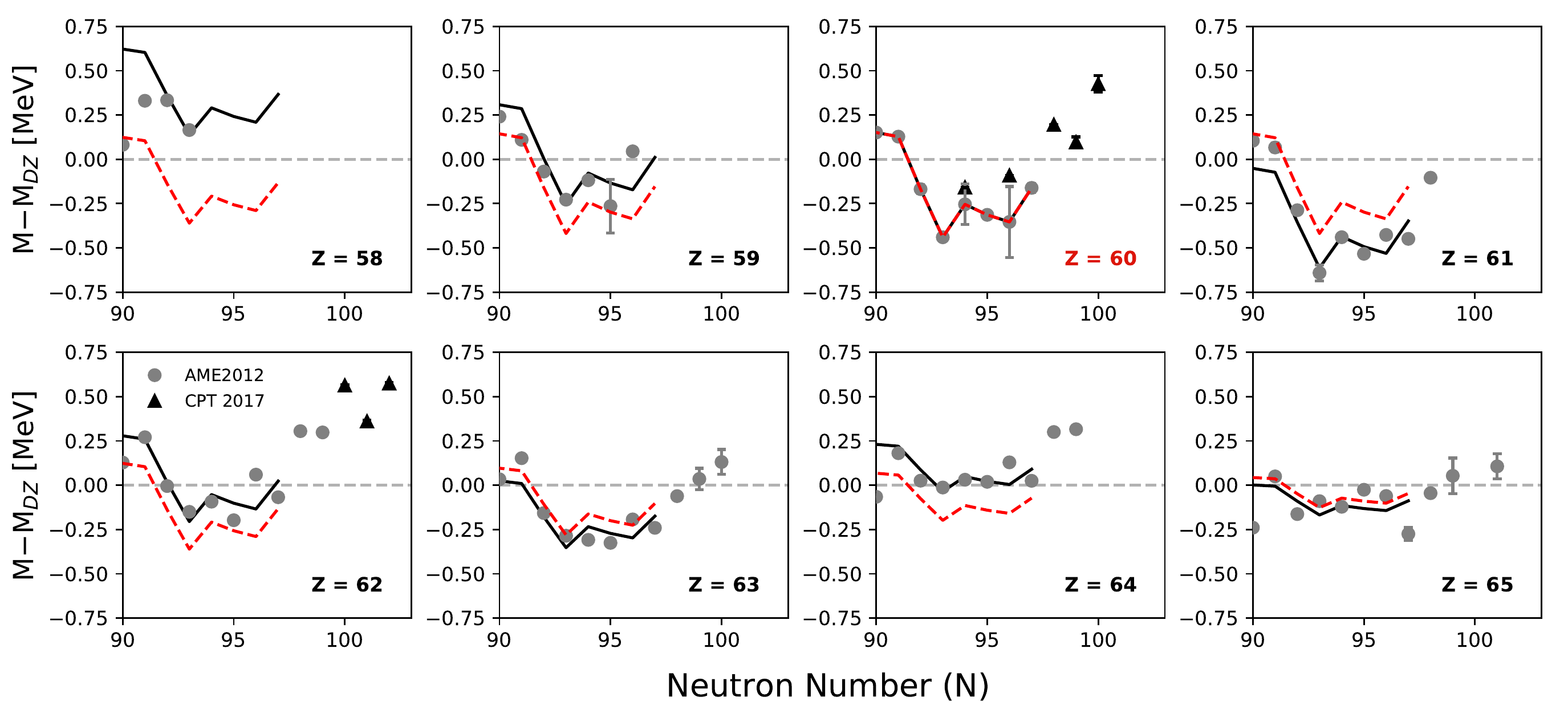}
\end{center}
\caption{The mass values for nearby isotopic chains determined by our mass parameterization formula using solely the AME2012 data for neodymium ($Z=60$) (red dashed line). We use the differences between the average of the overall trend from parameterization values and experimental values when adjusting the scale of mass predictions (as shown by the black solid line).}
\label{fig:getShifty}
\end{figure*} 

Because relative differences in masses between isotopes are most important for rare-earth peak formation, the parameterization we use is able to find the relevant trends in the mass surface. However, for an absolute mass scale, our MCMC procedure can only predict the values for the isotopic chain at which we center the calculation, typically neodymium ($Z=60$). Therefore, when we display our mass predictions for other isotopic chains, we pin the predicted trend to AME2012 data. To do so we add a correction term to the parameterization that was applied by the Monte Carlo and use the modified version of

\begin{equation}
M(Z,N) = M_{DZ}(Z,N) + a_N e^{-(Z-C)^2/2f} + \delta(Z).
\end{equation}

\noindent In order to determine $\delta(Z)$, we use AME2012 data to evaluate how well the parameterization fits with this experimental data. When centered at $C=60$, we set $\delta(60)=0$ to find the $a_N$ values, $a_{N,exp}$, at which $M_{DZ}(60,N)+a_{N,exp} =  M_{AME12}(60,N)$. We find  $a_{90,exp}=0.151$ MeV, $a_{91,exp}=0.128$ MeV, $a_{92,exp}=-0.169$ MeV, $a_{93,exp}=-0.440$ MeV, $a_{94,exp}=-0.254$ MeV, $a_{95,exp}=-0.314$ MeV, $a_{96,exp}=-0.354$ MeV, and $a_{97,exp}=-0.162$ MeV. The overall adjustment to the mass of each isotopic chain, $\delta(Z)$, is then based on the average difference between AME2012 and the prediction from the parameterization, that is $\delta (Z)= \left<M_{AME12}(Z,N)- (M_{DZ}(Z,N)+a_{N,exp}e^{-(Z-C)^2/2f})\right>$. Since we model $N=93$ and above, we base our adjustments on neodymium data for $N\ge92$ which gives $\delta(58)= 0.498$ MeV, $\delta(59)= 0.164$ MeV, $\delta(60)= 0.0$, $\delta(61)= -0.195$ MeV, $\delta(62)= 0.155$ MeV, $\delta(63)= -0.072$ MeV, $\delta(64)= 0.163$ MeV, and $\delta(65)= -0.042$ MeV. This procedure is illustrated in Figure~\ref{fig:getShifty} which shows the prediction from the mass parameterization (both before and after considering $\delta(Z)$ adjustments) for multiple isotopic chains given the AME2012 mass data for neodymium. We note that we have checked that applying masses that include $\delta(Z)$ in our $r$-process calculations shows the MCMC solution found to still produce a rare-earth peak. 

\bibliographystyle{yahapj}
\bibliography{REMMrefs}{}

\begin{thebibliography}{}
\providecommand\natexlab[1]{#1}
\providecommand\JournalTitle[1]{#1}

\bibitem[{{Abbott} {et~al.}(2017){Abbott}, {Abbott}, {Abbott}, \&
  et~al.}]{AbbottApJL}
{Abbott}, B.~P., {Abbott}, R., {Abbott}, T.~D., \& et~al. 2017,
  \href{http://dx.doi.org/10.3847/2041-8213/aa91c9}{\JournalTitle{ApJL}, 848,
  L12}

\bibitem[{{Abbott} {et~al.}(2018){Abbott}, {Abbott}, {Abbott}, \&
  et~al.}]{AbbottPRL}
---. 2018,
  \href{http://dx.doi.org/10.1103/PhysRevLett.121.161101}{\JournalTitle{\prl},
  121, 161101}

\bibitem[{{Abbott et al.}(2017)}]{AbbottGW170817}
{Abbott et al.}, B.~P. 2017,
  \href{http://dx.doi.org/10.1103/PhysRevLett.119.161101}{\JournalTitle{\prl},
  119, 161101}

\bibitem[{{Arnould} {et~al.}(2007){Arnould}, {Goriely}, \&
  {Takahashi}}]{Arnould07}
{Arnould}, M., {Goriely}, S., \& {Takahashi}, K. 2007,
  \href{http://dx.doi.org/10.1016/j.physrep.2007.06.002}{\JournalTitle{\physrep},
  450, 97}

\bibitem[{Audi {et~al.}(2017)Audi, Kondev, Wang, Huang, \& Naimi}]{NUBASE2016}
Audi, G., Kondev, F., Wang, M., Huang, W., \& Naimi, S. 2017,
  \href{http://stacks.iop.org/1674-1137/41/i=3/a=030001}{\JournalTitle{Chinese
  Physics C}, 41, 030001}

\bibitem[{Audi {et~al.}(2012)Audi, Wang, Wapstra, Kondev, MacCormick, Xu, \&
  Pfeiffer}]{AME2012}
Audi, G., Wang, M., Wapstra, {\relax A.H}., {et~al.} 2012,
  \href{http://stacks.iop.org/1674-1137/36/i=12/a=002}{\JournalTitle{Chinese
  Physics C}, 36, 1287}

\bibitem[{{Barat} {et~al.}(2007){Barat}, {Dautremer}, \&
  {Montagu}}]{Barat+2007}
{Barat}, E., {Dautremer}, T., \& {Montagu}, T. 2007,
  \href{http://dx.doi.org/10.1109/NSSMIC.2007.4436469}{in 2007 IEEE Nuclear
  Science Symposium Conference Record, Vol.~1}, 880

\bibitem[{{Barlow}(2003)}]{Barlow2003}
{Barlow}, R. 2003, in PHYSTAT2003: Statistical Problems in Particle Physics,
  Astrophysics, and Cosmology, ed. L.~{Lyons}, R.~{Mount}, \& R.~{Reitmeyer},
  250

\bibitem[{{Barlow}(2004)}]{Barlow2004}
{Barlow}, R. 2004, \JournalTitle{arXiv e-prints}, physics/0406120

\bibitem[{{Barnes} {et~al.}(2016){Barnes}, {Kasen}, {Wu}, \&
  {Mart{\'{\i}}nez-Pinedo}}]{Barnes+16}
{Barnes}, J., {Kasen}, D., {Wu}, M.-R., \& {Mart{\'{\i}}nez-Pinedo}, G. 2016,
  \href{http://dx.doi.org/10.3847/0004-637X/829/2/110}{\JournalTitle{ApJ}, 829,
  110}

\bibitem[{{Bartel} {et~al.}(1982){Bartel}, {Quentin}, {Brack}, {Guet}, \&
  {H{\r{a}}kansson}}]{DFTSkM}
{Bartel}, J., {Quentin}, P., {Brack}, M., {Guet}, C., \& {H{\r{a}}kansson},
  H.~B. 1982,
  \href{http://dx.doi.org/10.1016/0375-9474(82)90403-1}{\JournalTitle{\nphysa},
  386, 79}

\bibitem[{{Berg}(2004)}]{MCMCbookBerg}
{Berg}, B.~A. 2004, Markov Chain Monte Carlo Simulations and Their Statistical
  Analysis (World Scientific)

\bibitem[{{Brooks} {et~al.}(2011){Brooks}, {Gelman}, {Jones}, \&
  ~}]{MCMCHandbook}
{Brooks}, S., {Gelman}, A., {Jones}, G., \& ~, {Meng}, X.-L., eds. 2011,
  Handbook of Markov Chain Monte Carlo, Handbooks of Modern Statistical Methods
  (Chapman \& Hall / CRC Press)

\bibitem[{{Capano} {et~al.}(2020){Capano}, {Tews}, {Brown}, {Margalit}, {De},
  {Kumar}, {Brown}, {Krishnan}, \& {Reddy}}]{Capano+2020}
{Capano}, C.~D., {Tews}, I., {Brown}, S.~M., {et~al.} 2020,
  \href{http://dx.doi.org/10.1038/s41550-020-1014-6}{\JournalTitle{Nature
  Astronomy}, 4, 625}

\bibitem[{Carlson {et~al.}(2015)Carlson, Gandolfi, Pederiva, Pieper,
  Schiavilla, Schmidt, \& Wiringa}]{Carlson+2015}
Carlson, J., Gandolfi, S., Pederiva, F., {et~al.} 2015,
  \href{http://dx.doi.org/10.1103/RevModPhys.87.1067}{\JournalTitle{Reviews of
  Modern Physics}, 87, 1067}

\bibitem[{{Chabanat} {et~al.}(1998){Chabanat}, {Bonche}, {Haensel}, {Meyer}, \&
  {Schaeffer}}]{DFTSLy4}
{Chabanat}, E., {Bonche}, P., {Haensel}, P., {Meyer}, J., \& {Schaeffer}, R.
  1998,
  \href{http://dx.doi.org/10.1016/S0375-9474(98)00180-8}{\JournalTitle{\nphysa},
  635, 231}

\bibitem[{{C{\^o}t{\'e}} {et~al.}(2017){C{\^o}t{\'e}}, {O'Shea}, {Ritter},
  {Herwig}, \& {Venn}}]{BenoitMCMC}
{C{\^o}t{\'e}}, B., {O'Shea}, B.~W., {Ritter}, C., {Herwig}, F., \& {Venn},
  K.~A. 2017,
  \href{http://dx.doi.org/10.3847/1538-4357/835/2/128}{\JournalTitle{ApJ}, 835,
  128}

\bibitem[{{C{\^o}t{\'e}} {et~al.}(2018){C{\^o}t{\'e}}, {Fryer}, {Belczynski},
  {Korobkin}, {Chru{\'s}li{\'n}ska}, {Vassh}, {Mumpower}, {Lippuner},
  {Sprouse}, {Surman}, \& {Wollaeger}}]{CoteGW170817}
{C{\^o}t{\'e}}, B., {Fryer}, C.~L., {Belczynski}, K., {et~al.} 2018,
  \href{http://dx.doi.org/10.3847/1538-4357/aaad67}{\JournalTitle{ApJ}, 855,
  99}

\bibitem[{{Cowperthwaite et al.}(2017)}]{Cowperthwaite2017}
{Cowperthwaite et al.}, P.~S. 2017,
  \href{http://stacks.iop.org/2041-8205/848/i=2/a=L17}{\JournalTitle{ApJL},
  848, L17}

\bibitem[{{deBoer} {et~al.}(2014){deBoer}, {G{\"o}rres}, {Smith}, {Uberseder},
  {Wiescher}, {Kontos}, {Imbriani}, {Di Leva}, \& {Strieder}}]{deBoer+2014}
{deBoer}, R.~J., {G{\"o}rres}, J., {Smith}, K., {et~al.} 2014,
  \href{http://dx.doi.org/10.1103/PhysRevC.90.035804}{\JournalTitle{\prc}, 90,
  035804}

\bibitem[{{Drischler} {et~al.}(2020){Drischler}, {Furnstahl}, {Melendez}, \&
  {Phillips}}]{Drischler+2020}
{Drischler}, C., {Furnstahl}, R.~J., {Melendez}, J.~A., \& {Phillips}, D.~R.
  2020,
  \href{http://dx.doi.org/10.1103/PhysRevLett.125.202702}{\JournalTitle{\prl},
  125, 202702}

\bibitem[{{Duflo} \& {Zuker}(1995)}]{DufloZuker}
{Duflo}, J., \& {Zuker}, A.~P. 1995,
  \href{http://dx.doi.org/10.1103/PhysRevC.52.R23}{\JournalTitle{\prc}, 52,
  R23}

\bibitem[{{Efron}(1977)}]{EfronCoxLikelihood}
{Efron}, B. 1977,
  \href{http://dx.doi.org/10.1080/01621459.1977.10480613}{\JournalTitle{Journal
  of the American Statistical Association}, 72, 557}

\bibitem[{{Eichler} {et~al.}(2015){Eichler}, {Arcones}, {Kelic}, {Korobkin},
  {Langanke}, {Marketin}, {Martinez-Pinedo}, {Panov}, {Rauscher}, {Rosswog},
  {Winteler}, {Zinner}, \& {Thielemann}}]{Eichler15}
{Eichler}, M., {Arcones}, A., {Kelic}, A., {et~al.} 2015,
  \href{http://dx.doi.org/10.1088/0004-637X/808/1/30}{\JournalTitle{ApJ}, 808,
  30}

\bibitem[{{Fern{\'a}ndez} {et~al.}(2015){Fern{\'a}ndez}, {Kasen}, {Metzger}, \&
  {Quataert}}]{Fernandez+15}
{Fern{\'a}ndez}, R., {Kasen}, D., {Metzger}, B.~D., \& {Quataert}, E. 2015,
  \href{http://dx.doi.org/10.1093/mnras/stu2112}{\JournalTitle{\mnras}, 446,
  750}

\bibitem[{{Frebel}(2018)}]{Frebel}
{Frebel}, A. 2018,
  \href{http://dx.doi.org/10.1146/annurev-nucl-101917-021141}{\JournalTitle{Annual
  Review of Nuclear and Particle Science}, 68, 237}

\bibitem[{{Gelman} \& {Rubin}(1992)}]{GelmanRubin}
{Gelman}, A., \& {Rubin}, D.~B. 1992,
  \href{http://dx.doi.org/10.1214/ss/1177011136}{\JournalTitle{Statistical
  Science}, 7, 457}

\bibitem[{{Goodwin} {et~al.}(2019){Goodwin}, {Galloway}, {Heger}, {Cumming}, \&
  {Johnston}}]{Goodwin+2019}
{Goodwin}, A.~J., {Galloway}, D.~K., {Heger}, A., {Cumming}, A., \& {Johnston},
  Z. 2019,
  \href{http://dx.doi.org/10.1093/mnras/stz2638}{\JournalTitle{\mnras}, 490,
  2228}

\bibitem[{{Goriely}(1999)}]{goriely99}
{Goriely}, S. 1999, \JournalTitle{\aap}, 342, 881

\bibitem[{{Goriely}(2015)}]{GorielySPY}
---. 2015,
  \href{http://dx.doi.org/10.1140/epja/i2015-15022-3}{\JournalTitle{European
  Physical Journal A}, 51, 22}

\bibitem[{{Goriely} {et~al.}(2010){Goriely}, {Chamel}, \& {Pearson}}]{HFB21}
{Goriely}, S., {Chamel}, N., \& {Pearson}, J.~M. 2010,
  \href{http://dx.doi.org/10.1103/PhysRevC.82.035804}{\JournalTitle{\prc}, 82,
  035804}

\bibitem[{{Goriely} \& {Mart{\'{\i}}nez Pinedo}(2015)}]{GorielyGMPGEF}
{Goriely}, S., \& {Mart{\'{\i}}nez Pinedo}, G. 2015,
  \href{http://dx.doi.org/10.1016/j.nuclphysa.2015.07.020}{\JournalTitle{\nphysa},
  944, 158}

\bibitem[{{Gulam Razul} {et~al.}(2003){Gulam Razul}, Fitzgerald, \&
  Andrieu}]{GulamRazul+2003}
{Gulam Razul}, S., Fitzgerald, W., \& Andrieu, C. 2003,
  \href{http://dx.doi.org/https://doi.org/10.1016/S0168-9002(02)01807-7}{\JournalTitle{Nuclear
  Instruments and Methods in Physics Research Section A: Accelerators,
  Spectrometers, Detectors and Associated Equipment}, 497, 492 }

\bibitem[{{Horowitz} {et~al.}(2019){Horowitz}, {Arcones}, {C{\^o}t{\'e}},
  {Dillmann}, {Nazarewicz}, {Roederer}, {Schatz}, {Aprahamian}, {Atanasov},
  {Bauswein}, {Beers}, {Bliss}, {Brodeur}, {Clark}, {Frebel}, {Foucart},
  {Hansen}, {Just}, {Kankainen}, {McLaughlin}, {Kelly}, {Liddick}, {Lee},
  {Lippuner}, {Martin}, {Mendoza-Temis}, {Metzger}, {Mumpower}, {Perdikakis},
  {Pereira}, {O{\textquoteright}Shea}, {Reifarth}, {Rogers}, {Siegel},
  {Spyrou}, {Surman}, {Tang}, {Uesaka}, \& {Wang}}]{HorowitzRIB2018}
{Horowitz}, C.~J., {Arcones}, A., {C{\^o}t{\'e}}, B., {et~al.} 2019,
  \href{http://dx.doi.org/10.1088/1361-6471/ab0849}{\JournalTitle{Journal of
  Physics G Nuclear Physics}, 46, 083001}

\bibitem[{Iliadis {et~al.}(2016)Iliadis, Anderson, Coc, Timmes, \&
  Starrfield}]{Iliadis+2016}
Iliadis, C., Anderson, K.~S., Coc, A., Timmes, F.~X., \& Starrfield, S. 2016,
  \href{http://dx.doi.org/10.3847/0004-637x/831/1/107}{\JournalTitle{\apj},
  831, 107}

\bibitem[{{Just} {et~al.}(2015){Just}, {Bauswein}, {Pulpillo}, {Goriely}, \&
  {Janka}}]{Just+15}
{Just}, O., {Bauswein}, A., {Pulpillo}, R.~A., {Goriely}, S., \& {Janka}, H.-T.
  2015, \href{http://dx.doi.org/10.1093/mnras/stv009}{\JournalTitle{MNRAS},
  448, 541}

\bibitem[{{Kasen} {et~al.}(2017){Kasen}, {Metzger}, {Barnes}, {Quataert}, \&
  {Ramirez-Ruiz}}]{Kasen}
{Kasen}, D., {Metzger}, B., {Barnes}, J., {Quataert}, E., \& {Ramirez-Ruiz}, E.
  2017, \href{http://dx.doi.org/10.1038/nature24453}{\JournalTitle{\nat}, 551,
  80}

\bibitem[{{Kawano} {et~al.}(2016){Kawano}, {Capote}, {Hilaire}, \& {Chau
  Huu-Tai}}]{Kawano2016}
{Kawano}, T., {Capote}, R., {Hilaire}, S., \& {Chau Huu-Tai}, P. 2016,
  \href{http://dx.doi.org/10.1103/PhysRevC.94.014612}{\JournalTitle{\prc}, 94,
  014612}

\bibitem[{{Kortelainen} {et~al.}(2010){Kortelainen}, {Lesinski}, {Mor{\'e}},
  {Nazarewicz}, {Sarich}, {Schunck}, {Stoitsov}, \& {Wild}}]{DFTUNEDF0}
{Kortelainen}, M., {Lesinski}, T., {Mor{\'e}}, J., {et~al.} 2010,
  \href{http://dx.doi.org/10.1103/PhysRevC.82.024313}{\JournalTitle{\prc}, 82,
  024313}

\bibitem[{Marazzi \& Yohai(2004)}]{Marazzi2004}
Marazzi, A., \& Yohai, V.~J. 2004,
  \href{http://dx.doi.org/https://doi.org/10.1016/j.jspi.2003.06.011}{\JournalTitle{Journal
  of Statistical Planning and Inference}, 122, 271 }, contemporary Data
  Analysis: Theory and Methods in

\bibitem[{{Marti} \& {Suess}(1988)}]{MartiSuess}
{Marti}, K., \& {Suess}, H.~E. 1988,
  \href{http://dx.doi.org/10.1007/BF00793201}{\JournalTitle{Astrophysics and
  Space Science}, 144, 507}

\bibitem[{{Meeker} \& {Escobar}(1998)}]{CensTruncDataBook}
{Meeker}, W.~Q., \& {Escobar}, L.~A., eds. 1998, Statisitcal Methods for
  Reliability Data, Wiley Series in Probability and Statistics (John Wiley \&
  Sons, Inc.)

\bibitem[{{Mendoza-Temis} {et~al.}(2015){Mendoza-Temis}, {Wu}, {Langanke},
  {Mart{\'\i}nez-Pinedo}, {Bauswein}, \& {Janka}}]{Mendoza-Temis+15}
{Mendoza-Temis}, J. d.~J., {Wu}, M.-R., {Langanke}, K., {et~al.} 2015,
  \href{http://dx.doi.org/10.1103/PhysRevC.92.055805}{\JournalTitle{\prc}, 92,
  055805}

\bibitem[{{Metzger} {et~al.}(2008){Metzger}, {Thompson}, \&
  {Quataert}}]{Metzger+2008}
{Metzger}, B.~D., {Thompson}, T.~A., \& {Quataert}, E. 2008,
  \href{http://dx.doi.org/10.1086/526418}{\JournalTitle{ApJ}, 676, 1130}

\bibitem[{{Miller} {et~al.}(2020){Miller}, {Chirenti}, \&
  {Lamb}}]{MillerLamb+2020}
{Miller}, M.~C., {Chirenti}, C., \& {Lamb}, F.~K. 2020,
  \href{http://dx.doi.org/10.3847/1538-4357/ab4ef9}{\JournalTitle{ApJ}, 888,
  12}

\bibitem[{{Miller} {et~al.}(2019){Miller}, {Lamb}, {Dittmann}, {Bogdanov},
  {Arzoumanian}, {Gendreau}, {Guillot}, {Harding}, {Ho}, {Lattimer}, {Ludlam},
  {Mahmoodifar}, {Morsink}, {Ray}, {Strohmayer}, {Wood}, {Enoto}, {Foster},
  {Okajima}, {Prigozhin}, \& {Soong}}]{MillerLambNICER}
{Miller}, M.~C., {Lamb}, F.~K., {Dittmann}, A.~J., {et~al.} 2019,
  \href{http://dx.doi.org/10.3847/2041-8213/ab50c5}{\JournalTitle{ApJL}, 887,
  L24}

\bibitem[{{M{\"o}ller} {et~al.}(1997){M{\"o}ller}, {Nix}, \&
  {Kratz}}]{MollerSd0}
{M{\"o}ller}, P., {Nix}, J.~R., \& {Kratz}, K.-L. 1997,
  \href{http://dx.doi.org/10.1006/adnd.1997.0746}{\JournalTitle{Atomic Data and
  Nuclear Data Tables}, 66, 131}

\bibitem[{{M{\"o}ller} {et~al.}(2016){M{\"o}ller}, {Sierk}, {Ichikawa}, \&
  {Sagawa}}]{FRDM2012}
{M{\"o}ller}, P., {Sierk}, A.~J., {Ichikawa}, T., \& {Sagawa}, H. 2016,
  \href{http://dx.doi.org/10.1016/j.adt.2015.10.002}{\JournalTitle{At.\ Data
  Nucl.\ Data Tables}, 109, 1}

\bibitem[{{Mumpower} {et~al.}(2016{\natexlab{a}}){Mumpower}, {Kawano}, \&
  {M{\"o}ller}}]{Mumpower+16}
{Mumpower}, M.~R., {Kawano}, T., \& {M{\"o}ller}, P. 2016{\natexlab{a}},
  \href{http://dx.doi.org/10.1103/PhysRevC.94.064317}{\JournalTitle{\prc}, 94,
  064317}

\bibitem[{{Mumpower} {et~al.}(2018){Mumpower}, {Kawano}, {Sprouse}, {Vassh},
  {Holmbeck}, {Surman}, \& {M{\"o}ller}}]{BDFrp}
{Mumpower}, M.~R., {Kawano}, T., {Sprouse}, T.~M., {et~al.} 2018,
  \href{http://dx.doi.org/10.3847/1538-4357/aaeaca}{\JournalTitle{ApJ}, 869,
  14}

\bibitem[{{Mumpower} {et~al.}(2012){Mumpower}, {McLaughlin}, \&
  {Surman}}]{Matt12}
{Mumpower}, M.~R., {McLaughlin}, G.~C., \& {Surman}, R. 2012,
  \href{http://dx.doi.org/10.1103/PhysRevC.85.045801}{\JournalTitle{\prc}, 85,
  045801}

\bibitem[{{Mumpower} {et~al.}(2016{\natexlab{b}}){Mumpower}, {McLaughlin},
  {Surman}, \& {Steiner}}]{REMM1}
{Mumpower}, M.~R., {McLaughlin}, G.~C., {Surman}, R., \& {Steiner}, A.~W.
  2016{\natexlab{b}},
  \href{http://dx.doi.org/10.3847/1538-4357/833/2/282}{\JournalTitle{\apj},
  833, 282}

\bibitem[{{Mumpower} {et~al.}(2017){Mumpower}, {McLaughlin}, {Surman}, \&
  {Steiner}}]{REMM2}
---. 2017,
  \href{http://dx.doi.org/10.1088/1361-6471/44/3/034003}{\JournalTitle{Journal
  of Physics G Nuclear Physics}, 44, 034003}

\bibitem[{{Mumpower} {et~al.}(2015){Mumpower}, {Surman}, {Fang}, {Beard},
  {M{\"o}ller}, {Kawano}, \& {Aprahamian}}]{Mumpower+15}
{Mumpower}, M.~R., {Surman}, R., {Fang}, D.~L., {et~al.} 2015,
  \href{http://dx.doi.org/10.1103/PhysRevC.92.035807}{\JournalTitle{\prc}, 92,
  035807}

\bibitem[{{N{\"a}ttil{\"a}} {et~al.}(2016){N{\"a}ttil{\"a}}, {Steiner},
  {Kajava}, {Suleimanov}, \& {Poutanen}}]{NattilaSteiner+2016}
{N{\"a}ttil{\"a}}, J., {Steiner}, A.~W., {Kajava}, J.~J.~E., {Suleimanov},
  V.~F., \& {Poutanen}, J. 2016,
  \href{http://dx.doi.org/10.1051/0004-6361/201527416}{\JournalTitle{A\&A},
  591, A25}

\bibitem[{Neufcourt {et~al.}(2019)Neufcourt, Cao, Nazarewicz, Olsen, \&
  Viens}]{Neufcourt+drip}
Neufcourt, L., Cao, Y., Nazarewicz, W., Olsen, E., \& Viens, F. 2019,
  \href{http://dx.doi.org/10.1103/PhysRevLett.122.062502}{\JournalTitle{\prl},
  122, 062502}

\bibitem[{Neufcourt {et~al.}(2018)Neufcourt, Cao, Nazarewicz, \&
  Viens}]{Neufcourt+extrap}
Neufcourt, L., Cao, Y., Nazarewicz, W., \& Viens, F. 2018,
  \href{http://dx.doi.org/10.1103/PhysRevC.98.034318}{\JournalTitle{\prc}, 98,
  034318}

\bibitem[{{Orford} {et~al.}(2018){Orford}, {Vassh}, {Clark}, {McLaughlin},
  {Mumpower}, {Savard}, {Surman}, {Aprahamian}, {Buchinger}, {Burkey},
  {Gorelov}, {Hirsh}, {Klimes}, {Morgan}, {Nystrom}, \&
  {Sharma}}]{OrfordVassh2018}
{Orford}, R., {Vassh}, N., {Clark}, J.~A., {et~al.} 2018,
  \href{http://dx.doi.org/10.1103/PhysRevLett.120.262702}{\JournalTitle{\prl},
  120, 262702}

\bibitem[{Pastore {et~al.}(2018)Pastore, Baroni, Carlson, Gandolfi, Pieper,
  Schiavilla, \& Wiringa}]{Pastore+2018}
Pastore, S., Baroni, A., Carlson, J., {et~al.} 2018,
  \href{http://dx.doi.org/10.1103/PhysRevC.97.022501}{\JournalTitle{\prc}, 97,
  022501}

\bibitem[{{Perego} {et~al.}(2014){Perego}, {Rosswog}, {Cabez{\'o}n},
  {Korobkin}, {K{\"a}ppeli}, {Arcones}, \& {Liebend{\"o}rfer}}]{Perego+14}
{Perego}, A., {Rosswog}, S., {Cabez{\'o}n}, R.~M., {et~al.} 2014,
  \href{http://dx.doi.org/10.1093/mnras/stu1352}{\JournalTitle{\mnras}, 443,
  3134}

\bibitem[{Piarulli {et~al.}(2018)Piarulli, Baroni, Girlanda, Kievsky, Lovato,
  Lusk, Marcucci, Pieper, Schiavilla, Viviani, \& Wiringa}]{Piarulli+2018}
Piarulli, M., Baroni, A., Girlanda, L., {et~al.} 2018,
  \href{http://dx.doi.org/10.1103/PhysRevLett.120.052503}{\JournalTitle{\prl},
  120, 052503}

\bibitem[{{Planck Collaboration} {et~al.}(2020){Planck Collaboration},
  {Aghanim}, {Akrami}, {Ashdown}, {Aumont}, {Baccigalupi}, {Ballardini},
  {Banday}, {Barreiro}, {Bartolo}, {Basak}, {Battye}, {Benabed}, {Bernard},
  {Bersanelli}, {Bielewicz}, {Bock}, {Bond}, {Borrill}, {Bouchet}, {Boulanger},
  {Bucher}, {Burigana}, {Butler}, {Calabrese}, {Cardoso}, {Carron},
  {Challinor}, {Chiang}, {Chluba}, {Colombo}, {Combet}, {Contreras}, {Crill},
  {Cuttaia}, {de Bernardis}, {de Zotti}, {Delabrouille}, {Delouis}, {Di
  Valentino}, {Diego}, {Dor{\'e}}, {Douspis}, {Ducout}, {Dupac}, {Dusini},
  {Efstathiou}, {Elsner}, {En{\ss}lin}, {Eriksen}, {Fantaye}, {Farhang},
  {Fergusson}, {Fernandez-Cobos}, {Finelli}, {Forastieri}, {Frailis},
  {Fraisse}, {Franceschi}, {Frolov}, {Galeotta}, {Galli}, {Ganga},
  {G{\'e}nova-Santos}, {Gerbino}, {Ghosh}, {Gonz{\'a}lez-Nuevo}, {G{\'o}rski},
  {Gratton}, {Gruppuso}, {Gudmundsson}, {Hamann}, {Handley}, {Hansen},
  {Herranz}, {Hildebrandt}, {Hivon}, {Huang}, {Jaffe}, {Jones}, {Karakci},
  {Keih{\"a}nen}, {Keskitalo}, {Kiiveri}, {Kim}, {Kisner}, {Knox},
  {Krachmalnicoff}, {Kunz}, {Kurki-Suonio}, {Lagache}, {Lamarre}, {Lasenby},
  {Lattanzi}, {Lawrence}, {Le Jeune}, {Lemos}, {Lesgourgues}, {Levrier},
  {Lewis}, {Liguori}, {Lilje}, {Lilley}, {Lindholm}, {L{\'o}pez-Caniego},
  {Lubin}, {Ma}, {Mac{\'\i}as-P{\'e}rez}, {Maggio}, {Maino}, {Mandolesi},
  {Mangilli}, {Marcos-Caballero}, {Maris}, {Martin}, {Martinelli},
  {Mart{\'\i}nez-Gonz{\'a}lez}, {Matarrese}, {Mauri}, {McEwen}, {Meinhold},
  {Melchiorri}, {Mennella}, {Migliaccio}, {Millea}, {Mitra},
  {Miville-Desch{\^e}nes}, {Molinari}, {Montier}, {Morgante}, {Moss}, {Natoli},
  {N{\o}rgaard-Nielsen}, {Pagano}, {Paoletti}, {Partridge}, {Patanchon},
  {Peiris}, {Perrotta}, {Pettorino}, {Piacentini}, {Polastri}, {Polenta},
  {Puget}, {Rachen}, {Reinecke}, {Remazeilles}, {Renzi}, {Rocha}, {Rosset},
  {Roudier}, {Rubi{\~n}o-Mart{\'\i}n}, {Ruiz-Granados}, {Salvati}, {Sandri},
  {Savelainen}, {Scott}, {Shellard}, {Sirignano}, {Sirri}, {Spencer},
  {Sunyaev}, {Suur-Uski}, {Tauber}, {Tavagnacco}, {Tenti}, {Toffolatti},
  {Tomasi}, {Trombetti}, {Valenziano}, {Valiviita}, {Van Tent}, {Vibert},
  {Vielva}, {Villa}, {Vittorio}, {Wandelt}, {Wehus}, {White}, {White},
  {Zacchei}, \& {Zonca}}]{PlanckMCMC}
{Planck Collaboration}, {Aghanim}, N., {Akrami}, Y., {et~al.} 2020,
  \href{http://dx.doi.org/10.1051/0004-6361/201833910}{\JournalTitle{\aap},
  641, A6}

\bibitem[{{Raaijmakers} {et~al.}(2019){Raaijmakers}, {Riley}, {Watts}, {Greif},
  {Morsink}, {Hebeler}, {Schwenk}, {Hinderer}, {Nissanke}, {Guillot},
  {Arzoumanian}, {Bogdanov}, {Chakrabarty}, {Gendreau}, {Ho}, {Lattimer},
  {Ludlam}, \& {Wolff}}]{RaaijmakersNICER}
{Raaijmakers}, G., {Riley}, T.~E., {Watts}, A.~L., {et~al.} 2019,
  \href{http://dx.doi.org/10.3847/2041-8213/ab451a}{\JournalTitle{ApJL}, 887,
  L22}

\bibitem[{{Radice} {et~al.}(2018){Radice}, {Perego}, {Hotokezaka}, {Fromm},
  {Bernuzzi}, \& {Roberts}}]{Radice18}
{Radice}, D., {Perego}, A., {Hotokezaka}, K., {et~al.} 2018,
  \href{http://dx.doi.org/10.3847/1538-4357/aaf054}{\JournalTitle{ApJ}, 869,
  130}

\bibitem[{{Riley} {et~al.}(2019){Riley}, {Watts}, {Bogdanov}, {Ray}, {Ludlam},
  {Guillot}, {Arzoumanian}, {Baker}, {Bilous}, {Chakrabarty}, {Gendreau},
  {Harding}, {Ho}, {Lattimer}, {Morsink}, \& {Strohmayer}}]{RileyNICER}
{Riley}, T.~E., {Watts}, A.~L., {Bogdanov}, S., {et~al.} 2019,
  \href{http://dx.doi.org/10.3847/2041-8213/ab481c}{\JournalTitle{ApJL}, 887,
  L21}

\bibitem[{{Robert}(1998)}]{MCMCbookCRobert}
{Robert}, C.~P., ed. 1998, Lecture Notes in Statistics, Vol. 135,
  Discretization and MCMC Convergence Assessment (Springer)

\bibitem[{{Roberts} {et~al.}(2011){Roberts}, {Kasen}, {Lee}, \&
  {Ramirez-Ruiz}}]{RobertsWahl}
{Roberts}, L.~F., {Kasen}, D., {Lee}, W.~H., \& {Ramirez-Ruiz}, E. 2011,
  \href{http://dx.doi.org/10.1088/2041-8205/736/1/L21}{\JournalTitle{ApJL},
  736, L21}

\bibitem[{Sangaline \& Pratt(2016)}]{SangalinePratt}
Sangaline, E., \& Pratt, S. 2016,
  \href{http://dx.doi.org/10.1103/PhysRevC.93.024908}{\JournalTitle{\prc}, 93,
  024908}

\bibitem[{{Shibagaki} {et~al.}(2016){Shibagaki}, {Kajino}, {Mathews}, {Chiba},
  {Nishimura}, \& {Lorusso}}]{Shibagaki}
{Shibagaki}, S., {Kajino}, T., {Mathews}, G.~J., {et~al.} 2016,
  \href{http://dx.doi.org/10.3847/0004-637X/816/2/79}{\JournalTitle{ApJ}, 816,
  79}

\bibitem[{{Sneden} {et~al.}(2008){Sneden}, {Cowan}, \& {Gallino}}]{Sneden}
{Sneden}, C., {Cowan}, J.~J., \& {Gallino}, R. 2008,
  \href{http://dx.doi.org/10.1146/annurev.astro.46.060407.145207}{\JournalTitle{\araa},
  46, 241}

\bibitem[{{Sprouse} {et~al.}(2020){Sprouse}, {Navarro Perez}, {Surman},
  {Mumpower}, {McLaughlin}, \& {Schunck}}]{TrevorDFT}
{Sprouse}, T.~M., {Navarro Perez}, R., {Surman}, R., {et~al.} 2020,
  \href{http://dx.doi.org/10.1103/PhysRevC.101.055803}{\JournalTitle{\prc},
  101, 055803}

\bibitem[{{Steiner} {et~al.}(2018){Steiner}, {Heinke}, {Bogdanov}, {Li}, {Ho},
  {Bahramian}, \& {Han}}]{SteinerGlobClus}
{Steiner}, A.~W., {Heinke}, C.~O., {Bogdanov}, S., {et~al.} 2018,
  \href{http://dx.doi.org/10.1093/mnras/sty215}{\JournalTitle{\mnras}, 476,
  421}

\bibitem[{{Steiner} {et~al.}(2010){Steiner}, {Lattimer}, \&
  {Brown}}]{SteinerLattimer+2010}
{Steiner}, A.~W., {Lattimer}, J.~M., \& {Brown}, E.~F. 2010,
  \href{http://dx.doi.org/10.1088/0004-637X/722/1/33}{\JournalTitle{ApJ}, 722,
  33}

\bibitem[{{Strolger} {et~al.}(2020){Strolger}, {Rodney}, {Pacifici}, {Narayan},
  \& {Graur}}]{DTDSn1MCMC}
{Strolger}, L.-G., {Rodney}, S.~A., {Pacifici}, C., {Narayan}, G., \& {Graur},
  O. 2020,
  \href{http://dx.doi.org/10.3847/1538-4357/ab6a97}{\JournalTitle{ApJ}, 890,
  140}

\bibitem[{{Surman} {et~al.}(1997){Surman}, {Engel}, {Bennett}, \&
  {Meyer}}]{Reb97}
{Surman}, R., {Engel}, J., {Bennett}, J.~R., \& {Meyer}, B.~S. 1997,
  \href{http://dx.doi.org/10.1103/PhysRevLett.79.1809}{\JournalTitle{\prl}, 79,
  1809}

\bibitem[{{Surman} {et~al.}(2008){Surman}, {McLaughlin}, {Ruffert}, {Janka}, \&
  {Hix}}]{Surman+08}
{Surman}, R., {McLaughlin}, G.~C., {Ruffert}, M., {Janka}, H.~T., \& {Hix},
  W.~R. 2008, \href{http://dx.doi.org/10.1086/589507}{\JournalTitle{\apjl},
  679, L117}

\bibitem[{Utama \& Piekarewicz(2017)}]{UtamaPiekarewiczMass}
Utama, R., \& Piekarewicz, J. 2017,
  \href{http://dx.doi.org/10.1103/PhysRevC.96.044308}{\JournalTitle{\prc}, 96,
  044308}

\bibitem[{{Utama} {et~al.}(2016){Utama}, {Piekarewicz}, \&
  {Prosper}}]{UtamaPiekarewiczCrust}
{Utama}, R., {Piekarewicz}, J., \& {Prosper}, H.~B. 2016,
  \href{http://dx.doi.org/10.1103/PhysRevC.93.014311}{\JournalTitle{\prc}, 93,
  014311}

\bibitem[{Van~Schelt {et~al.}(2012)Van~Schelt, Lascar, Savard, Clark, Caldwell,
  Chaudhuri, Fallis, Greene, Levand, Li, Sharma, Sternberg, Sun, \&
  Zabransky}]{jonprc}
Van~Schelt, J., Lascar, D., Savard, G., {et~al.} 2012,
  \href{http://dx.doi.org/10.1103/PhysRevC.85.045805}{\JournalTitle{\prc}, 85,
  045805}

\bibitem[{Van~Schelt {et~al.}(2013)Van~Schelt, Lascar, Savard, Clark, Bertone,
  Caldwell, Chaudhuri, Levand, Li, Morgan, Orford, Segel, Sharma, \&
  Sternberg}]{jonprl}
---. 2013,
  \href{http://dx.doi.org/10.1103/PhysRevLett.111.061102}{\JournalTitle{\prl},
  111, 061102}

\bibitem[{{Vassh} {et~al.}(2020){Vassh}, {Mumpower}, {McLaughlin}, {Sprouse},
  \& {Surman}}]{VasshFRLDM2019}
{Vassh}, N., {Mumpower}, M.~R., {McLaughlin}, G.~C., {Sprouse}, T.~M., \&
  {Surman}, R. 2020,
  \href{http://dx.doi.org/10.3847/1538-4357/ab91a9}{\JournalTitle{\apj}, 896,
  28}

\bibitem[{{Vassh} {et~al.}(2019){Vassh}, {Vogt}, {Surman}, {Randrup},
  {Sprouse}, {Mumpower}, {Jaffke}, {Shaw}, {Holmbeck}, {Zhu}, \&
  {McLaughlin}}]{VasshGEF2019}
{Vassh}, N., {Vogt}, R., {Surman}, R., {et~al.} 2019,
  \href{http://dx.doi.org/10.1088/1361-6471/ab0bea}{\JournalTitle{Journal of
  Physics G Nuclear Physics}, 46, 065202}

\bibitem[{{Vilen} {et~al.}(2018){Vilen}, {Kelly}, {Kankainen}, {Brodeur},
  {Aprahamian}, {Canete}, {Eronen}, {Jokinen}, {Kuta}, {Moore}, {Mumpower},
  {Nesterenko}, {Penttil{\"a}}, {Pohjalainen}, {Porter}, {Rinta-Antila},
  {Surman}, {Voss}, \& {{\'n}yst{\"o}}}]{JYFLTRAP}
{Vilen}, M., {Kelly}, J.~M., {Kankainen}, A., {et~al.} 2018,
  \href{http://dx.doi.org/10.1103/PhysRevLett.120.262701}{\JournalTitle{\prl},
  120, 262701}

\bibitem[{{Villar} {et~al.}(2017){Villar}, {Guillochon}, {Berger}, {Metzger},
  {Cowperthwaite}, {Nicholl}, {Alexand er}, {Blanchard}, {Chornock},
  {Eftekhari}, {Fong}, {Margutti}, \& {Williams}}]{Villar}
{Villar}, V.~A., {Guillochon}, J., {Berger}, E., {et~al.} 2017,
  \href{http://dx.doi.org/10.3847/2041-8213/aa9c84}{\JournalTitle{ApJL}, 851,
  L21}

\bibitem[{{Wang} {et~al.}(2017){Wang}, {Audi}, {Kondev}, {Huang}, {Naimi}, \&
  {Xu}}]{AME2016}
{Wang}, M., {Audi}, G., {Kondev}, F.~G., {et~al.} 2017,
  \href{http://dx.doi.org/10.1088/1674-1137/41/3/030003}{\JournalTitle{Chinese
  Physics C}, 41, 030003}

\bibitem[{{Wu} \& {MacFadyen}(2018)}]{WuMacFadyen}
{Wu}, Y., \& {MacFadyen}, A. 2018,
  \href{http://dx.doi.org/10.3847/1538-4357/aae9de}{\JournalTitle{ApJ}, 869,
  55}

\bibitem[{{Zeng} \& {Lin}(2007)}]{ZengCensData}
{Zeng}, D., \& {Lin}, D.~Y. 2007,
  \href{http://dx.doi.org/10.1111/j.1369-7412.2007.00606.x}{\JournalTitle{Journal
  of the Royal Statistical Society: Series B (Statistical Methodology)}, 69,
  507}

\bibitem[{{Zhu} {et~al.}(2018){Zhu}, {Wollaeger}, {Vassh}, {Surman}, {Sprouse},
  {Mumpower}, {M{\"o}ller}, {McLaughlin}, {Korobkin}, {Kawano}, {Jaffke},
  {Holmbeck}, {Fryer}, {Even}, {Couture}, \& {Barnes}}]{Cfpaper}
{Zhu}, Y., {Wollaeger}, R.~T., {Vassh}, N., {et~al.} 2018,
  \href{http://dx.doi.org/10.3847/2041-8213/aad5de}{\JournalTitle{ApJL}, 863,
  L23}

\bibitem[{{Zimmerman} {et~al.}(2020){Zimmerman}, {Carson}, {Schumacher},
  {Steiner}, \& {Yagi}}]{Zimmerman2020}
{Zimmerman}, J., {Carson}, Z., {Schumacher}, K., {Steiner}, A.~W., \& {Yagi},
  K. 2020, \JournalTitle{arXiv e-prints}, arXiv:2002.03210

\end{thebibliography}

\end{document}